\documentclass[10pt]{svjour3-new}                     
\smartqed  
\usepackage{graphicx}
\usepackage{mathptmx}      
\usepackage{bbm,multicol}
\usepackage[english]{babel}
\usepackage{graphics}
\usepackage{latexsym}
\usepackage{amsmath}
\usepackage{amsfonts}
\usepackage{amssymb}
\usepackage{enumerate}
\usepackage{epsf}
\usepackage{xypic}
\newcommand{\ignore}[1]{}
\newcommand{\Proof}{\noindent {\it Proof\/}:\ }
\newcommand{\QED}{\hspace*{\fill}$\Box$}

\journalname{~~~}
\begin{document}
\title{Locality and applications to subsumption testing and interpolation 
in $\mathcal{EL}$ and some of its extensions}\thanks{This work was partly
  supported by the German Research
  Council (DFG) as part of the Transregional
  Collaborative Research Center ``Automatic
  Verification and Analysis of Complex
  Systems'' (SFB/TR 14 AVACS). See
  \texttt{www.avacs.org} for more
  information.}


\titlerunning{Locality and applications to subsumption testing}        

\author{Viorica Sofronie-Stokkermans} 


\institute{Viorica Sofronie-Stokkermans \\
              University Koblenz-Landau, Koblenz, Germany and \\
              Max-Planck-Institut f{\"u}r Informatik, Saarbr{\"u}cken, Germany \\
          \email{sofronie@uni-koblenz.de}}


\date{} 

\maketitle

\vspace{-2mm}
\begin{abstract}
In this paper we show that subsumption problems in lightweight 
description logics (such as $\mathcal{EL}$ and $\mathcal{EL}^+$) 
can be expressed as uniform 
word problems in classes of semilattices with monotone operators. 
We use possibilities of efficient local reasoning in such classes of algebras, 
to obtain uniform PTIME decision procedures for CBox subsumption in 
$\mathcal{EL}$, $\mathcal{EL}^+$ and extensions thereof. 
These locality considerations allow us to present a new family of 
(possibly many-sorted) logics 
which extend $\mathcal{EL}$ and $\mathcal{EL}^+$ with 
$n$-ary roles and/or numerical domains. 
As a by-product, this allows us to show that the algebraic 
models of  ${\cal EL}$ and ${\cal EL}^+$ have ground interpolation 
and thus that ${\cal EL}$, ${\cal EL}^+$, 
and their extensions studied in this paper have interpolation.
We also show how these ideas can be used for the description logic 
$\mathcal{EL}^{++}$.
\end{abstract}

\vspace{-2mm}
\section{Introduction}
\label{intro}
Description logics are logics for knowledge representation used in 
databases and ontologies. They provide a logical basis for modeling 
and reasoning about objects, classes of objects (concepts), and 
relationships between them (roles).  
Recently, tractable 
description logics such as $\mathcal{EL}$ \cite{Baader2003} have attracted 
much interest. 
Although they have restricted expressivity, this expressivity 
is sufficient for formalizing the type of knowledge used in 
widely used ontologies such as the medical ontology SNOMED 
\cite{snomed1,snomed2}. 
Several papers were dedicated to studying the properties of $\mathcal{EL}$ 
and its extensions 
$\mathcal{EL}^+$ \cite{Baader-2005,Baader-dl-2006} and 
$\mathcal{EL}^{++}$ \cite{Baader-ijcai-2005},   
and to understanding the limits of tractability 
in extensions of $\mathcal{EL}$. Undecidability results for  
extensions of $\mathcal{EL}$  are obtained in \cite{Baader-dl-2003}
using a reduction to the word problem for semi-Thue systems. 

In this paper we show that the subsumption problem in 
$\mathcal{EL}$ and $\mathcal{EL}^+$ can be expressed as 
a uniform word problem in certain varieties of semilattices with 
monotone operators. We identify a large class of such algebras for which
the uniform word problem is decidable in PTIME. For this, we use results
on so-called {\em local theory extensions} which we introduced in 
\cite{Sofronie-cade-05} and further developed in \cite{Sofronie-ijcar-06,Sofronie-lmcs-08,sofronie-ihlemann-ismvl-07}. In \cite{Jacobs-Sofronie-pdpar-entcs07,ihlemann-jacobs-sofronie-tacas08,sofronie-ki08} 
we proved that local theory extensions occur in a natural 
way in  verification (especially in program verification, and in the verification of parametric systems) and in mathematics.
The purpose of this paper is to show 
that the concept of local theory extension turns out to be useful also 
for identifying and studying tractable extensions of ${\cal EL}$. 
General results on local theories allow us 
to:
\begin{itemize}
\item uniformly present extensions of 
$\mathcal{EL}$ and $\mathcal{EL}^+$ with $n$-ary roles (and 
concrete domains); 
\item provide uniform complexity analysis for $\mathcal{EL}$ and $\mathcal{EL}^+$  and their extensions; 
\item analyze interpolation in the corresponding algebraic models and its consequences. 
\end{itemize}
\begin{figure}[t]
\centering
\caption{Constructors considered in this paper and their semantics}

\medskip
\begin{tabular}{|l|l|l|l|}
\hline
& Constructor name & Syntax & Semantics \\
\hline 
\hline
        & bottom & $\perp$ & $\emptyset$ \\
\cline{2-4}
        & top & $\top$ & $D$ \\
\cline{2-4}
Concept & conjunction & $C_1 \sqcap C_2$ & $C_1^{\mathcal I} \cap C_2^{\mathcal I}$ \\
\cline{2-4}
constructors & existential restriction & $\exists r.C$ & $\{ x \mid \exists y ((x,y) \in 
r^{\mathcal I} \mbox{  and } y \in  C^{\mathcal I}) \}$ \\
\cline{2-4}
& existential restriction & $\exists r.(C_1, \dots, C_n)$ & $\{ x \mid \exists y_1, \dots, y_n ((x,y_1, \dots, y_n) \in 
r^{\mathcal I}$ \\
& for $n$-ary roles & & $~~~~~~\mbox{  and } y_i \in  C_i^{\mathcal I}) \text{ for } 1 \leq i \leq n \}$ \\
\hline
\hline 
Role & existential restriction & $\exists r.(i, C)$ & $\{ (x, y_1, \dots, y_{i-1}, y_{i+1}, \dots, y_n) \mid \exists y_i$ \\
constructors & for $n$-ary roles & $2 \leq i \leq n$ & $~~~~~~~~ ((x,y_1, \dots, y_n) \in 
r^{\mathcal I} \mbox{  and } y_i \in  C^{\mathcal I})  \}$ \\[1ex]
\hline  
\end{tabular}

\bigskip
\begin{tabular}{|l|l|l|}
\hline
 Role inclusions & Syntax & Semantics \\
\hline 
Simple role inclusions  & $r \sqsubseteq s$ & $r^{\cal I} \subseteq s^{\cal I}$ \\
\cline{2-3}
& $r_1 \circ r_2 \sqsubseteq s$ & $r_1^{\cal I} \circ r_2^{\cal I} \subseteq s^{\cal I}$ \\
\cline{2-3}
& $r_1 \circ r_2 \sqsubseteq id$ & $r_1^{\cal I} \circ r_2^{\cal I} \subseteq id^{\cal I}$ \\
\hline 
Guarded role inclusions & $(r \sqsubseteq s)_C$ & $\forall x, y ~~(y \in C \wedge (x, y) \in r^{\cal I} \rightarrow (x, y) \in  s^{\cal I})$ \\
\cline{2-3}
 & $(r_1 \circ r_2 \sqsubseteq s)_C$ & $\forall x, y ~~(y \in C \wedge (x, y) \in r_1^{\cal I} \circ r_2^{\cal I} \rightarrow  (x, y) \in s^{\cal I})$ \\
 & $(r_1 \circ r_2 \sqsubseteq id)_C$ & $\forall x, y ~~(y \in C \wedge (x, y) \in r_1^{\cal I} \circ r_2^{\cal I} \rightarrow  x = y)$ ~~~~~~~~~~~~~~~~~~~~~\\
\hline 
\end{tabular}

\medskip
(Similar constructions also for $n$-ary roles, cf.\ Sect.~\ref{dl-alg-sem}.)
\label{fig-summary}
\end{figure}
The concept constructors, role constructors and role inclusions 
we can consider are summarized in Figure~\ref{fig-summary}. 
The main contributions of the paper are: 
\begin{itemize}
\item We show that
the subsumption problem in $\mathcal{EL}$ (resp.\ $\mathcal{EL}^+$)   can be 
expressed as a uniform word problem in classes of semilattices with 
monotone operators (possibly satisfying certain composition laws).
\item We show that the corresponding classes of semilattices with operators
have local presentations and we use methods for 
efficient reasoning in local theories or in local theory extensions  
in order to obtain PTIME decision procedures for  
$\mathcal{EL}$ and $\mathcal{EL}^+$. 
\item These locality considerations allow us to 
present new families of PTIME  logics with $n$-ary roles 
(and possibly also concrete domains) 
which extend $\mathcal{EL}$ and $\mathcal{EL}^+$.
\item In particular, we identify a PTIME extension of $\mathcal{EL}$ with two sorts, ${\sf concept}$ 
and ${\sf num}$, where the concepts of sort ${\sf num}$ are interpreted 
as elements in the ORD-Horn, convex fragment of Allen's interval algebra.
\item We notice that the axioms which correspond, at an algebraic 
level, to the role inclusions in ${\cal EL}^+$ are exactly of the 
type studied in the context of hierarchical interpolation 
in \cite{Sofronie-ijcar-06}. As a by-product, we thus show that the algebraic 
models of  ${\cal EL}$ and ${\cal EL}^+$ have the ground 
interpolation property and infer that ${\cal EL}$, ${\cal EL}^+$, 
and their extensions studied in this paper have interpolation.
\item We end the paper with some considerations on possibilities of 
handling ${\cal EL}^{++}$ constructors and ABoxes. 
\end{itemize}
Some of the results of this paper were reported -- in preliminary form -- 
in \cite{sofronie-dl08,sofronie-aiml08}. 
At that time we could only prove a weak locality property in the presence
of role inclusions. In this paper we considerably improve the results 
presented in \cite{sofronie-dl08,sofronie-aiml08} by showing that 
${\cal EL}$, ${\cal EL}^+$ as well as some of their extensions 
enjoy {\em the same type} of locality property, which allows to reduce, 
ultimately,  CBox subsumption checking to checking the satisfiability of 
ground clauses in the theory of partially-ordered sets. 
We thus obtain a cubic time decision procedures for CBox subsumption in   
a class of extensions of ${\cal EL}$. 
New contributions of this paper are 
also (i) the applications of our results on interpolation in local theory 
extensions \cite{Sofronie-ijcar-06,Sofronie-lmcs-08} to interpolation in 
${\cal EL}^+$ and (ii) the presentation of PTIME results in 
${\cal EL}^{++}$ in the framework of locality. 

\medskip
\noindent 
{\em Structure of the paper.} In Sect.\ \ref{dl-gen} we present 
generalities on description logic and 
introduce the description logics $\mathcal{EL}$ and $\mathcal{EL}^+$. 
In Sect.\ \ref{algebra} we provide the notions from algebra and 
correspondence theory needed in the paper. In Sect.~\ref{dl-alg-sem} 
we show that for many extensions of ${\cal EL}$ 
CBox subsumption can be expressed 
as a uniform word problem in the class of semilattices with monotone 
operators satisfying certain composition axioms. 
In Sect.~\ref{locality} we present general definitions and results on 
local theory extensions and 
in Sect.~\ref{complexity} we show that the algebraic models of 
$\mathcal{EL}$ and $\mathcal{EL}^+$ have  local 
presentations, 
thus providing 
an alternative proof of the fact that CBox subsumption in 
$\mathcal{EL}$ and $\mathcal{EL}^+$
is decidable in PTIME. 
Locality results for more general classes of semilattice with operators 
are used in Sect.~\ref{sect-extensions} for defining 
extensions of $\mathcal{EL}$ and $\mathcal{EL}^+$ with a subsumption problem 
decidable in PTIME. In Sect.~\ref{interpolation} we use these results 
for obtaining interpolation results for ${\cal EL}$ and its extensions. 
The results in Sect.~\ref{el++} show that also PTIME decidability 
of CBox subsumption in ${\cal EL}^{++}$ can be explained within the framework 
of locality.

\vspace{-2mm}
\section{Description logics: generalities}
\label{dl-gen}
The central notions in description logics are concepts and roles.
In any description logic a set $N_C$ of 
{\em concept names} and a set $N_R$ of {\em roles} is assumed to be given.
Complex concepts are defined starting with the concept names in $N_C$, 
with the help of a set of {\em concept constructors}. 
The available constructors determine the expressive power of a description
logic. 
The semantics of description logics is defined in terms of interpretations
${\mathcal I} = (D^{\mathcal I}, \cdot^{\mathcal I})$, where $D^{\mathcal I}$  
is a non-empty set, and the function 
$\cdot^{\mathcal I}$ maps each concept name $C \in N_C$ to a set 
$C^{\mathcal I} \subseteq D^{\mathcal I}$ and each 
role name $r \in N_R$ to a binary relation
$r^{\mathcal I} \subseteq D^{\mathcal I} \times D^{\mathcal I}$. 
Fig.~2 shows the constructor names used in the description logic 
${\mathcal A}{\mathcal L}{\mathcal C}$ and their semantics.
The extension of $\cdot^{\mathcal I}$ to concept descriptions is 
inductively defined using the semantics of the constructors.
\begin{figure}[t]
\centering
\caption{${\cal ALC}$ constructors and their semantics}

\bigskip
\begin{tabular}{|l|l|l|}
\hline
Constructor name & Syntax & Semantics \\
\hline
\hline
bottom & $\perp$ & $\emptyset$ \\
\hline 
top & $\top$ & $D$ \\
\hline 
negation & $\neg C$ & $D^{\mathcal I} \backslash C^{\mathcal I}$ \\
\hline
conjunction & $C_1 \sqcap C_2$ & $C_1^{\mathcal I} \cap C_2^{\mathcal I}$ \\
\hline
disjunction & $C_1 \sqcup C_2$ & $C_1^{\mathcal I} \cup C_2^{\mathcal I}$ \\
\hline
existential restriction & $\exists r.C$ & $\{ x \mid \exists y ((x,y) \in 
r^{\mathcal I} \mbox{  and } y \in  C^{\mathcal I}) \}$ \\
\hline
universal restriction & $\forall r.C$ & $\{ x \mid \forall y ((x,y) \in 
r^{\mathcal I} \longrightarrow  y \in  C^{\mathcal I}) \}$ \\
\hline
\end{tabular}
\label{table-dl-constr}
\end{figure}

\medskip
\noindent 
\begin{definition}[Terminology] A {\em terminology}\/ (or TBox, for short) is a finite 
set consisting of {\em primitive concept definitions} of the form $C \equiv D$, where $C$ is a concept name and $D$ a concept description; and 
{\em general concept inclusions} (GCI) of the form 
$C \sqsubseteq D$, where $C$ and $D$ are concept descriptions. 
\end{definition}
\begin{definition}[Interpretation] 
An interpretation ${\mathcal I}$ is a model 
of a TBox ${\mathcal T}$ if  it satisfies: 
\begin{itemize}
\item all concept definitions in 
${\mathcal T}$, i.e.\ $C^{\mathcal I} {=} D^{\mathcal I}$ for all definitions 
$C {\equiv} D \in {\mathcal T}$; 
\item all general concept inclusions in ${\mathcal T}$, i.e.\ 
$C^{\mathcal I} {\subseteq} D^{\mathcal I}$ for every 
$C {\sqsubseteq} D \in {\mathcal T}$. 
\end{itemize}
\end{definition}
Since definitions can be expressed as double inclusions, 
in what follows we will only refer to TBoxes consisting of 
general concept inclusions (GCI) only.
\begin{definition}[TBox subsumption]
Let ${\mathcal T}$ be a TBox, and $C_1, C_2$ two concept descriptions.
 $C_1$ is subsumed by $C_2$ w.r.t.\ ${\mathcal T}$ 
(for short, $C_1 \sqsubseteq_{\mathcal T} C_2$) 
if and only if $C_1^{\mathcal I} \subseteq C_2^{\mathcal I}$ for every model  
${\mathcal I}$ of ${\mathcal T}$.
\end{definition}

\vspace{-4mm}
\subsection{The description logics $\mathcal{EL}$, $\mathcal{EL}^+$ and some 
extensions} 
\label{sect-el}
By restricting the type of allowed concept constructors
less expressive but tractable description logics can be defined. 
If we only allow intersection and existential restriction
as concept constructors, we obtain the description logic 
${\mathcal E}{\mathcal L}$ \cite{Baader2003}, a logic used in 
terminological reasoning in medicine \cite{snomed1,snomed2}. 
In \cite{Baader-2005,Baader-dl-2006}, the 
extension ${\mathcal E}{\mathcal L}^+$ of ${\mathcal E}{\mathcal L}$ 
with role inclusion axioms is studied. 
Relationships between concepts and roles are described using CBoxes. 
\begin{definition}[Constraint box] 
A CBox consists of a terminology ${\mathcal T}$ 
and a set $RI$ of role inclusions 
of the form $r_1 {\circ} {\dots} {\circ} r_n \sqsubseteq s$.
Since terminologies can be expressed as sets of general 
concept inclusions,  we will view CBoxes as 
unions $GCI {\cup} RI$ of a set $GCI$ of general concept inclusions and 
a set $RI$ of role inclusions 
of the form $r_1 {\circ} {\dots} {\circ} r_n \sqsubseteq s$, with 
$n {\geq} 1$.
\end{definition}
\begin{definition}[Models of CBoxes] 
An interpretation ${\mathcal I}$ is a model 
of the CBox ${\mathcal C} = GCI \cup RI$ if it is a model of $GCI$ and 
satisfies all role inclusions in ${\mathcal C}$, i.e.\ 
$r_1^{\mathcal I} \circ \dots \circ r_n^{\mathcal I} \subseteq s^{\mathcal I}$ 
for all $r_1 \circ \dots \circ r_n \subseteq s \in RI$.
\end{definition}
\begin{definition}[CBox subsumption]
If ${\mathcal C}$ is a CBox, and $C_1, C_2$ are concept descriptions
then $C_1 \sqsubseteq_{\mathcal C} C_2$  
if and only if $C_1^{\mathcal I} \subseteq C_2^{\mathcal I}$ for every model  
${\mathcal I}$ of ${\mathcal C}$.
\end{definition}

\medskip
\noindent In \cite{Baader-2005} it was shown that 
subsumption w.r.t.\ CBoxes in $\mathcal{EL}^+$ can be reduced in 
linear time to subsumption w.r.t.\ {\em normalized}  CBoxes, 
in which all GCIs have one of the forms: 
$C \sqsubseteq D, C_1 \sqcap C_2 \sqsubseteq D, C \sqsubseteq \exists r.D,
\exists r.C \sqsubseteq D$, where $C, C_1, C_2, D$ are concept names, 
and all role inclusions are of the form 
$r \sqsubseteq s$ or $r_1 \circ r_2 \sqsubseteq r$. 
Therefore, in what follows, we  consider w.l.o.g.\ that 
CBoxes only contain role inclusions of the form 
$r \sqsubseteq s$ and $r_1 \circ r_2 \sqsubseteq r$. 

\medskip
\noindent In \cite{Baader-ijcai-2005}, 
the extension $\mathcal{EL}^{++}$ of $\mathcal{EL}^+$
is introduced. In addition to the constructions in  $\mathcal{EL}^+$, 
$\mathcal{EL}^{++}$ can be parameterized by one or more concrete 
domains ${\mathcal D}_1, \dots, {\mathcal D}_m$, which correspond to 
standard data types and permit reference to concrete data objects 
such as strings and integers. Formally, a concrete domain is a 
pair ${\cal D} = (D^{\cal D}, {\cal P}^{\cal D})$, where 
$D^{\cal D}$ is a set and ${\cal P}^{\cal D}$ is a family of 
predicate names with given (strictly positive) arity, and given 
interpretations as relations on $D^{\cal D}$. The link between 
the description logic and the concrete domains is established 
by means of a set of {\em feature names} ${\sf N}_F$, interpreted as maps 
$f : D \rightarrow D_i$, where $D$ is the universe of the interpretation 
${\cal I}$ of the description logic and $D_i$ is the universe of a 
concrete domain ${\cal D}_i$. 
TBoxes can contain constraints referring to features and 
concrete domains, of the form
$$\begin{array}{|c|c|c|}
\hline 
\text{ Name } & \text{ Syntax } & \text{ Semantics } \\
\hline 
~{\sf concrete}~ & ~p(f_1, \dots, f_n)~ & ~\{ x \in D^{\cal I} \mid \exists y_1, \dots, y_k {\in} D^{\cal D}: f^{\cal I}_i(x) {=} y_i \text{ for } 1 {\leq} i {\leq k}~  \\
~{\sf domains}~ & ~p \in {\cal P}^{\cal D}_i, f_i \in {\sf N}_F~ & ~~~~~~~~~~~~~~~~~~~~~ ~~~~~~~~~~~~~~~~~~~~~\text{ and } p^D(y_1, \dots, y_k) \}\\
\hline 
\end{array}$$  

\noindent 
In this paper we show that CBox subsumption for 
$\mathcal{EL}$ and $\mathcal{EL}^+$ 
can be expressed as a uniform 
word problem for classes of semilattices with monotone operators.
We then analyze various other types of axioms leading to 
extensions of $\mathcal{EL}$ and $\mathcal{EL}^+$, including 
a variant of $\mathcal{EL}^{++}$ without ABoxes.

\medskip 
\noindent We start by presenting the necessary notions from 
algebra.

\vspace{-2mm}
\section{Algebra: preliminaries} 
\label{algebra}
We assume known notions such as partially-ordered set and order filter/ideal
in a partially-ordered set. For further information cf.\ \cite{Priestley90}.
In what follows we will use one-sorted as well as many-sorted algebraic
structures. 

\smallskip
\noindent 
Let $\Sigma$ be a (one-sorted) signature consisting of a 
set of function symbols, together with an arity function $a : \Sigma
\rightarrow {\mathbb N}$ which associates with every function symbol 
its arity. An algebraic structure (over $\Sigma$) is a tuple 
${\cal A} = (A, \{ f_A \}_{f \in \Sigma})$, where $A$ is a non-empty set 
(the universe of ${\cal A}$) and for every $f \in \Sigma$, 
if $a(f) = n$ then $f_A : A^n \rightarrow A$. 

\smallskip
\noindent
Let $(S, \Sigma)$ be a many-sorted signature consisting of a 
set $S$ of sorts and a set $\Sigma$ of function symbols, 
together with an arity function $a : \Sigma \rightarrow (S^* \rightarrow S)$ 
which associates with every function symbol $f$ its arity 
$a(f) = s_1, \dots, s_n \rightarrow s$ (which specifies the sorts of the 
$n$ arguments of $f$ and the sort of the output). 
A (many-sorted) algebraic structure (over $(S, \Sigma)$) is a tuple 
${\cal A} = ( \{A_s\}_{s \in S}, \{ f_A \}_{f \in \Sigma})$, where 
for every $s \in S$, $A_s$ is a non-empty set 
(the universe of ${\cal A}$ of sort $s$) and for every $f \in \Sigma$, 
if $a(f) = s_1\dots s_n \rightarrow s$ then $f_A : A_{s_1} \times \dots \times
A_{s_n} \rightarrow A_s$. 

\vspace{-2mm}
\subsection{Semilattices, (distributive) lattices, Boolean algebras}

An algebraic structure  $(L, \wedge)$ consisting of 
a non-empty set $L$ together with a binary operation $\wedge$ 
is called {\em semilattice} if $\wedge$ is associative, 
commutative and idempotent. 
An algebraic  structure $(L, \vee, \wedge)$ consisting of 
a non-empty set $L$ together with two binary operations $\vee$ and $\wedge$
on $L$ is called {\em lattice} if $\vee$ and $\wedge$ are associative, 
commutative and idempotent and satisfy the absorption laws.
A {\em distributive lattice}  is a lattice 
that satisfies either of the distributive laws $(D_{\wedge})$ or $(D_{\vee})$, 
which are equivalent in a lattice. 
\begin{eqnarray*}
(D_{\wedge}) & \quad \quad \quad \forall x, y, z ~~~~ x \wedge (y \vee z) = 
(x \wedge y) \vee (x \wedge z) \\
(D_{\vee})   & \quad \quad \quad \forall x, y, z ~~~~ x \vee (y \wedge z) = 
(x \vee y) \wedge (x \vee z)
\end{eqnarray*}
In any semilattice $(L, \wedge)$ or lattice $(L, \vee, \wedge)$ 
an order can be defined in a canonic way by 
$$x \leq y \text{ if and only if } x \wedge y = x.$$ 
An element $0$ which is smaller than all other elements w.r.t.\ $\leq$
is called first element; an element $1$ which is larger 
than all other elements w.r.t.\ $\leq$ is called last element.  
A lattice having both a first and a last element is called {\em bounded}.
A Boolean algebra is a structure $(B, \vee, \wedge, \neg, 0, 1)$, 
such that $(B, \vee, \wedge, 0, 1)$ is a bounded distributive lattice 
and $\neg$ is a unary operation that satisfies: 
\begin{eqnarray*}
\vspace{-6mm}
{\sf (Complement)} & \quad \forall x & \neg x \vee x = 1  
\quad \quad \quad \forall x ~~~  \neg x \wedge x = 0 
\end{eqnarray*}

\vspace{-2mm}
\noindent Let ${\mathcal V}$ be a class of algebras. 
The {\em universal Horn theory}  of ${\mathcal V}$ is the collection of those 
closed formulae valid in  ${\mathcal V}$ which are of the form 
\begin{eqnarray}
\vspace{-4mm}
\forall x_1 \dots \forall x_n (\bigwedge_{i=1}^n s_{i1} = s_{i2} 
\rightarrow t_{1} = t_{2})
\label{horn-1}
\end{eqnarray}

\vspace{-4mm}
\noindent 
The formula~(\ref{horn-1}) above is valid in ${\mathcal V}$ if 
for each algebra ${\mathcal A} \in {\mathcal V}$ with universe $A$ 
and for each assignment $v$ of values in $A$ to 
the variables, if $v(s_{i1}) = v(s_{i2})$ for all 
$i \in \{ 1, \dots, n \}$ then 
$v(t_{1}) = v(t_{2})$.\footnote{If ${\cal A}$ is an algebra with universe $A$ 
and $v : X \rightarrow A$ an assignment, then $v$ extends in a 
canonical way to a homomorphism ${\overline v}$ from the algebra of terms 
with variables $X$ to ${\cal A}$. For every term $t$ with variables in $X$ 
we will, for the sake of simplicity, write $v(t)$ instead of 
${\overline v}(t)$.} 
The problem of deciding the validity of universal Horn sentences in a class 
${\mathcal V}$ of algebras is also called the {\em uniform word problem}
for  ${\mathcal V}$. It is known that the uniform word problem is decidable 
for the following classes of algebras:  
The class ${\sf SL}$ of semilattices (in PTIME),  
the class ${\sf DL}$ of distributive lattices (coNP-complete), and 
the class ${\sf Bool}$ of Boolean algebras (NP-complete). 

\vspace{-4mm}
\subsection{Boolean algebras with operators}
In what follows we will consider the following class of Boolean algebras 
with operators:
\begin{definition} 
Let ${\sf BAO}({\Sigma})$ be the class of Boolean algebras  
with operators in $\Sigma$, of the form 
$(B, \vee, \wedge, \neg, 0, 1, 
\{ f_B \}_{f \in \Sigma})$,  
such that for every $f \in \Sigma$ of arity $n = a(f)$, 
$f_B : B^{n} \rightarrow B$ is a join-hemimorphism, i.e.\ 

$\begin{array}{lrcl}
\forall x_1, \dots, x_i, x'_i, \dots, x_n~~~  & f(x_1, \dots, x_i \vee x'_i, \dots, x_n) & = &  f(x_1, \dots, x_i, \dots, x_n)  \vee f(x_1, \dots, x'_i, \dots, x_n)\\
\forall x_1, \dots, \dots, x_n ~~ & f(x_1,~ \dots~, ~0~ ,~ \dots~, x_n) & = & 0.
\end{array}$
\end{definition}
With every join-hemimorphism on a Boolean algebra $B$, 
$f_B : B^n \rightarrow B$ we can associate 
a map $g_B : B^n \rightarrow B$ defined for every $(x_1, \dots, x_n) \in B^n$ 
by
$ g_B(x_1, \dots, x_n) = \neg f_B(\neg x_1, \dots, \neg x_n).$
The map $g_B$ is a meet-hemimorphism in every argument, i.e.\ it satisfies, 
for every $1 \leq i \leq n$: 

$\begin{array}{lrcl}
\forall x_1, \dots, x_i, x'_i, \dots, x_n ~~~ & g(x_1, \dots, x_i \wedge x'_i, \dots, x_n) & = &  g(x_1, \dots, x_i, \dots, x_n)  \wedge g(x_1, \dots, x'_i, \dots, x_n)\\
\forall x_1, \dots, \dots, x_n  & g(x_1,~ \dots~, ~1~ ,~ \dots~, x_n) & = & 1.
\end{array}$

\noindent In relationship with ${\cal EL}$ and ${\cal EL}^+$  we will also use 
the following types of algebras:
\begin{itemize}
\item ${\sf DLO}({\Sigma})$ the class of bounded distributive 
lattices with operators  
$(L, \vee, \wedge, 0, 1, 
\{ f_L \}_{f \in \Sigma})$,  
such that 
$f_L : L^n \rightarrow L$ is a join-hemimorphism of arity $n = a(f)$;
\item ${\sf SLO}({\Sigma})$ the class of all $\wedge$-semilattices 
with operators  
$(S, \wedge, 0, 1, \{ f_S \}_{f \in \Sigma})$,  
such that 
$f_S$ is monotone and $f_S(0) = 0$.
\end{itemize}
In what follows we will denote join-hemimorphisms by $f_{\exists}$ 
and the associated meet-he\-mi\-morphisms by $f_{\forall}$. The 
reason for this notation will become clear in Section~\ref{correspondence}, 
and especially in Section~\ref{dl-alg-sem}.

\vspace{-2mm}
\subsection{Correspondence theory}
\label{correspondence} 
We now present some links between axioms satisfied in Boolean algebras 
with operators and properties of relational spaces.\footnote{Most
  calculations in the results presented here are simple; the correspondence
  results presented here could be also obtained as a consequence of 
  a general result in algebraic logic, namely Sahlqvist's theorem.} 
\begin{definition}[Duals of Boolean algebras with operators]
Let ${\bf B} = (B, \wedge, \vee, \neg, 0, 1, \{ f_{\exists}\}_{f \in \Sigma})$ 
be a Boolean algebra with operators having the property that for every 
$f \in \Sigma$, $f_{\exists} : B^{a(f)} \rightarrow B$ is a join-hemimorphism in every argument, and let 
$f_{\forall} : B^{a(f)} \rightarrow B$ be 
defined by $f_{\forall}(x_1, \dots, x_n) = \neg f_{\exists}(\neg x_1, \dots, \neg x_n)$ for every $x_i \in B^{a(f)}$ (a meet-hemimorphism in every argument).

The {\em Stone dual of ${\bf B}$} is the topological relational space  
$D({\bf B}) = ({\cal F}_p(B), \{ r_f \}_{f \in N_R}, \tau)$ 
having as support the set ${\cal F}_p(B)$ of all prime filters of 
$B$ with the Stone topology,  
and relations associated with the operators of $B$ in a canonical 
way by: 
$$r_{f}(F, F_1, \dots, F_n) \text{ iff } f_{\exists}(F_1, \dots, F_n) \subseteq F.$$
\end{definition}
\begin{definition}[Canonical extension of a Boolean algebra with operators]
The {\em canonical extension} of ${\bf B}$ is the Boolean algebra of subsets 
of the Stone dual $D({\bf B})$ of ${\bf B}$, 
 ${\cal P}(D({\bf B})) = ({\cal P}({\cal F}_p(B)), \cap, \cup, \emptyset,{\cal F}_p(B), \{ f_{\exists r_f} \}_{f \in \Sigma})$, 
where 
$$ f_{\exists r_f}(U_1, \dots, U_{n}) = \{ F \mid \exists F_1, \dots, F_n \in {\cal F}_p(B), r_f(F, F_1, \dots, F_n) \}$$
\end{definition}

\vspace{-2mm}
\subsubsection{From algebras to relational spaces}
\label{corresp-bao-rel}
We now analyze the link between properties of Boolean algebras with operators 
and properties of their duals. We focus on the properties related to the role 
inclusions considered in the study of ${\cal EL}^+$. We consider
slightly more general {\em guarded role inclusions} of the form: 
$$ \begin{array}{ll} 
\forall x & (x \in C \wedge r(x, y) \rightarrow s(x, y)) \\
\forall x, y & (x \in C \wedge r_1 \circ s(x, y) \rightarrow r_2(x, y)) \\
\forall x, y & (x \in C \wedge r_1 \circ s(x, y) \rightarrow x = y) 
\end{array}$$ 
\begin{theorem}
Let $B \in BAO(\Sigma)$, let $f, g, h \in \Sigma$ be 
unary join-hemimorphisms on $B$; and let $c$ be a constant and $C$ be the 
predicate associated in a canonical way with $c$ in $D(B)$ by 
$$C(F) \text{ iff } c \in F.$$ 
\begin{itemize}
\item[~(1)] If $B \models \forall x (x \leq c \rightarrow g(x) \leq h(x))$ 
then $D(B) \models \forall x,y (y \in C \wedge r_g(x, y)  \rightarrow r_h(x, y))$.\item[(2)] If $B \models \forall x (x \leq c \rightarrow f(g(x)) \leq h(x))$ then 
$D(B) ~\models~ \forall x, y ~(y \in C ~\wedge$ $r_f \circ r_g(x, y) \rightarrow r_h(x, y))$.
\item[(3)] If $B \models \forall x (x \leq c \rightarrow f(g(x)) \leq x)$ then 
$D(B) \models \forall x, y (y \in C \wedge r_f \circ r_g(x, y)  \rightarrow x = y)$.
\end{itemize}
\label{bao-rel-guards}
\end{theorem}
\Proof (1) Assume that 
$B \models \forall x (x \leq c \rightarrow g(x) \leq h(x))$. 
Let $F, G \in {\cal F}_p(B)$. Assume that $G \in C$ and $r_g(F, G)$. 
Then $c \in G$ and $g(G) \subseteq F$. 
We show that $h(G) \subseteq F$. Let $x \in G$. 
Then $c \wedge x \in G$. As $c \wedge x \leq x$, 
$g(c \wedge x) \leq h(c \wedge x) \leq h(x)$. Thus, $h(x) \in F$, 
i.e.\ $h(G) \subseteq F$. 
Hence, $(F, G) \in r_h$.

(2) Let $F, G \in {\cal F}_p(B)$. Assume that $G \in C$ (i.e.\ $c \in G$) 
and $(F, G) \in r_f \circ r_g$. 
Then there exists $H \in {\cal F}_p(B)$ such that $(F, H) \in r_f$ and 
$(H, G) \in r_g$, i.e.\ such that $f(H) \subseteq F$ and $g(G) \subseteq H$. 
Then $f(g(G)) \subseteq f(H) \subseteq F$. Let $x \in G$. 
Then $c \wedge x \in G$. Hence,  
$f(g(c \wedge x)) \in F$. As $c \wedge x \leq x$,
for every $x \in G$, 
$f(g(c \wedge x)) \leq h(c \wedge x) \leq h(x)$, so $h(x) \in F$. 
This shows that $h(G) \subseteq F$, i.e.\ 
$(F, G) \in r_h$. 
The proof of (3) is analogous to that of (2). \QED

\medskip
\noindent In the particular case when $c = 1$ we obtain the following correspondence result:
\begin{corollary}
Let $B \in BAO(\Sigma)$, and let $f, g, h \in \Sigma$ be unary 
join-hemimorphisms on $B$.
\begin{itemize}
\item[(1)] If $B \models g(x) \leq h(x)$ then in 
$D(B)$,  $r_g  \subseteq r_h$.
\item[(2)] If $B \models f(g(x)) \leq h(x)$ then in 
$D(B)$,  $r_f \circ r_g \subseteq r_h$.
\item[(3)] If $B \models f(g(x)) \leq x$ then in 
$D(B)$,  $r_f \circ r_g  \subseteq id$, where $id = \{ (x, x) \mid x \in {\cal F}_p(B) \}$. 
\end{itemize}
\label{bao-rel}
\end{corollary}
Analogons of Theorem~\ref{bao-rel-guards} and Corollary~\ref{bao-rel} 
can also be proved for operators with higher arity:
\begin{theorem}
Let $B \in BAO(\Sigma)$, and let $f, g, g_1, \dots, g_n, h \in \Sigma$ 
such that $f,g$ are $n$-ary,  $g_i$ are $n_i$-ary, and $h$ is an $m$-ary  
join-hemimorphism on $B$, and let
$c_i$ (resp.\ $c^i_j$) be constants and $C_i$ (resp.\ $C^i_j$) the 
predicates associated in a canonical way with $c_i$ in $D(B)$ as 
explained above. Then: 
\begin{itemize}
\item[(1)] If $n = m$ and $B \models \forall x_1, \dots, x_n (\bigwedge_i x_i \leq c_i \rightarrow g(x_1, \dots, x_n) \leq h(x_1, \dots, x_n))$ then 
\[D(B) \models \forall x, x_1, \dots, x_n (x_1 \in C_1 \wedge \dots \wedge x_n \in C_n \wedge r_g(x, x_1, \dots, x_n)    \rightarrow r_h(x, x_1, \dots, x_n)).\]
\item[(2)] If $B {\models} \forall \overline{x_1}, {\dots}, \overline{x_n} 
(\bigwedge_{i = 1}^n  x^i_1 {\leq} c^i_1 \wedge \dots \wedge x^i_{n_i} {\leq}
  c^i_{n_i} \rightarrow f(g_1({\overline x}_1), {\dots}, g_n({\overline x}_n))
  {\leq} h({\overline x}_1, {\dots}, {\overline x}_n))$ (where $\sum_{i=1}^n
  n_i = m$) then 
\[D(B) \models \forall x, \overline{x_1}, \dots, \overline{x_n} (\bigwedge_{i = 1}^n x^i_1 \in C^i_1 \wedge \dots, x^i_{n_i} \in C^i_{n_i} \wedge r_f(x, r_{g_1}({\overline x_1}), \dots, r_{g_n}(\overline{x_n})) \rightarrow r_h({\overline x}_1, \dots, {\overline x}_n)).\]
\item[(3)] If $g_i$ are unary and 
$B \models  \forall x (x \leq c \rightarrow f(g_1(x), \dots, g_n(x)) \leq x)$ 
then
\[D(B) \models \forall y (y \in C \wedge r_f(x, r_{g_1}(y), \dots, r_{g_n}(y))   \rightarrow x = y).\]
\end{itemize}
\label{bao-rel-n-guards}
\end{theorem}
\Proof The proof of (1) is similar to the proof of item (1) in 
Theorem~\ref{bao-rel-guards}.
(2) Let $F \in {\cal F}_p(B)$ and 
${\overline F}_1, \dots, {\overline F}_n$ be tuples of prime 
filters such that ${\overline F}_i$'s length corresponds to the arity 
of $g_i$. Assume that $F^i_j \in C^i_j$ (i.e.\ $c^i_j \in F^i_j$) 
and that 
$(F, {\overline F}_1, \dots, {\overline F}_n) \in r_f \circ (r_{g_1}, \dots, r_{g_n})$. Then there exist $F_1, \dots, F_n \in {\cal F}_p(B)$ such that 
$(F, F_1, \dots, F_n) \in r_f$ and $(F_i, {\overline F}_i) \in r_{g_i}$. 
Then $f(F_1, \dots, F_n) \subseteq F$ and $g_i({\overline F}_i) \subseteq F_i$.
It follows that $f(F,g_1({\overline F}_1), \dots,  g_n({\overline F}_n)) \subseteq F$. As in the proof of (2) in Theorem~\ref{bao-rel-guards} 
we can then conclude that 
$(F, {\overline F}_1, \dots, {\overline F}_n) \in r_h$.
The proof of (3) is similar. \QED

\begin{corollary}
Let $B \in BAO(\Sigma)$, and let $f, g, g_1, \dots, g_n, h \in \Sigma$ 
be such that $f, g$ are $n$-ary,  $g_i$ are $n_i$-ary, and $h$ is an $m$-ary  
join-hemimorphism on $B$. Then: 
\begin{itemize}
\item[(1)] If $n = m$ and $B \models g(\overline{x}) \leq h(\overline{x})$ 
then in $D(B)$,  $r_g  \subseteq r_h$.
\item[(2)] If $B \models f(g_1({\overline x}_1), \dots, g_n({\overline x}_n))
  \leq h({\overline x}_1, \dots, {\overline x}_1)$ (where $\sum n_i = m$) 
then in 
$D(B)$,  $$r_f \circ (r_{g_1}, \dots, r_{g_n}) \subseteq r_h.$$
\item[(3)] If $g_i$ are unary and 
$B \models f(g_1(x), \dots, g_n(x)) \leq x$ then in 
$D(B)$,  $r_f \circ (r_{g_1}, \dots, r_{g_n})  \subseteq id$, 
where $id = \{ (x, x) \mid x \in {\cal F}_p(B) \}$ is the identity relation.
\end{itemize}
\label{bao-rel-n}
\end{corollary}

\vspace{-2mm}
\subsubsection{From relational spaces to algebras}
We now consider relational spaces, i.e.\ 
structures of the form ${\bf D} = (D, \{ r_D \}_{r \in \Sigma})$, 
where $D$ is a set and for every $r \in \Sigma$, $r_D$ is a relation 
on $D$. The dual of a Boolean algebra (if we ignore the 
topology) is a relational space. 
The canonical extension associated with a Boolean algebra $B$ 
is the Boolean algebra 
$$({\cal P}({\cal F}_p(B)), \cup, \cap, \neg, \emptyset, {\cal F}_p, \{ f_{\exists r_f} \}_{f \in \Sigma})$$
of subsets of ${\cal F}_p(B)$, with operators $f_{\exists r_f}, f_{\forall r_f}$ 
defined from the relations 
$r_f$ by: 

\smallskip
$\begin{array}{lll} 
f_{\exists r_f}(U_1, \dots, U_n) & = & 
\{ x \mid \exists y_1, \dots, y_n ((x,y_1, \dots, y_n) \in r_f \mbox{  and } y_i \in U_i \text { for } 1 \leq i \leq n)\} \label{1}\\
 f_{\forall r}(U_1, \dots, U_n)  & =  & \{ x \mid \forall y_1, \dots, y_n ((x,y_1, \dots, y_n) \in r_f \Rightarrow y_i \in U_i \text { for } 1 \leq i \leq n)\}. \label{2}
\end{array}$
 
\smallskip
\noindent 
With every relational space one can associate a Boolean algebra, 
with the universe consisting of all subsets of $D$. 
\begin{theorem}
Let ${\bf D} = (D, \{ r_D \}_{r \in \Sigma})$ be a relational space, let
$C \subseteq D$ and let ${\cal P}({\bf D}) $ be the Boolean algebra with 
operators 
$({\cal P}(D), \cup, \cap, \emptyset, D, \{ f_{\exists r}\}_{r \in \Sigma})$, where $f_{\exists r}$ is as in definition~(\ref{1}) above, and $c$ be a constant symbol with interpretation $C$. 
Then the following hold: 
\begin{itemize}
\item[(1)]  If~ ${\bf D} \models \forall x, y ( y \in C \wedge r_1(x, y) \rightarrow r_2(x, y))$ then ${\cal P}({\bf D}) \models \forall x ~~(x \leq c \rightarrow f_{\exists r_1}(x) \leq f_{\exists r_2}(x)).$ 
\item[(2)] If~ ${\bf D} \models \forall x, y (y {\in} C \wedge r_1 {\circ} s(x, y) {\rightarrow} r_2(x, y))$ then ${\cal P}({\bf D}) \models \forall x ~~( x {\leq} c {\rightarrow} f_{\exists r_1}(f_{\exists s}(x)) {\leq} f_{\exists r_2}(x)).$
\item[(3)] If~ ${\bf D} \models \forall x, y (y \in C \wedge r_1 \circ s(x, y) \rightarrow x = y)$ then 
${\cal P}({\bf D}) \models \forall x (x \leq c \rightarrow f_{\exists r_1}(f_{\exists s}(x)) \leq x).$
\end{itemize}
\label{rel-to-bao-guards}
\end{theorem}
\Proof Clearly, ${\mathcal P}({\bf D}) \in {\sf BAO}(\Sigma)$. 
Let $r_1, r_2, r {\in} N_R$ and $U {\in} {\mathcal P}(D)$ with $U \subseteq C$. 

(1) Assume that 
${\bf D} \models \forall x, y ( y \in C \wedge r_1(x, y) \rightarrow r_2(x, y))$. 
Let $x \in f_{\exists r_1}(U)$. Then there exists $y \in U$ such that 
$r_1(x, y)$. As $U \subseteq C$, $y \in C$ so $r_2(x, y)$. 

(2) Assume that 
${\bf D} \models 
\forall x, y (y \in C \wedge r_1 \circ s(x, y) \rightarrow r_2(x, y))$. 
Let $x \in f_{\exists r_1 \circ s}(U)$. Then 
there exists $y \in U$ such that $(x, y) \in r_1 \circ r_2$. As before, 
 $y \in C$ so $r_2(x, y)$. The proof of (3) is similar. \QED
\begin{corollary}
Let ${\bf D} = (D, \{ r_D \}_{r \in \Sigma})$ be a relational space and let 
${\cal P}({\bf D}) $ be the Boolean algebra with operators 
$({\cal P}(D), \cup, \cap, \emptyset, D, \{ f_{\exists r}\}_{r \in \Sigma})$, where $f_{\exists r}$ is as in definition~(\ref{1}) above. The following hold: 
\begin{itemize}
\item[(1)]  If~ ${\bf D} \models r_1 \subseteq r_2$ then ${\cal P}({\bf D}) \models \forall x ~~ f_{\exists r_1}(x) \leq f_{\exists r_2}(x)$. 
\item[(2)] If~ ${\bf D} \models r_1 \circ s \subseteq r_2$ then ${\cal P}({\bf D}) \models \forall x ~~ f_{\exists r_1}(f_{\exists s}(x)) \leq f_{\exists r_2}(x)$. 
\item[(3)] If~ ${\bf D} \models r_1 \circ s \subseteq id$ then ${\cal P}({\bf D}) \models \forall x ~~ f_{\exists r_1}(f_{\exists s}(x)) \leq x$.
\end{itemize}
\label{rel-to-bao}
\end{corollary}

\noindent Similar results hold also for $n$-ary relations. 
\begin{theorem}
Let ${\bf D} = (D, \{ r_D \}_{r \in \Sigma})$ be a relational space 
and let 
${\cal P}({\bf D}) $ be the Boolean algebra with operators 
$({\cal P}(D), \cup, \cap, \emptyset, D, \{ f_{\exists r}\}_{r \in \Sigma}) \in BAO(\Sigma)$, where for every $r \in \Sigma, 
f_{\exists r}$ is defined as in formula~(\ref{1}) above. 
Let $r_1, s_1, \dots, s_n, r_2 \in \Sigma$ such that $r_1$ is an $n+1$-ary,  
$s_i$ are $n_i +1$-ary, and $r_2$ an $m+1$-ary relations. Let $C_i, C^j_k \subseteq D$ 
and let $c_i, c^j_k$ be constant symbols which are interpreted as 
$C_i, C^j_K$ respectively. 
The following hold: 
\begin{itemize}
\item[(1)]  If $n = m$ and ${\bf D} \models \forall x, {\overline y} (\bigwedge_i y_i \in C_i \wedge r_1(x, {\overline y}) \rightarrow r_2(x, {\overline y}))$ then 
\[{\cal P}({\bf D}) \models \forall x_1, \dots, x_n ~~(\bigwedge_i x_i \leq c_i \rightarrow f_{\exists r_1}(x_1, \dots, x_n) \leq f_{\exists r_2}(x_1, \dots, x_n)).\] 
\vspace{-4mm}
\item[(2)] If ${\bf D} \models \forall x, {\overline y_1}, {\dots},{\overline y_n} (\bigwedge_{i = 1}^n \bigwedge_{j = 1}^{n_i} y^j_i {\in} C^j_i \wedge r_1 {\circ} (s_1, \dots, s_n)(x, {\overline y_1}, {\dots},{\overline y_n}) \rightarrow r_2(x, {\overline y_1}, {\dots},{\overline y_n}))$ 
then \[{\cal P}({\bf D}) \models \forall {\overline x^1}, \dots, {\overline x^n} ~~ \bigwedge_{i,j} x^i_j \leq c^i_j \rightarrow f_{\exists r_1}(f_{\exists s_1}({\overline x^1}), \dots, f_{\exists s_n}({\overline x^n})) \leq f_{\exists r_2}({\overline x^1}, \dots, {\overline x^n}).\] 
\vspace{-4mm}
\item[(3)] If $s_i$ are binary and ${\bf D} \models \forall x, y (y \in C \wedge 
r_1 \circ (s_1, \dots, s_n)(x, y) \subseteq x = y)$ then 
\[{\cal P}({\bf D}) \models \forall x ~~ x \leq c \rightarrow  f_{\exists r_1}(f_{\exists s_1}(x), \dots, f_{\exists s_n}(x)) \leq x.\]
\end{itemize}
\label{rel-to-bao-n}
\end{theorem}
\Proof Analogous to the proof of Theorem~\ref{rel-to-bao-guards}. \QED

\medskip
\noindent If all $C^i_j$ are equal to $D$ all guards disappear 
and we obtain an $n$-ary analogon of Corollary~\ref{rel-to-bao}. 

\vspace{-2mm}
\section{Algebraic semantics for description logics} 
\label{dl-alg-sem}
A translation of concept descriptions into terms in a signature naturally 
associated with the set of constructors can be defined as follows.
For every role name $r$, we introduce unary function symbols, 
$f_{\exists r}$ and $f_{\forall r}$. The renaming is inductively defined by: 
\begin{itemize}
\item ${\overline C} = C$ for every concept name  $C$; 
\item $\overline{\neg C} = \neg {\overline C}$; $\quad \overline{C_1 \sqcap C_2} =  {\overline C_1} \wedge {\overline C_2}$,$\quad \overline{C_1 \sqcup C_2}  = {\overline C_1} \vee {\overline C_2}$;
\item $\overline{\exists r. C} = f_{\exists r}({\overline C})$, $\quad \overline{\forall r. C} = f_{\forall r}({\overline C})$.
\end{itemize}
There exists a one-to-one correspondence 
between interpretations 
${\mathcal I} = (D, \cdot^{\mathcal I})$ 
and Boolean algebras of sets with additional operators, 
$({\mathcal P}(D), \cup, \cap, \neg, \emptyset, D, 
\{ f_{\exists r}, f_{\forall r}\}_{r \in N_R})$, 
together with valuations $v : N_C \rightarrow  {\mathcal P}(D)$, 
where $f_{\exists r}, f_{\forall r}$ are 
defined, for every $U \subseteq D$, by:
\begin{eqnarray*}
f_{\exists r}(U) & = & 
\{ x \mid \exists y ((x,y) \in r^{\mathcal I} \mbox{  and } y \in U)\} \\
 f_{\forall r}(U)  & =  & \{ x \mid \forall y ((x,y) \in r^{\mathcal I} \Rightarrow y \in U)\}.
\end{eqnarray*} 
It is easy to see that, with these definitions:
\begin{itemize}
\item $f_{\exists r}$ is a join-hemimorphism,  i.e.\ $f_{\exists r}(x \vee y) = f_{\exists r}(x) \vee f_{\exists r}(y)$, $f_{\exists r}(0) = 0$; 
\item $f_{\forall r}$ is a meet-hemimorphism, i.e.\
$f_{\forall r}(x \wedge y) {=} f_{\forall r}(x) \wedge f_{\forall r}(y)$, 
$f_{\forall r}(1) {=} 1$;
\item $f_{\forall r}(x) = \neg f_{\exists r}(\neg x)$ for every $x \in B$.
\end{itemize}
Let $v :  N_C \rightarrow  {\mathcal P}(D)$ with
$v(A) = A^{\mathcal I}$ for all $A \in N_C$, and let ${\overline v}$ 
be the (unique) homomorphic extension of $v$ to  terms.
Let $C$ be a concept description and $\overline{C}$ 
be its associated term. Then $C^{\mathcal I} = 
{\overline v}({\overline C})$ (denoted by ${\overline C}^{\mathcal I}$).

\medskip
\noindent The TBox subsumption problem for 
the description logic ${\cal ALC}$ (which was defined in Section~\ref{dl-gen}) 
can be expressed as uniform 
word problem for Boolean algebras with suitable operators. 
\begin{theorem}
If ${\mathcal T}$ is an ${\mathcal A}{\mathcal L}{\mathcal C}$ TBox consisting of general concept inclusions between concept terms formed from concept names 
$N_C = \{ C_1, \dots, C_n \}$, and $D_1, D_2$ are concept descriptions, 
the following are equivalent: 
\begin{itemize}
\item[(1)] $D_1 {\sqsubseteq}_{\mathcal T} D_2$.  
\item[(2)] ${\bf {\mathcal P}(D)} \models \forall C_1 ... C_n \left(\left( \bigwedge_{C {\sqsubseteq} D \in {\mathcal T}}
\overline{C} {\leq} \overline{D} \right) \rightarrow \overline{D_1} {\leq} \overline{D_2}\right)$ 
for all in\-ter\-pre\-ta\-tions ${\mathcal I} = (D, \cdot^{\mathcal I})$, \\
where ${\bf {\mathcal P}(D)} = ({\mathcal P}(D), \cup, \cap, \neg, \emptyset, D, 
\{ f_{\exists r}, f_{\forall r}\}_{r \in N_R})$.  
\item[(3)] ${\sf BAO}_{N_R} {\models} \forall C_1 ... C_n \left(\left( \bigwedge_{C {\sqsubseteq} D \in {\mathcal T}} 
\overline{C} {\leq} \overline{D} \right) \rightarrow \overline{D_1} {\leq} \overline{D_2}\right).$
\end{itemize}
\label{bao}
\end{theorem}
\Proof The equivalence of (1) and (2) follows from the 
definition of $D_1 \sqsubseteq_{\mathcal T} D_2$. $(3) \Rightarrow (2)$ is 
immediate. $(2) \Rightarrow (3)$ 
follows from the fact that 
every algebra in ${\sf BAO}_{N_R}$ homomorphically 
embeds into a Boolean algebra of sets, its canonical extension.\QED


\medskip
\noindent An analogon of Theorem~\ref{bao} can be used for more general 
description logics in which in addition to the TBoxes also 
properties of roles need to be taken into account. We consider 
properties $R$ of roles which can be expressed by sets $R_a$ of clauses 
at an algebraic level. The main restriction we impose is that the 
sets of clauses $R_a$ are 
preserved when taking canonical extensions of Boolean algebras.
We denote by $BAO_{N_R}(R_a)$ the family of 
all algebras in $BAO_{N_R}$ which satisfy the axioms in $R_a$.
\begin{theorem}
Let ${\mathcal T}$ be an ${\mathcal A}{\mathcal L}{\mathcal C}$ TBox 
consisting of general concept inclusions between concept terms formed 
from concept names 
$N_C = \{ C_1, \dots, C_n \}$, and let $R$ be a 
family of general (e.g.\ guarded) {\em role inclusions}  
with the additional property 
that there exists a set $R_a$ of clauses in the signature of $BAO_{N_R}$
such that:
\begin{itemize}
\item[(i)] For each in\-ter\-pre\-ta\-tion 
${\mathcal I} = (D, \cdot^{\mathcal I})$,
which satisfies the constraints on roles 
in $R$, we have that ${\bf {\mathcal P}(D)} \models R_a,$
where 
${\bf {\mathcal P}(D)}$ stands for 
$({\mathcal P}(D), \cup, \cap, \neg, \emptyset, D, 
\{ f_{\exists r}, f_{\forall r}\}_{r \in N_R})$.  
\item[(ii)] Every  $B  \in BAO_{N_R}(R)$ embeds into an algebra of sets of the form ${\bf {\cal P}(D)}$ (defined as above), where 
$(D, \{ r \}_{r \in N_R})$ satisfies $R$.  
\end{itemize}
Then for any concept descriptions $D_1, D_2$ 
the following are equivalent: 
\begin{itemize}
\item[(1)] $D_1 {\sqsubseteq}_{{\mathcal T} \cup R} D_2$.
\item[(2)] ${\bf {\cal P}(D)} \models \forall C_1 ... C_n \left(\left( \bigwedge_{C {\sqsubseteq} D \in {\mathcal T}}
\overline{C} {\leq} \overline{D} \right) \rightarrow \overline{D_1} {\leq} \overline{D_2}\right)$ 
for all in\-ter\-pre\-ta\-tions ${\mathcal I} = (D, \cdot^{\mathcal I})$ \\
which are models of $R$.  
\item[(3)] ${\sf BAO}_{N_R}(R_a) {\models} \forall C_1 ... C_n \left(\left( \bigwedge_{C {\sqsubseteq} D \in {\mathcal T}} 
\overline{C} {\leq} \overline{D} \right) \rightarrow \overline{D_1} {\leq} \overline{D_2}\right).$
\end{itemize}
\label{bao-ext-ax}
\end{theorem}
\Proof (1) $\Leftrightarrow$ (2) Let 
${\mathcal I} = (D, \cdot^{\mathcal I})$ be an interpretation 
which is a model of $R$. Let $v : N_C \rightarrow {\cal P}(D)$ be a 
valuation with the property that ${\overline v}(C) \subseteq {\overline v}(D)$ 
for all $C {\sqsubseteq} D \in {\mathcal T}$. Since 
$D_1 {\sqsubseteq}_{{\mathcal T} \cup R} D_2$, it follows that 
${\overline v}(D_1) \subseteq {\overline v}(D_2)$.
(3) $\Rightarrow$ (2) follows from the fact that, by assumption (i), 
${\bf {\mathcal P}(D)} = ({\mathcal P}(D), \cup, \cap, \neg, \emptyset, D, 
\{ f_{\exists r}, f_{\forall r}\}_{r \in N_R}) \in {\sf BAO}_{N_R}(R_a)$.
(2) $\Rightarrow$ (3) follows from the fact that, by Assumption (ii), 
for every Boolean algebra $B$ with operators there exists a relational space 
$D$ which satisfies $R$, 
such that $B$ homomorphically embeds into a 
Boolean algebra of sets of the form ${\cal P}(D)$ which  
satisfies the conditions in (2). Hence,  
${\cal P}(D) \models \forall C_1 ... C_n \left(\left( \bigwedge_{C {\sqsubseteq} D \in {\mathcal T}} 
\overline{C} {\leq} \overline{D} \right) \rightarrow \overline{D_1} {\leq} \overline{D_2}\right)$. As $B$ is isomorphic to a subalgebra of ${\cal P}(D)$, it 
follows that
$B \models  \forall C_1 ... C_n \left(\left( \bigwedge_{C {\sqsubseteq} D \in {\mathcal T}} 
\overline{C} {\leq} \overline{D} \right) \rightarrow \overline{D_1} {\leq} \overline{D_2}\right)$. \QED

\begin{example}
Assume that $R$ consists of concept inclusions of the form 
$r \sqsubseteq s$ and $r_1
\circ s \sqsubseteq r_2$ and $r_1 \circ s \sqsubseteq id$ and 
$R_a$ consists of the corresponding axioms 
$\forall x (f_{\exists r}(x) \leq f_{\exists
  s}(x))$,   $\forall x (f_{\exists r_1}(f_{\exists r}(x)) \leq f_{\exists
  r_2}(x))$, and $\forall x (f_{\exists r_1}(f_{\exists r}(x)) \leq x)$. Then, 
by Corollaries~\ref{rel-to-bao} and~\ref{bao-rel} premises (i) and (ii) 
of Theorem~\ref{bao-ext-ax} hold, hence the CBox subsumption problem 
can be expressed as a uniform word problem in ${\sf BAO}_{N_R}(R_a)$. 
\end{example}

\vspace{-4mm}
\subsection{Algebraic semantics for $\mathcal{EL}$, $\mathcal{EL}^+$ and extensions thereof}
\label{alg-sem-el}

\vspace{-2mm} 
In \cite{Sofronie-amai-07} we studied the link between 
TBox subsumption in  $\mathcal{EL}$ 
and uniform word problems in the corresponding classes 
of semilattices with monotone functions. 
We now show that these results naturally extend to 
the description logic $\mathcal{EL}^+$. 
We will 
consider the following classes of algebras:
\begin{itemize} 
\item ${\sf BAO}^{\exists}_{N_R}$: the class of boolean algebras  
with operators  
$(B, \vee, \wedge, \neg, 0, 1, 
\{ f_{\exists r} \}_{r \in N_R})$,  
such that 
$f_{\exists r}$ is a unary join-hemimorphism;
\item ${\sf DLO}^{\exists}_{N_R}$: the class of bounded distributive 
lattices with operators  
$(L, \vee, \wedge, 0, 1, 
\{ f_{\exists r} \}_{r \in N_R})$,  
such that 
$f_{\exists r}$ is a unary join-hemimorphism; 

\item ${\sf SLO}^{\exists}_{N_R}$: the class of all $\wedge$-semilattices 
with operators  
$(S, \wedge, 0, 1, \{ f_{\exists R} \}_{R \in N_R})$,  
such that 
$f_{\exists R}$ is a  monotone unary function and $f_{\exists R}(0) = 0$.
\footnote{For the sake of simplicity, in this paper we assume that the 
description logics $\mathcal{EL}$ and $\mathcal{EL}^+$ contain 
the additional constructors $\perp, \top$, which will be interpreted
as $0$ and $1$. Similar considerations can be used to show that the 
algebraic semantics for variants of $\mathcal{EL}$ and $\mathcal{EL}^+$
having  only $\top$ (or $\perp$) is given by semilattices with $1$ (resp.\ 0).}
\end{itemize}

\vspace{-5mm}
\subsection{Algebraic semantics for $\mathcal{EL}^+$}
\label{sect:slo-el+}
In ${\cal EL}^+$ the following types of role inclusions are considered:
$$ r \sqsubseteq s \quad \quad \text{ and } \quad \quad r_1 \circ s \sqsubseteq r_2.$$
In \cite{Baader-2008} it is proved that subsumption w.r.t.\ $GCI$'s 
in the extension
$\mathcal{ELI}$ of $\mathcal{EL}$ with inverse roles is ExpTime complete.
It is also proved that subsumption w.r.t.\ general TBoxes  
in the extension
$\mathcal{EL}^{\sf sym}$ of $\mathcal{EL}$ with symmetric roles is 
ExpTime complete. We will now start by considering also CBoxes containing 
role inclusion axioms which describe weaker, left- and 
right-inverse properties of roles, of the form:
$r \circ s \subseteq id.$

\medskip
\noindent Let $RI$ be a set of axioms of the form $r \sqsubseteq s$, 
$r_1 \circ r_2 \sqsubseteq r$, and $r_1 \circ r_2 \sqsubseteq id$ 
with $r_1, r_2, r \in N_R. $
We associate with $RI$ the following set $RI_a$ of axioms: 
\begin{eqnarray*}
 RI_a & = & \{\forall x ~~  f_{\exists r}(x) \leq f_{\exists s}(x) \mid r \sqsubseteq s \in RI\} \cup \\
& &  \{ \forall x ~~(f_{\exists r_2} \circ f_{\exists r_1})(x) \leq f_{\exists r}(x) \mid r_1 \circ r_2 \sqsubseteq r \in RI \} \cup \\
& &  \{ \forall x ~~(f_{\exists r_2} \circ f_{\exists r_1})(x) \leq x \mid r_1 \circ r_2 \sqsubseteq id \in RI \}. \end{eqnarray*}
Let 
${\sf BAO}^{\exists}_{N_R}(RI)$ (resp.\ ${\sf DLO}^{\exists}_{N_R}(RI)$, 
${\sf SLO}^{\exists}_{N_R}(RI)$) be the subclass of 
${\sf BAO}^{\exists}_{N_R}$ (resp.\ ${\sf DLO}^{\exists}_{N_R}$, ${\sf SLO}^{\exists}_{N_R}$) consisting of those algebras which satisfy $RI_a$.

\begin{lemma}
Let ${\mathcal I} = (D, \cdot^{\mathcal I})$ be a model of an 
$\mathcal{EL}^+$ CBox  ${\mathcal C} = GCI \cup RI$. Then the 
algebra 
${\cal P}({\bf D})_{|(\wedge, 0, 1)} = 
({\mathcal P}(D), \cap, \emptyset, D, \{ f_{\exists r} \}_{r \in N_R})$  is a
semilattice with operators in ${\sf SLO}^{\exists}_{N_R}(RI)$.
\end{lemma}
\Proof Clearly, $({\mathcal P}(D), \cap, \emptyset, D, \{ f_{\exists r} \}_{r \in N_R}) \in {\sf SLO}^{\exists}_{N_R}$. The proof of the second part uses exactly the same arguments as the proof of Theorem~\ref{rel-to-bao-guards} 
and Corollary~\ref{rel-to-bao}. \QED 

\medskip
\noindent We will now show that every algebra in 
${\sf SLO}^{\exists}_{N_R}(RI)$ embeds into (the bounded semilattice reduct of)
an algebra in ${\sf BAO}^{\exists}_{N_R}(RI)$. We start with a more 
general lemma, which will be important also for proving 
the locality results in Section~\ref{complexity}. 

\begin{lemma}
For every structure 
${\cal S} = (S, \wedge, 0, 1, \{ f_S \}_{f \in \Sigma})$ in which $f_S$ 
are partial functions, if properties (i), (ii) and (iii) below hold, 
then ${\cal S}$ embeds into a semilattice with 
operators in ${\sf SLO}^{\exists}_{N_R}(RI)$. 
\begin{itemize}
\item[(i)]  $(S, \wedge, 0, 1)$ is a bounded semilattice; $\leq$ the partial order on $S$ defined by $x {\leq} y$ iff $x {\wedge} y = y$.
\item[(ii)] For every $f \in \Sigma$ with arity $n$, 
$f_S$ is a partial $n$-ary function on $S$ which satisfies the monotonicity 
axiom ${\sf Mon}(f)$ whenever all terms are defined. 
$${\sf Mon}(f) ~~~ \forall x, y (x \leq y \rightarrow f(x) \leq f(y))$$
\item[(iii)] There exists a set $RI^{\sf flat}$ of axioms of the form\footnote{These axioms are logically equivalent with those discussed before; the reason 
for preferring the flat version will become apparent in 
Section~\ref{complexity}.}:
\begin{eqnarray*}
\forall x ~~~& & g(x)  \leq h(x) \\
\forall x, y ~~~& x \leq g(y) \rightarrow & f(x) \leq h(y) \\
\forall x, y ~~~& x \leq g(y) \rightarrow & f(x) \leq y 
\end{eqnarray*} 
such that:
\begin{itemize}
\item if $g, h$ appear in a rule as above and $g_S(s)$ is defined then also 
$h_S(s)$ is defined; 
\item for every $\beta : \{ x, y \} \rightarrow S$, and every axiom 
$D \in RI^{\sf flat}$ if all terms in ${\overline \beta}(D)$ are defined, then 
${\overline \beta}(D)$ is true in $S$ (where ${\overline \beta}$ is the 
canonical extension of $\beta$ to formulae). 
\end{itemize}
\end{itemize}
\label{local-alg-el+}
\end{lemma}
\Proof Let ${\bf S} = (S, \wedge, 0, 1, \{ f_S \}_{f \in \Sigma})$ 
be a 0,1 semilattice, and let $f_S, g_S : S \rightarrow S$ be  
partially defined functions which satisfy the conditions above.
Consider the lattice of 
all order-ideals of $S$, 
 ${\cal OI}({\bf S}) = ({\cal OI}(S), \cap, \cup, \{ 0 \}, S, 
\{ {\overline f}_S \}_{f \in \Sigma})$, 
where join is set union, 
meet is set intersection, and the additional operators in $\Sigma$ are 
defined, for every order ideal $U$ of $S$,  by
$${\overline f}_S(U) = 
{\downarrow} \{ f_S(u) \mid u \in U, f_S(u) \mbox{ defined} \}.$$
Note that ${\overline f}_A(\{ 0 \}) = \{ 0 \}$ and 
${\overline f}_S(U_1 \cup U_2) ~=~ {\downarrow} f_S(U_1 \cup U_2) =
{\downarrow}(f_S(U_1) \cup f_S(U_2)) = 
{\downarrow} f_S(U_1) \cup {\downarrow} f_S(U_2)$.
Thus,  
${\cal OI}({\bf S}) \in {\sf DLO}^{\exists}_{N_R}$.
\footnote{A similar construction can be made starting from 
$\wedge$-semilattices with monotone operators which have only 1 (resp.\ 0) 
or neither 0 nor 1.}
Moreover, $\eta : {\bf S} \rightarrow {\cal OI}({\bf S})$ defined by 
$\eta(x) := {\downarrow} x$ is an injective homomorphism 
w.r.t.\ the bounded semilattice operations 
and 
$\eta(f_S(x)) =  {\downarrow} f_S(x) = {\overline f}_S({\downarrow} x)$.
We prove that ${\overline f}_S, 
{\overline g}_S, {\overline h}_S$ 
satisfy the axioms in $RI^{\sf flat}$. Consider first the axiom:
\begin{eqnarray}
\forall x, y ~  & (y \leq g(x)  \rightarrow & f(y) \leq x) \label{ax1}
\end{eqnarray}
Let $U, V \in {\overline S}$ be such that $U \subseteq{\overline g}(V)$. 
Let $x \in {\overline f}_S(U)$. Then there exists $u \in U$ such that 
$f_S(u)$ is defined and $x \leq f_S(u)$. 
Since $U \subseteq g(V)$, we know that there exists $v \in V$ with 
$g_S(v)$ defined and $u \leq g(v)$. Since $S$ satisfies Axiom~(\ref{ax1}), 
and $g_S(v), f_S(u)$ are defined and $u \leq g_S(v)$ it follows that 
$f_S(u) \leq v.$ Thus, $x \leq f_S(u) \leq v \in V$, so $x \in V$. 
This shows that for all $U, V \in {\overline S}$:
$$ U \leq {\overline g}(V) \rightarrow {\overline f}(U) \subseteq V.$$
We now check preservation of the axioms of the form:
\begin{eqnarray}
\forall x  ~ &  & g(x)  \leq  h(x) \label{ax2} \\
\forall x, y ~  & (y \leq g(x)  \rightarrow & f(y) \leq h(x)) \label{ax3}
\end{eqnarray}
We assume that $S$ has the property that $h_S(a)$ is defined whenever 
$g_S(a)$ is defined. We have to show that 
if $f_S, g_S, h_S$ are monotone whenever defined and satisfy 
one of the axioms above (say (\ref{ax3}); the case of Axiom~(\ref{ax2}) is similar) whenever defined 
then ${\overline f}_S, {\overline g}_S$ and ${\overline h}_S$ 
satisfy (\ref{ax3}).

Let $U, V \in {\overline S}$ be such that $U \subseteq{\overline g}_S(V)$. 
Let $x \in {\overline f}_S(U)$. Then there exists $u \in U$ such that 
$f_S(u)$ is defined and $x \leq f_S(u)$. 
Since $U \subseteq {\overline g}_S(V)$, we know that there exists $v \in V$ with 
$g_S(v)$ defined and $u \leq g_S(v)$. Due to the first condition in (iii), 
$h_S(v)$ must be defined as well. 
Since $S$ satisfies Axiom~(\ref{ax3}) and $g_S(v), f_S(u), h_S(v)$ 
are defined and $u \leq g_S(v)$ it follows that 
$f_S(u) \leq h_S(v).$ Thus, there exists $v \in V$ such that 
$x \leq f_S(u) \leq h_S(v)$, so $x \in {\overline h}_S(V)$. 
This shows that for all $U, V \in {\overline S}$:
$$ U \leq {\overline g}(V) \rightarrow {\overline f}(U) \subseteq {\overline h}(V).$$ \QED
\begin{lemma}
Every ${\bf S} \in {\sf SLO}^{\exists}_{N_R}(RI)$  
embeds into (the bounded semilattice reduct of) a 
lattice in ${\sf DLO}^{\exists}_{N_R}(RI)$.
Every lattice in ${\sf DLO}^{\exists}_{N_R}(RI)$ 
embeds into (the bounded lattice reduct of) an algebra in 
${\sf BAO}^{\exists}_{N_R}(RI)$.
\label{embeddings-slo-dlo-bao}
\end{lemma}
\Proof The first part follows from Lemma~\ref{local-alg-el+}. 
The second statement is a consequence of Priestley duality for 
distributive lattices. Let ${\bf L} \in {\sf DLO}^{\exists}_{N_R}(RI)$. 
Let ${\mathcal F}_p$ be the set of prime filters of 
$L$, and  $B({\bf L}) = ({\mathcal P}({\mathcal F}_p), \cup, \cap, 
\{ {\overline f}_{\exists r} \}_{r \in N_r})$, where for $r \in R$, 
${\overline f}_{\exists r}$ is defined by 
$$ {\overline f}_{\exists r}(U) = \{ F \in {\mathcal F}_p \mid \exists G \in U: f_{\exists r}(G) \subseteq F \}.$$
Let $i : {\bf L} \rightarrow   B({\bf L})$ be defined by 
$i(x) = \{ F \in {\mathcal F}_p \mid x \in F \}$. 
Obviously, $i$ is a lattice homomorphism. 
We show that $i(f_{\exists r}(x))  = {\overline f}_{\exists r}(i(x))$. 
\begin{eqnarray*}
{\overline f}_{\exists r}(i(x)) & = & \{ F \in {\mathcal F}_p \mid \exists G \in i(x): f_{\exists r}(G) \subseteq F \} \\
 & = & \{ F \in {\mathcal F}_p \mid \exists G: x \in G \text{ and } f_{\exists r}(G) \subseteq F \} \\
 & \subseteq & \{ F \in {\mathcal F}_p \mid f_{\exists r}(x) \in F \} = i(f_{\exists r}(x)). 
\end{eqnarray*}
To prove the converse inclusion, let $F \in i(f_{\exists r}(x))$.
Then $F \in {\mathcal F}_p$ and $f_{\exists r}(x) \in F$.  Let 
$G = f_{\exists r}^{-1}(F)$.
As $F$ is a prime filter, and 
$f_{\exists r}$ is a join-hemimorphism, 
$G$ is a prime filter with 
$x \in G$ and  $f_{\exists r}(G) \subseteq F$, 
so $F \in {\overline f}_{\exists r}(i(x))$. 
Finally, we show that $B({\bf L})$ satisfies the axioms in $RI_a$. Let 
$U \in B({\bf L})$. By definition, 
\begin{eqnarray*}
{\overline f}_{\exists r_1}(U) & = & \{ F \in {\mathcal F}_p \mid \exists G_1 \in U: f_{\exists r_1}(G_1) \subseteq F \},  \\
{\overline f}_{\exists r_2}({\overline f}_{\exists r_1}(U)) & = & \{ F \in {\mathcal F}_p \mid \exists G_1 \in {\overline f}_{\exists r_1}(U): f_{\exists r_2}(G_1) \subseteq F \} \\
 & = & \{ F \in {\mathcal F}_p \mid \exists G_1, \exists G_2 \in U: f_{\exists r_1}(G_2) \subseteq G_1 \text{ and } f_{\exists r_2}(G_1) \subseteq F \} \\
 & \subseteq & \{ F \in {\mathcal F}_p \mid \exists G_2 \in U: f_{\exists r_2}(f_{\exists r_1}(G_2)) \subseteq F \}. 
\end{eqnarray*}
Assume that $r_1 \sqsubseteq r \in RI$. 
We know that ${\bf L} \models \forall x$, $f_{\exists r_1}(x) \leq f_{\exists r}(x)$. 
Let $F \in {\overline f}_{\exists r_1}(U)$. Then $f_{\exists r_1}(G_1) \subseteq F$ for some $G_1 \in U$, so also $f_{\exists r}(G_1) \subseteq F$. 
Hence, ${\overline f}_{\exists r_1}(U) \subseteq {\overline f}_{\exists r}(U)$.
Similarly we can prove that if $r_1 \circ r_2 \sqsubseteq r \in RI$ then 
${\overline f}_{\exists r_2}({\overline f}_{\exists r_1}(U)) \subseteq 
{\overline f}_{\exists r}(U)$ and that if $r_1 \circ r_2 \sqsubseteq id \in RI$
then ${\overline f}_{\exists r_2}({\overline f}_{\exists r_1}(U)) \subseteq U$.
\QED

\begin{theorem}
If the only concept constructors are 
intersection and existential restriction, then 
for all concept descriptions $D_1, D_2$ and every $\mathcal{EL}^+$ CBox 
${\mathcal C} {=} GCI {\cup} RI$, with concept names 
$N_C = \{ C_1, \dots, C_n \}$ the following are equivalent:
\begin{itemize}
\item[(1)] $D_1 {\sqsubseteq}_{\mathcal C} D_2$.  
\item[(2)] ${\sf SLO}^{\exists}_{N_R}(RI) \models 
\forall C_1 \dots C_n \left(\left( \bigwedge_{C {\sqsubseteq} D \in GCI} 
\overline{C} {\leq} \overline{D} \right) \rightarrow \overline{D_1} {\leq} \overline{D_2}\right).$
\end{itemize}
\label{el-slat}
\end{theorem}
\Proof  
We know that 
$C_1 \sqsubseteq_{\mathcal C} C_2$ iff 
$C_1^{\mathcal I} \subseteq C_2^{\mathcal I}$ for every model  
${\mathcal I}$ of the CBox ${\mathcal C}$.
Assume first that 
(2) holds. Let ${\mathcal I} = (D, \cdot^{\mathcal I})$ be an interpretation 
that satisfies ${\mathcal C}$. Then 
${\cal P}({\bf D})_{|\wedge} = ({\mathcal P}(D), \cap, \emptyset, D, \{ f_{\exists r} \}_{r \in N_R}) \in 
{\sf SLO}^{\exists}_{N_R}(RI)$, hence 
${\cal P}({\bf D})_{|\wedge}
 \models 
\left( \bigwedge_{C \sqsubseteq D \in GCI} 
\overline{C} \leq \overline{D} \right) \rightarrow 
\overline{D_1} \leq \overline{D_2}.$ 
 As ${\mathcal I}$ is a model of 
$GCI$, $\overline{C}^{\mathcal I} \subseteq \overline{D}^{\mathcal I}$ for all 
$C \sqsubseteq D \in GCI$, so  
$D_1^{\mathcal I} \, {=} \, \overline{D_1}^{\mathcal I} \, {\subseteq} \, \overline{D_2}^{\mathcal I} \, {=} \,
D_2^{\mathcal I}.$ 
To prove $(1) \Rightarrow (2)$ 
note first that in this case 
the premises of Thm.\ \ref{bao-ext-ax} are fulfilled. 
By Thm.\ \ref{bao-ext-ax}, 
if $D_1 \sqsubseteq_{\mathcal C} D_2$ then 
${\sf BAO}_{N_R}(RI) \models \left( \bigwedge_{C \sqsubseteq D \in {\mathcal C}} 
{\overline C} \leq {\overline D} \right) \rightarrow \overline{D_1} \leq \overline{D_2}.$
Let ${\bf S} \in {\sf SLO}^{\exists}_{N_R}(RI)$.  By 
Lemma~\ref{embeddings-slo-dlo-bao}, ${\bf S}$ embeds into an 
algebra in ${\sf BAO}^{\exists}_{N_R}$ which satisfies 
$RI_a$. Therefore,   
${\bf S} \models \left( \bigwedge_{C \sqsubseteq D \in GCI} 
\overline{C} \leq \overline{D} \right) \rightarrow 
\overline{C_1} \leq \overline{C_2}.$ \QED

\medskip
We will show that the word problem for the class of algebras 
${\sf SLO}^{\exists}_{N_R}(RI)$ is decidable in PTIME. For this we will 
prove that ${\sf SLO}^{\exists}_{N_R}(RI)$ has a ``local'' presentation.
The general locality definitions, as well as methods for recognizing local 
presentations are given in Sect.~\ref{locality}. The 
application to the class of models for $\mathcal{EL}$ and $\mathcal{EL}^+$
are given in Sect.~\ref{complexity}. Before doing this, we present 
some additional types of constraints on the roles which can be 
handled similarly. This will allow us to obtain a new tractable 
extension of $\mathcal{EL}^+$.

\medskip

\subsection{Guarded role inclusions} 
In applications it may be interesting  to consider 
role inclusions guarded by membership to a certain concept, 
i.e.\ role inclusions of the form: 
\begin{eqnarray}
\forall x, y & (y \in C \wedge r(x, y) & \rightarrow r'(x, y)) \label{c1}\\
\forall x, y & (y \in C \wedge r \circ s (x, y) & \rightarrow r'(x, y)) \label{c2}\\
\forall x, y & (y \in C \wedge r \circ s (x, y) & \rightarrow x = y). \label{c3}
\end{eqnarray}
The corresponding axioms at the algebra level we consider are: 
\begin{eqnarray}
\forall x & (x \leq C & \rightarrow f_r(x) \leq f_{r'}(x)) \label{d1}\\
\forall x & (x \leq C & \rightarrow f_r(f_s(x)) \leq f_{r'}(x)) \label{d2}\\
\forall x & (x \leq C & \rightarrow f_r(f_s(x)) \leq f_{r'}(x)). \label{d3}
\end{eqnarray}
\begin{theorem}
Assume that the only concept constructors are 
intersection and existential restriction. 
Let 
${\mathcal C} {=} GCI {\cup} RI {\cup} GRI$ be a  CBox containing a set $GCI$ of general concept inclusions, a set $RI$ 
of role inclusions of the type considered in Sect.~\ref{sect:slo-el+} 
and a set $GRI$ of guarded role inclusions 
of the form~(\ref{c1})--(\ref{c3}),  
with concept names $N_C = \{ C_1, \dots, C_n \}$. 
Then 
for all concept descriptions $D_1, D_2$ the following are equivalent:
\begin{itemize}
\item[(1)] $D_1 {\sqsubseteq}_{\mathcal C} D_2$. 
\item[(2)] $GRI(C_1, \dots, C_n) \wedge 
\left( \bigwedge_{C {\sqsubseteq} D \in GCI} 
\overline{C} {\leq} \overline{D} \right) \wedge 
\overline{D_1} {\not\leq} \overline{D_2}$ is unsatisfiable w.r.t.\ 
${\sf BAO}^{\exists}_{N_R}(RI)$.  
\item[(3)] $GRI(C_1, \dots, C_n) \wedge 
\left( \bigwedge_{C {\sqsubseteq} D \in GCI} 
\overline{C} {\leq} \overline{D} \right) \wedge 
\overline{D_1} {\not\leq} \overline{D_2}$ is unsatisfiable w.r.t.\ 
${\sf SLO}^{\exists}_{N_R}(RI)$. 
\end{itemize}
\end{theorem}
\Proof The proof is analogous to that of Theorem~\ref{el-slat} and 
uses the results in Theorems~\ref{bao-rel-guards} 
and~\ref{rel-to-bao-guards}, as well as an analogon of 
Theorem~\ref{embeddings-slo-dlo-bao}. \QED

\vspace{-2mm}
\subsection{Extensions of $\mathcal{EL}^+$ with $n$-ary roles and concrete domains} 
\label{extensions-of-el-n-ary}
We now present a possibility of 
extending $\mathcal{EL}^+$ with concrete domains, 
which is a natural generalization of the extension in 
Section~\ref{alg-sem-el}. This extension is  
different from the extensions with 
concrete domains and those with $n$-ary quantifiers studied in the 
description logic literature (cf.\ e.g.\ 
\cite{Baader-ijcai-2005,Baader2005-ki}).

Later, in Section~\ref{el++} we will present another extension 
(the one used in $\mathcal{EL}^{++}$).  
\begin{figure}[t]
\label{table:el-ext-constr}
\centering
\caption{Constructors for $\mathcal{EL}$ with $n$-ary roles and their semantics}
\medskip
\begin{tabular}{|l|l|l|}
\hline
Constructor & Syntax & Semantics \\
\hline
\hline
bottom & $\perp$ & $\emptyset$ \\
\hline 
top & $\top$  & $D$ \\
\hline 
conjunction & $C_1 \sqcap C_2$ & $C_1^{\mathcal I} \cap C_2^{\mathcal I}$ \\
\hline
existential & $\exists R.(C_1, \dots C_n)$ & $\{ x \mid \exists y_1, \dots, y_n \, (x, y_1, \dots, y_n) \in R^{\mathcal I} \mbox{  and } y_i \in  C_i^{\mathcal I} \}$ \\
\hline
\end{tabular}

\end{figure}

\medskip
\noindent We consider $n$-ary roles because 
in relational databases, relations of higher arity are often used. 
This is especially important 
when we need to express dependencies between several 
(not only two) individuals. 
\begin{example}
We would like to express, for instance, information about all the 
routes from cities in a set $C_1$ to cities in a set $C_2$ passing 
through cities in a set $C_3$. This could be done using ternary 
roles interpreted as ternary relations. 
\end{example}

\subsubsection{An extension of $\mathcal{EL}^+$ with $n$-ary roles}
\label{ext1-n-ary}
\noindent An extension of the description logic $\mathcal{ALC}$, 
containing $n$-ary roles instead of binary roles 
(interpreted as $n$-ary relations) can  easily be defined. 
The definition of TBox subsumption can be extended naturally  
to the $n$-ary case. In this paper we will restrict to ${\cal EL}$ (cf.\ Figure~\ref{table:el-ext-constr}), i.e.\ 
consider only existential restrictions,   
which are in this case $n$-ary -- of the form $\exists r.(C_1, \dots, C_n)$
-- and are  
interpreted in any interpretation ${\cal I} = (D, \cdot^{\cal I})$ as: 
$$\exists r.(C_1, \dots, C_n)^{\cal I} = \{ x \mid \exists y_1, \dots y_n (y_1 \in C_1 \wedge \dots \wedge y_n \in C_n \wedge r^{\cal I}(x, y_1, \dots, y_n)) \}.$$ 
A translation of concept descriptions into terms can be defined 
in a natural way also in this case, with the difference that 
for every role name $r$ with arity $n+1$, we introduce an $n$-ary 
function symbol $f_{\exists r}$. The renaming is inductively defined
as in the binary case, with the difference that: 
$$\overline{\exists r.(C_1, \dots, c_n)} = f_{\exists r}({\overline C_1},
\dots, {\overline C_n}).$$
Also in the $n$-ary case we denote by ${\sf BAO}^{\exists}_{N_R}$ the class of 
Boolean algebras with operators  
$(B, \vee, \wedge, \neg, 0, 1, 
\{ f_{\exists r} \}_{r \in N_R})$, such that for every $r \in N_r$ with 
arity $n+1$, 
$f_{\exists r}$ is a join-hemimorphism with arity $n$;
${\sf DLO}^{\exists}_{N_R}$ and ${\sf SLO}^{\exists}_{N_R}$ are defined 
similarly.
An extension of $\mathcal{EL}^+$ with $n$-ary roles can be obtained 
by allowing role inclusions of type: 

\vspace{-6mm}
\begin{eqnarray}
r_1 & \sqsubseteq & r_2 \label{n-ary-incl} \\
r_1 \circ (s_1, \dots, s_n) & \sqsubseteq & r_2 \label{n-ary} \\
r_1 \circ (s_1, \dots, s_n) & \sqsubseteq & id ~~~ \text{ for binary relations } s_i \label{n-ary-inv} 
\end{eqnarray}

\vspace{-2mm}
\noindent An interpretation ${\cal I} = (D, \cdot^{\cal I})$ satisfies a 
role inclusion type~(\ref{n-ary}) if it satisfies the formula: 
$$ \forall x, {\overline x_i}, {\overline y^k_j} ~~~(r_1(x, x_1, \dots, x_n) \wedge \bigwedge_{k = 1}^n s_k(x_k, y^k_1, \dots, y^k_{m_k})) \rightarrow r_2(x, y^1_1, \dots, y^1_{m_1}, \dots, y^n_1, \dots, y^n_{m_n}).$$
The truth of role inclusions of type~(\ref{n-ary-incl}) resp.~(\ref{n-ary-inv}) is defined in a similar way. 
As in the case of $\mathcal{EL}^+$ we can also prove that TBox subsumption 
can be expressed as a uniform word problem w.r.t.\ the class of 
semilattices with monotone operators associated with the roles, 
satisfying axioms corresponding in a natural way to the role inclusion laws above:

\vspace{-5mm}
$$\begin{array}{lrcl}
\forall x_1, \dots, x_n ~~~ & f_{\exists r_1}(x_1, \dots, x_n) & \leq & f_{\exists r_2}(x_1, \dots, x_n) \\
\forall {\overline y^k_j}~~~ & 
f_{\exists r_1}(f_{\exists s_1}(y^1_1, \dots, y^1_{m_1}), \dots, f_{\exists s_n}(y^n_1, \dots, y^n_{m_n})) & \leq & f_{\exists r_2}(y^1_1, \dots, y^1_{m_1},  \dots, y^n_1, \dots, y^n_{m_n}) \\
\forall x ~~~ & f_{\exists r_1}(f_{\exists s_1}(x), \dots, f_{\exists s_n}(x)) & \leq & x
\end{array}$$

\vspace{-1mm}
\noindent This type of inequalities are exactly of the form studied in 
Section~\ref{correspondence}. 
A straightforward generalization of Theorem~\ref{el-slat}, using the corresponding corrolaries of Theorem~\ref{bao-rel-n-guards} and~\ref{rel-to-bao-n},
 yields:
\begin{theorem}
If the only concept constructors are 
intersection and existential restriction, then 
for all concept descriptions $D_1, D_2$ and every $\mathcal{EL}^+$ CBox 
${\mathcal C} {=} GCI {\cup} RI$ -- where $RI$ consists of role inclusions of type~(\ref{n-ary-incl})--(\ref{n-ary-inv}) -- with concept names 
$N_C = \{ C_1, \dots, C_n \}$ the following are equivalent:
\begin{itemize}
\item[(1)] $D_1 {\sqsubseteq}_{\mathcal C} D_2$.  
\item[(2)]  ${\sf BAO}^{\exists}_{N_R}(RI) \models 
\forall C_1 \dots C_n \left(\left( \bigwedge_{C {\sqsubseteq} D \in GCI} 
\overline{C} {\leq} \overline{D} \right) \rightarrow \overline{D_1} {\leq} \overline{D_2}\right).$
\item[(3)] ${\sf SLO}^{\exists}_{N_R}(RI) \models 
\forall C_1 \dots C_n \left(\left( \bigwedge_{C {\sqsubseteq} D \in GCI} 
\overline{C} {\leq} \overline{D} \right) \rightarrow \overline{D_1} {\leq} \overline{D_2}\right).$
\end{itemize}
\label{el-+-slat-gen}
\end{theorem}
A similar result is obtained if we also consider guarded role inclusions.
\begin{theorem}
Assume that the only concept constructors are 
intersection and existential restriction. 
Let 
${\mathcal C} {=} GCI {\cup} RI {\cup} GRI$ be a  CBox containing a set $GCI$ of general concept inclusions, a set $RI$ 
of role inclusions and a set $GRI$ of guarded role inclusions 
of the form discussed above,  
with concept names $N_C = \{ C_1, \dots, C_n \}$. 
Then 
for all concept descriptions $D_1, D_2$ the following are equivalent:
\begin{itemize}
\item[(1)] $D_1 {\sqsubseteq}_{\mathcal C} D_2$. 
\item[(2)] $GRI(C_1, \dots, C_n) \wedge 
\left( \bigwedge_{C {\sqsubseteq} D \in GCI} 
\overline{C} {\leq} \overline{D} \right) \wedge 
\overline{D_1} {\not\leq} \overline{D_2}$ is unsatisfiable w.r.t.\ 
${\sf BAO}^{\exists}_{N_R}(RI)$. 
\item[(3)] $GRI(C_1, \dots, C_n) \wedge 
\left( \bigwedge_{C {\sqsubseteq} D \in GCI} 
\overline{C} {\leq} \overline{D} \right) \wedge 
\overline{D_1} {\not\leq} \overline{D_2}$ is unsatisfiable w.r.t.\ 
${\sf SLO}^{\exists}_{N_R}(RI)$. 
\end{itemize}
\end{theorem}

\subsubsection{$\mathcal{EL}^+$ with $n$-ary roles and concrete domains}
\label{ext1-n-ary-concrete}
A further extension is obtained by allowing for certain concrete sorts 
-- having the same support in all interpretations; or  additionally assuming 
that there exist specific concrete concepts which have a fixed semantics 
(or additional fixed properties) in all interpretations.

\begin{example}
\label{ex1}
Consider a description logic having a usual (${\sf concept}$) 
sort and a 'concrete' sort ${\sf num}$ with 
fixed domain ${\mathbb R}$. 
We may be interested in general concrete concepts of 
sort ${\sf num}$ (interpreted as subsets of ${\mathbb R}$) 
or in special concepts of sort ${\sf num}$
such as ${\uparrow} n$, ${\downarrow} n$, or $[n, m]$ for 
$m, n \in {\mathbb R}$. For 
any interpretation ${\mathcal I}$, 
${\uparrow} n^{\mathcal I} = \{ x \in {\mathbb R} \mid x \geq n \}$, 
${\downarrow} n^{\mathcal I} = \{ x \in {\mathbb R} \mid x \leq n \}$, and 
$[n, m]^{\mathcal I} = \{ x \in {\mathbb R} \mid n \leq x \leq m \}$.
We will denote the arities of roles using a many-sorted framework. 
Let $(D, {\mathbb R}, \cdot^{\mathcal I})$ be an interpretation with two 
sorts ${\sf concept}$ and ${\sf num}$. 
A role with arity $(s_1, \dots, s_n)$ is interpreted as a 
subset of $D_{s_1} \times \dots \times D_{s_n}$, where 
$D_{\sf concept} = D$ and $D_{\sf num} = {\mathbb R}$.

\begin{enumerate}
\item 
Let ${\sf price}$ be a binary role or arity $({\sf concept}, {\sf num})$, 
which associates with every element of sort ${\sf concept}$ 
its possible prices.  
The concept 

\smallskip
~~~~~~~~~~~~~~~~~~~~~~~$\exists {\sf price}.{\uparrow} n = \{ x \mid \exists k \geq n: 
{\sf price}(x, k)  \}$ 

\smallskip
\noindent represents the class of all individuals with 
some {\sf price} greater than or equal to $n$. 

\medskip
\item 
Let {\sf has-weight-price} be a role of arity 
$({\sf concept},{\sf num},{\sf num})$. $\!\!$
 The concept 

\smallskip
$
\exists \mbox{ {\sf has-weight-price}}.({\uparrow} {\sf y}, {\downarrow} {\sf p})  = \{ x \mid \exists y' {\geq} {\sf y}, \exists p' {\leq} {\sf p}
 \mbox{  and }   
\mbox{{\sf has-weight-price}}(x, y', p')  \}$

\noindent denotes the family of individuals for which 
a weight above  ${\sf y}$ and 
a price below ${\sf p}$ exist.
\end{enumerate}
\end{example}
The example below can be generalized by allowing a set of concrete sorts.
We discuss the algebraic semantics of this type of extensions
of $\mathcal{EL}$. 

\medskip
\noindent Let ${\sf SLO}^{\exists}_{N_R, S}$ denote the class of all 
structures $(S, {\mathcal P}(A_1), \dots, {\mathcal P}(A_n), 
\{ f_{\exists r} \mid r \in N_R \})$, 
where $S$ is a semilattice, $A_1, \dots, A_n$ are concrete domains, and 
$\{ f_{\exists r} \mid r \in N_R \}$ are $n$-ary monotone operators. 
We may allow constants of concrete sort, interpreted as 
sets in ${\mathcal P}(A_i)$.  
%

\begin{theorem}
If the only concept constructors are 
intersection and existential restriction, then 
for all concept descriptions $D_1, D_2,$ and every CBox 
${\cal C} = GCI \cup RI$
consisting of general concept inclusions $GCI$ 
with concrete domains as 
defined above,
and role inclusions $RI$ of the type considered in 
Sect.~\ref{sect:slo-el+} or Sect.~\ref{ext1-n-ary} 
the following are equivalent:
\begin{itemize}
\item[(1)] $D_1 \sqsubseteq_{\mathcal C} D_2$. 
\item[(2)] ${\sf SLO}^{\exists}_{N_R, S}(RI) \models 
\forall C_1, \dots, C_n \left(\left( \bigwedge_{C \sqsubseteq D \in GCI} 
\overline{C} \leq \overline{D} \right) \rightarrow \overline{D_1} \leq \overline{D_2} \right).$
\end{itemize}
\label{el-slat-ext}
\end{theorem}
\Proof Analogous to the proof of Theorem~\ref{el-slat}. \QED  

\medskip
\noindent We can also consider guarded role inclusions 
for 
$n$-ary many-sorted roles. All the previous 
results lift without 
problems. 

\vspace{-2mm}
\subsection{Existential restrictions for roles}
We will also consider relationships of the form 

\vspace{-3mm}
$$\{ (x, y_1, \dots, y_{i-1}, y_{i+1}, \dots, y_n) \mid \exists y_i \in C: r( x, y_1, \dots, y_n)\}.$$
\noindent In analogy to concept construction by existential restrictions, 
we can apply existential restriction to $n+1$-ary roles for  
obtaining $n$-ary roles. 
The syntax and semantics are: 
$$\begin{array}{|l|l|} 
\hline 
\text{ Role construction } & \text{ Semantics } \\
\hline 
\exists r. (j, C) ~~(1\leq j\leq n)& \exists r. (j, C)^{\cal I} = \{ (x, x_1, \dots, x_{j-1}, x_{j+1}, \dots, x_n) \mid \exists x_j \in C: r(x, x_1, \dots, x_n) \} \\
\hline 
\end{array}$$
\begin{example}
\label{ex2}
Consider a database where we can 
express relationships of the form:  
\begin{eqnarray*}
r_{\sf interm}(x, y, z) & & \text{(there exists 
a route from $x$ to $y$ passing through $z$)} \\ 
r(x, y) & & \text{(there exists 
a route from $x$ to $y$).}
\end{eqnarray*} 
We will also want to 
express relationships of the form 
{\em ``For all $x_1, x_2$, if there exists a route from $x_1$ to $x_2$ 
passing through some city in $C_3$, 
then there exists a route from $x_1$ to $x_2$.''} 
We need therefore to express a new relation $r'$ where $r'(x_1, x_2)$ 
stands for there exists a route from $x_1$ to $x_2$ 
passing through some city in    $C_3$.
For this we will need constructors of the type
$\exists r.(j, C)$. They help to formulate the property 
above as $\exists r_{\sf interm}(3, C_3) \sqsubseteq r$, interpreted as:
$$\{ (x_1, x_2) \mid \exists x_3 \in C_3: r_{\sf interm}(x_1, x_2, x_3) \} \subseteq r, \text{ ~~i.e.\ ~~}\forall x_1, x_2 ~~ \exists r_{\sf interm}(3, C_3)(x_1, x_2) \rightarrow r(x_1, x_2). $$
\end{example}
\begin{lemma} Assume that  $s = \exists r. (i, C)$. Then 

\vspace{-3mm}
\begin{eqnarray*}
f_{\exists s}(U_1, \dots, U_{i-1}, U_{i+1}, \dots, U_n) & = & \{ x \mid \exists x_i \in U_i, i \in \{ 1, \dots, n \} \backslash \{ i \}:  s(x, x_1, \dots, x_{i-1}, x_{i+1}, \dots, x_n) \} \\
& = &  \{ x \mid \exists x_i \in U_i, i \in \{ 1, \dots, n \} \backslash \{ i \}:  r(x, x_1,  \dots, x_n) \} \\
& = & f_{\exists r}(U_1, \dots, U_{i-1}, C, U_{i+1}, \dots, U_n).
\end{eqnarray*}
\end{lemma}

\vspace{-1mm}
\noindent The axioms which corresponds to role restrictions are of the type:

\vspace{-4mm}
\begin{eqnarray}f_{\exists (\exists r.(i,C))}(x_1, \dots, x_{i-1}, x_{i+1}, \dots, x_n) = f_{\exists r}(x_1, \dots, x_{i-1}, C,  x_{i+1}, \dots, x_n).\label{er}
\end{eqnarray} 

\vspace{-1mm}
\noindent All results established for  ${\cal EL}^+$ hold also if this kind of 
role constructions are considered.
\begin{theorem}
Assume the only concept constructors are 
intersection and existential restriction. 
Let ${\mathcal C} = GCI {\cup} RI {\cup} GRI {\cup ER}$ 
be a CBox containing general concept inclusions ($GCI$), (guarded) 
role inclusions ($RI$, resp.\ $GRI$) and a set $ER$
of definitions of roles by existential restrictions 
with concept names $N_C = \{ C_1, \dots, C_n \}$. 
Then for all concept descriptions $D_1, D_2$ the following are equivalent:
\begin{itemize}
\item[(1)] $D_1 {\sqsubseteq}_{\mathcal C} D_2$. 
\item[(2)] $GRI(C_1, \dots, C_n) \wedge 
\left( \bigwedge_{C {\sqsubseteq} D \in GCI} 
\overline{C} {\leq} \overline{D} \right) \wedge 
\overline{D_1} {\not\leq} \overline{D_2}$ is unsatisfiable w.r.t.\ 
${\sf SLO}^{\exists}_{N_R}(RI \cup ER)$. 
\end{itemize}
\end{theorem}
In addition, we may also need to express numerical information. 
\begin{example}
Consider a variant of Example~\ref{ex2}, in which we use a role with arity 4, 
$r_{il}$, where 
$r_{il}(x_1, x_2, x_3, n)$ expresses 
the fact that there exists a route from $x_1$ to $x_2$ passing through 
$x_3$ of length $n$. Also in this situation we would like to talk 
about all routes from $x_1$ to  $x_2$ passing through $x_3$ which are 
shorter than a certain length $l$. This can also be expressed using 
projections as the relation $\exists r_{il}(4, \downarrow l)$, where: 
$$\exists r_{il}(4, \downarrow l) = \{ (x_1, x_2, x_3) \mid \exists x_4 (x_4 \leq l \wedge r_{il}(x_1, x_2, x_3, x_4)) \}.$$
\end{example}

\medskip
\noindent 
We will show that the axioms describing the 
algebraic models for the extensions 
of ${\cal EL}^+$ we considered here are ``local'', a property which ensures 
that the uniform word problem (resp. the problem of checking the validity
of a set of ground unit clauses) is decidable in PTIME. We start by 
presenting a few important results on local theories and local theory extensions.

\vspace{-2mm}
\section{Local theories; local theory extensions} 
\label{locality}
First-order theories are sets of formulae (closed under logical consequence), 
typically the set of all consequences of a set of axioms. 
Alternatively, we may consider the set of all models of a theory. 
In this paper we consider theories specified by their sets of axioms. 
(At places, however, -- usually when talking about local extensions of a theory -- we will refer to a theory, 
and mean the set of all its models.) 

\smallskip
\noindent 
Before defining the notion of local theory and local theory extension 
we will introduce some 
preliminary notions on partial models of a theory. 

\begin{definition}[Partial and total models] 
Let $\Pi = (S, \Sigma, {\sf Pred})$ be a many-sorted signature
with set of sorts $S$, set of function symbol $\Sigma$ and set of predicates 
${\sf Pred}$. 
A partial $\Pi$-structure is a structure $(\{ A_s \}_{s \in S}, \{ f_A \}_{f \in \Sigma}, \{ P_A \}_{P \in {\sf Pred}})$ in which 
for some  function symbols $f \in \Sigma$, $f_A$ may be partial.
\end{definition} 
\begin{definition}
A {\em weak $\Pi$-embedding} between the partial structures 
$A = (\{ A_s \}_{s \in S},$ $\{ f_A \}_{f \in \Sigma},$ 
$\{ P_A \}_{P \in {\sf Pred}})$  
and $B = ( \{ B_s \}_{s \in S}, 
\{ f_B \}_{f \in \Sigma}, \{ P_B \}_{P \in {\sf Pred}})$
is a (many-sorted) family $i = (i_s)_{s \in S}$ 
of total maps $i_s : A_s \rightarrow B_s$ such that 
\begin{itemize}
\item[(i)] if 
$f_A(a_1, \dots, a_n)$ is defined (in $A$) then also 
$f_B(i_{s_1}(a_1), \dots, i_{s_n}(a_n))$ is defined (in $B$) and 
$i_s(f_A(a_1, \dots, a_n)) = f_B(i_{s_1}(a_1), \dots, i_{s_n}(a_n))$; 
\item[(ii)] for each sort $s$, $i_s$ is injective and an embedding w.r.t.\ 
${\sf Pred}$, i.e.\ for every $P \in {\sf Pred}$ 
with arity $s_1 \dots s_n$ and every $a_1, \dots, a_n$ where 
$a_i \in A_{s_i}$,  
$P_A(a_1, \dots, a_n)$ if and only if 
$P_B(i_{s_1}(a_1), \dots, i_{s_n}(a_n))$. 
\end{itemize}
In this case we say that $A$ {\em weakly embeds} 
into $B$. 
\end{definition}

\begin{definition}
If $A$ is a partial structure and $\beta : X \rightarrow A$
is a valuation 
then for every literal $L = (\neg) P(t_1, \dots, t_n)$ with 
$P \in {\sf Pred} {\cup} \{ = \}$ we say that 
$(A, \beta) \models_w L$ 
if:
\begin{itemize}
\item[(i)] either $\beta(t_i)$ are all defined and 
$(\neg) P_A(\beta(t_1), \dots, \beta(t_n))$ is true in $A$, 
\item[(ii)] or $\beta(t_i)$ is not defined for some argument $t_i$ of $P$. 
\end{itemize}
Weak satisfaction of clauses ($(A, \beta) \models_w C$) 
can then be defined in the usual way. We say that $A$ is a 
{\em weak partial model} of a set of clauses ${\mathcal K}$ if $(A, \beta)  \models_w C$ for every $\beta : X \rightarrow A$ and for every clause 
$C \in {\mathcal K}$. 
\end{definition}
The notion of {\em local theory} was introduced by 
Givan and McAllester \cite{GivanMcAllester92,McAllester-acm-tocl-02}.
They studied  sets of Horn clauses 
${\mathcal K}$ with the property that, for any  ground Horn clause $C$, 
${\mathcal K} \models C$ only if 
already ${\mathcal K}[C] \models C$ (where ${\mathcal K}[C]$ is the set of 
instances of ${\mathcal K}$ in which all terms are subterms of 
ground terms in either ${\mathcal K}$ or $C$).
Since the size of ${\mathcal K}[C]$ is polynomial in the size of $C$ for a fixed 
${\mathcal K}$ and 
satisfiability of sets of ground Horn clauses can be checked in 
linear time \cite{DowlingGallier}, it follows that for local theories, 
validity of ground Horn clauses can be checked in polynomial time. 
Givan and McAllester 
proved that every problem 
which is decidable in PTIME can be encoded as an entailment problem of 
ground clauses w.r.t.\ a local theory \cite{McAllester-acm-tocl-02}. 
The property above can easily be generalized to the notion of locality  
of a set of (Horn) clauses:
\begin{definition}
A {\em local theory} is a  set of Horn clauses 
${\mathcal K}$ such that, for any set $G$ of ground Horn clauses, 
${\mathcal K} \cup G \models \perp$ if and only if 
already ${\mathcal K}[G] \cup G \models \perp$,  
where ${\mathcal K}[G]$ is the set of 
instances of ${\mathcal K}$ in which all terms are subterms of 
ground terms in either ${\mathcal K}$ or $G$.
\end{definition}
In \cite{Ganzinger-01-lics}, Ganzinger established a link between 
proof theoretic and semantic concepts for polynomial time 
decidability of uniform word problems which had already been studied 
in algebra \cite{Skolem20,Burris95}. 

\vspace{-2mm}
\subsection{Local theory extensions}
\label{local-ext}
We will also consider extensions of theories, in which the 
signature is extended by new {\em function symbols} (i.e.\ we assume 
that the set of predicate symbols remains unchanged in the extension).
Let ${\mathcal T}_0$ be an arbitrary theory with signature 
$\Pi_0 = (S, \Sigma_0, {\sf Pred})$, where $S$ is a set of sorts, 
$\Sigma_0$ a set of function symbols, and ${\sf Pred}$ 
a set of predicate symbols. 
We consider extensions
${\mathcal T}_1$ of ${\mathcal T}_0$ with signature 
$\Pi = (S, {\Sigma}, {\sf Pred})$,
where the set of 
function symbols is $\Sigma = \Sigma_0 \cup \Sigma_1$
(i.e.\ the 
signature is extended by new function symbols).
We assume that ${\mathcal T}_1$ is obtained from ${\mathcal T}_0$ by 
adding a set ${\mathcal K}$ of (universally quantified) clauses in the 
signature $\Pi$.
Thus, ${\sf Mod}({\mathcal T}_1)$ consists of all $\Pi$-structures 
which are models of ${\mathcal K}$ and 
whose reduct to $\Pi_0$ is a model of ${\mathcal T}_0$.
In what follows, when referring to 
{\em (weak) partial models} of ${\mathcal T}_0 \cup {\mathcal K}'$, 
we mean (weak) partial 
models of ${\mathcal K}'$ whose reduct to $\Pi_0$ is a total model of 
${\mathcal T}_0$.

\vspace{-2mm}
\subsubsection{Locality of an extension}
In what follows, when we refer to sets $G$ of ground clauses 
we assume that they are in the signature 
$\Pi^c = (S, \Sigma \cup \Sigma_c, {\sf Pred})$, 
where $\Sigma_c$ is a set of new constants. 

\medskip
\noindent We will focus on the following type of locality of 
a theory extension 
${\mathcal T}_0 \subseteq {\mathcal T}_1$, where 
${\mathcal T}_1 = {\mathcal T}_0 \cup {\mathcal K}$ with ${\mathcal K}$ a set of (universally
quantified) clauses: 

\medskip
\noindent
\begin{tabular}{@{}ll}
${\sf (Loc)}$  & For every finite set $G$ of ground clauses 
                     ${\mathcal T}_1 {\cup} G \models \perp$ iff 
                     ${\mathcal T}_0 {\cup} {\mathcal K}[G] {\cup} G$ \\
& has no weak partial model with all terms in ${\sf st}({\mathcal K}, G)$ 
  defined. \\
\end{tabular}

\noindent 
Here, ${\sf st}({\mathcal K}, G)$ is the set of all ground terms occurring 
in ${\mathcal K}$ or $G$.

\medskip
\noindent 
We say that an extension ${\mathcal T}_0 \subseteq {\mathcal T}_1$ is 
{\em local} if it satisfies condition ${\sf (Loc)}$. 
(Note that a local equational theory \cite{Ganzinger-01-lics} 
is a local extension of the pure  theory of equality 
(with no function symbols).) 
A more general notion, namely $\Psi$-locality of an extension   
theory (in which the instances to be considered are described 
by a closure operation 
$\Psi$) is introduced in \cite{ihlemann-jacobs-sofronie-tacas08}. 
Let ${\mathcal K}$ be a set of clauses. Let $\Psi_{\mathcal K}$ be a 
function associating with any set $T$ of ground terms 
a set $\Psi_{\mathcal K}(T)$ of ground terms such that 
\begin{itemize}
\item[(i)] all ground subterms in ${\mathcal K}$ and $T$ are in 
$\Psi_{\mathcal K}(T)$;  
\item[(ii)] for all sets of ground terms 
$T, T'$ if $T \subseteq T'$ then 
$\Psi_{\mathcal K}(T) \subseteq \Psi_{\mathcal K}(T')$; 
\item[(iii)] for all sets of ground terms 
$T$,  $\Psi_{\mathcal K}(\Psi_{\mathcal K}(T)) \subseteq \Psi_{\mathcal K}(T)$; 
\item[(iv)] 
$\Psi$ is compatible with any map $h$ between constants, i.e.\ for any
map $h : C \rightarrow C$, 
$\Psi_{\mathcal K}({\overline h}(T)) = {\overline h}(\Psi_{\mathcal K}(T))$, 
where ${\overline h}$ is the unique extension of $h$ to terms. 
\end{itemize}
Let ${\mathcal K}{[\Psi_{\mathcal K}(G)]}$ be 
the set of instances of ${\mathcal K}$ where the variables are instantiated 
with terms in $\Psi_{\mathcal K}({\sf st}({\mathcal K}, G))$ 
(set denoted in what follows by $\Psi_{\mathcal K}(G)$),  where 
${\sf st}({\mathcal K}, G)$ is the set of all ground terms occurring 
in ${\mathcal K}$ or $G$. 
We say that ${\mathcal K}$ is $\Psi$-stably local 
if it satisfies: 

\smallskip
\noindent \begin{tabular}{@{}l@{}l}
$({\sf Loc}^{\Psi})~$ & 
for every finite set $G$ of ground clauses, ${\mathcal K} {\cup} G$
has a model which is a model of ${\cal T}_0$ \\
& iff ${\mathcal K}{[\Psi_{\mathcal K}(G)]} {\cup} G$ has a partial  model
which is a total model of ${\cal T}_0$ and in which all \\
& terms in $\Psi_{\mathcal K}(G)$ are defined.
\end{tabular}

\medskip
\noindent If
${\Psi}_{\mathcal K}(G) = {\sf st}({\mathcal K}, G)$ 
we recover the definition of local theory extension.

\smallskip
\noindent In $\Psi$-local theories and theory extensions hierarchical 
reasoning is possible. We present the ideas for the case of local theories.

\subsubsection{Hierarchical reasoning}
Consider a $\Psi$-local theory extension 
${\mathcal T}_0 \subseteq {\cal T}_1 = {\mathcal T}_0 \cup {\mathcal K}$.
The locality conditions defined above 
require that, for every set $G$ of ground clauses, 
${\mathcal T}_1 \cup G$ is satisfiable  if and only if 
${\mathcal T}_0 \cup {\mathcal K}[\Psi_{\cal K}(G)] \cup G$ has a weak partial model 
with additional properties. 
All clauses in the set ${\mathcal K}[\Psi_{\cal K}(G)] \cup G$ have the property that 
the function symbols in $\Sigma_1$ have as arguments only ground terms. 
Therefore, ${\mathcal K}[\Psi_{\cal K}(G)] \cup G$ can be flattened 
and purified (i.e.\ the function symbols in $\Sigma_1$ are separated from 
the other symbols)
by introducing, in a bottom-up manner, new  constants $c_t$ for
subterms $t = f(g_1, \dots, g_n)$ with $f \in \Sigma_1$, $g_i$ ground 
$\Sigma_0 \cup \Sigma_c$-terms (where $\Sigma_c$ is a set of constants 
which contains the constants introduced by flattening, resp.\ purification), 
together with corresponding definitions $c_t = t$.
The set of clauses thus obtained 
has the form ${\mathcal K}_0 \cup G_0 \cup {\sf Def}$, 
where ${\sf Def}$ is a set of ground unit clauses of the form 
$f(g_1, \dots, g_n) = c$, where $f \in \Sigma_1$, $c$ is a 
constant, $g_1, \dots, g_n$ are ground 
terms without function symbols in $\Sigma_1$,
and ${\mathcal K}_0$ and $G_0$ are clauses without function 
symbols in $\Sigma_1$.  
Flattening and purification
preserve both satisfiability and unsatisfiability w.r.t.\ 
total algebras, and also w.r.t.\ 
partial algebras in which all ground subterms 
which are flattened are defined \cite{Sofronie-cade-05}.

\medskip
\noindent
For the sake of simplicity in what follows we will always 
flatten and then purify ${\mathcal K}[\Psi_{\mathcal K}(G)] \cup G$. Thus we 
ensure that  ${\sf Def}$ consists of ground unit clauses of the form 
$f(c_1, \dots, c_n) = c$, where $f \in \Sigma_1$, and 
$c_1, \dots, c_n, c$ are constants.
\begin{theorem}[\cite{Sofronie-cade-05,ihlemann-jacobs-sofronie-tacas08}]
Let ${\mathcal K}$ be a set of clauses. 
Assume that ${\mathcal T}_0 \subseteq {\cal T}_1 = {\mathcal T}_0 \cup {\mathcal K}$ is a 
$\Psi$-local theory extension, and that for every finite set $T$ of terms 
$\Psi_{\cal K}(T)$ is finite. For any set $G$ of ground clauses, 
let ${\mathcal K}_0 \cup G_0 \cup {\sf Def}$ 
be obtained from ${\mathcal K}[\Psi_{\cal K}(G)] \cup G$ by flattening and purification, 
as explained above. 
Then the following are equivalent:
\begin{itemize}
\item[(1)] $G$ is satisfiable w.r.t.\ ${\cal T}_1$.
\item[(2)] ${\mathcal T}_0 {\cup} {\mathcal K}[\Psi_{\cal K}(G)] {\cup} G$ has a partial 
model with all terms in ${\sf st}({\mathcal K}, G)$ defined.
\item[(3)] ${\mathcal T}_0 {\cup} {\mathcal K}_0 {\cup} G_0 {\cup} {\sf Def}$ has a partial 
model with all terms in ${\sf st}({\mathcal K}, G)$ defined. 
\item[(4)] ${\mathcal T}_0 \cup {\mathcal K}_0 \cup G_0 \cup {\sf Con}[G]_0$ 
has a (total) model, where 

$\displaystyle{~~~ {\sf Con}[G]_0  = \{ \bigwedge_{i = 1}^n c_i = d_i \rightarrow c = d \mid 
f(c_1, \dots, c_n) = c, f(d_1, \dots, d_n) = d \in {\sf Def} \}}.$
\end{itemize}
\label{lemma-rel-transl}
\end{theorem} 
\subsubsection{Parameterized decidability and complexity}
Theorem~\ref{lemma-rel-transl} allows us to show that:
\begin{itemize} 
\item decidability 
of checking satisfiability in a $\Psi$-local extension of a theory 
${\cal T}_0$ is a consequence of the decidability of the 
problem of checking the 
satisfiability of ground clauses in ${\cal T}_0$, and 
\item the 
complexity of the task of checking the satisfiability of sets of ground clauses 
w.r.t.\ a $\Psi$-local extension of a base theory ${\cal T}_0$ can be expressed 
as a function of the complexity of checking the satisfiability of sets of ground clauses in ${\cal T}_0$.
\end{itemize} 
\begin{theorem}[\cite{Sofronie-cade-05}]
Assume that the theory extension 
${\mathcal T}_0 \subseteq {\mathcal T}_1$ satisfies 
 condition ${\sf (Loc)}$.  
If all variables in the clauses in ${\mathcal K}$ occur below some
function symbol\footnote{This requirement ensures that all variables are
  instantiated in ${\mathcal K}[G]$, and that therefore the satisfiability
  problem can be reduced without problems to testing the satisfiability of 
a set of ground clauses.}  from $\Sigma_1$ 
and if testing satisfiability of ground clauses in 
${\mathcal T}_0$ is decidable, 
then testing satisfiability of ground clauses in ${\mathcal T}_1$ is decidable.

Assume in addition that the complexity of testing the satisfiability of 
a set of ground clauses of size $m$ w.r.t.\ ${\cal T}_0$ can be described 
by a function $g(m)$. Let $G$ be a set of ${\cal T}_1$-clauses of size 
$n$. 
Then the complexity of checking the satisfiability of $G$ w.r.t.\ ${\cal T}_1$
is of order $g(n^k)$, where $k$ is the maximum number of free 
variables in a clause in ${\cal K}$, at least $2$. 
\label{complex}
\end{theorem}
\Proof 
This follows from the fact that:
\begin{itemize} 
\item the number of clauses in ${\cal K}_0$ is polynomial in the size of 
$\Psi_{\cal K}(G)$, where the degree $d$ of the polynomial is at most 
the maximum number of free variables in a clause in ${\cal K}$;  
\item the number of clauses in $G_0$ is linear in the size of $G$; 
\item the number of clauses in ${\sf Con}[G]_0$ is quadratic in the size of $G$.\QED
\end{itemize}

\subsubsection{Recognizing local theory extensions}
The locality of an extension 
can be recognized by proving embeddability of partial models into 
total models 
\cite{Sofronie-cade-05,sofronie-ihlemann-ismvl-07,ihlemann-jacobs-sofronie-tacas08}. We will use the following notation: 

\medskip
\noindent \begin{tabular}{@{}ll}
${\sf PMod^{\Psi}_w}({\Sigma_1}, {\mathcal T}_1)$ & is  
the class of all weak partial models $A$ of ${\mathcal T}_1 = {\cal T}_0 \cup {\cal K}$ in which the \\
& $\Sigma_1$-functions are partial, the $\Sigma_0$-functions are total, and the 
set of terms \\
& $\{ f(a_1, \dots, a_n) \mid f_A(a_1, \dots, a_n) \text{ defined} \}$ is closed under $\Psi_{\cal K}$.
\end{tabular}

\medskip
\noindent For extensions 
${\mathcal T}_0 \subseteq  {\mathcal T}_1 = {\mathcal T}_0 \cup {\mathcal K}$,
where ${\mathcal K}$ is a set of clauses, 
we consider the 
condition:

\medskip
\noindent \begin{tabular}{@{}ll}
${\sf (Emb^{\Psi}_w)}$  & Every 
$A \in {\sf PMod^{\Psi}_w}({\Sigma_1}, {\mathcal T}_1)$ 
weakly embeds into a total model of ${\mathcal T}_1$. 
\end{tabular}

\medskip
\noindent In what follows we say that a non-ground clause is $\Sigma_1$-{\em flat} 
if function symbols (including constants) do not occur 
as arguments of function symbols in $\Sigma_1$.
A $\Sigma_1$-flat non-ground clause is called $\Sigma_1$-{\em linear} 
if whenever a variable occurs in two terms in the clause 
which start with function symbols in $\Sigma_1$, 
the two terms are identical, and if 
no term which starts with a function symbol 
in $\Sigma_1$ contains two occurrences of the same variable.

Flatness and linearity are important because for flat and linear sets
of axioms locality can be checked using semantic means. It is easy to 
see that every set of clauses can be flattened and linearized. 
Please note however that after flattening and linearization 
the set of instances in ${\cal K}[G]$ (resp.\ ${\cal K}[\Psi(G)]$ 
usually changes.
\begin{theorem}[\cite{ihlemann-jacobs-sofronie-tacas08}]
Let ${\mathcal K}$ be a set of $\Sigma$-flat and $\Sigma$-linear clauses.
If the extension 
${\mathcal T}_0 \subseteq {\mathcal T}_1 = {\cal T}_0 \cup {\cal K}$ 
satisfies ${\sf (Emb^{\Psi}_w)}$ 
-- where $\Psi$ satisfies conditions (i)--(iv) in Section~\ref{local-ext} -- 
then the extension satisfies ${\sf (Loc^{\Psi})}$.
\label{rel-loc-embedding}
\end{theorem}
\Proof Assume that ${\mathcal T}_0 \cup {\mathcal K}$ is not a
$\Psi$-local extension of ${\mathcal T}_0$. Then there exists 
a set $G$ of ground clauses (with additional constants) 
such that 
${\mathcal T}_0 \cup {\mathcal K} \cup G \models \perp$ but 
${\mathcal T}_0 \cup {\mathcal K}[\Psi_{\cal K}(G)] \cup G$ 
has a weak partial model $P$ 
in which all terms in $\Psi_{\cal K}(G)$ are defined.
We assume w.l.o.g.\ that $G = G_0 \cup G_1$, 
where $G_0$ contains no function symbols in 
$\Sigma_1$ and $G_1$ consists of ground unit clauses of the form
$f(c_1, \dots, c_n) \approx c,$
where 
$c_i, c$ are constants in $\Sigma_0 \cup \Sigma_c$
and $f \in \Sigma_1$.\footnote{All 
results below hold if only purified goals are considered;
flattening and linearity of goals is not absolutely necessary.} 

We construct another structure,  $A$, having the same support as $P$, 
which inherits all relations in ${\sf Pred}$ and 
all maps in $\Sigma_0 \cup \Sigma_c$ from $P$, but on which  
the domains of definition of the $\Sigma_1$-functions are restricted 
as follows: for every $f \in \Sigma_1$, 
$f_A(a_1, \dots, a_n)$ is defined if and only if 
there exist constants $c^1, \dots, c^n$ such that 
$f(c^1, \dots, c^n)$ is in $\Psi_{\cal K}(G)$ and 
$a^i = c^i_P$ for all $i \in \{ 1, \dots, n \}$. 
In this case we define $f_A(a_1, \dots, a_n) := f_P(c^1_P, \dots, c^n_P)$.
The reduct of $A$ to $(\Sigma_0 \cup \Sigma_c, {\sf Pred})$ 
coincides with that of $P$. 
Thus, $A$ is a model of ${\mathcal T}_0 \cup G_0$. 
By the way the operations in $\Sigma_1$ are defined in $A$ it is 
clear that $A$ satisfies $G_1$, so $A$ satisfies $G$. 

To show that $A \models_w {\mathcal K}$ we use the fact that
if $D$ is a clause in ${\mathcal K}$ and $\beta : X \rightarrow A$
is an assignment in which $\beta(t)$ is defined
for every term $t$ occurring in $D$, 
then (by the way $\Sigma_1$-functions are defined in $A$) 
we can construct a substitution $\sigma$ with 
$\sigma(D) \in {\mathcal K}[G]$ and $\beta \circ \sigma = \beta$. 
As $(P, \beta) \models_w \sigma(D)$ we can infer $(A, \beta) \models_w D$.

We now show that $D(A) = \{ f(a_1, \dots, a_n) \mid f_A(a_1, \dots, a_n) \text{ defined} \}$ is closed under $\Psi_{\cal K}$. 
By definition, $f(a_1, \dots, a_n) \in D(A)$ iff $\exists \text{ constants } c_1, \dots, c_n$ with ${c_i}_A = a_i$ for all $i$ and $f(c_1, \dots, c_n) \in \Psi_{\cal K}(G)$. Thus, 
$$\begin{array}{rll}
D(A) & =  \{ f(a_1, \dots, a_n) \mid f_A(a_1, \dots, a_n) \text{ defined} \} & \\
      & =  \{ f({c_1}_A, \dots, {c_n}_A) \mid c_i \text{ constants with } f(c_1, \dots, c_n) \in \Psi_{\cal K}(G) \} & \\
      & =  {\overline h}(\Psi_{\cal K}(G)) & \!\!\!\!\!\!\!\!\!\!\!\!\!\!\!\!\!\!\!\!\!\!\!\!\!\!\!\!\!\!\!\!\!\!\!\!\text{ where } h(c_i) = a_i \text{ for all } i \\
\Psi_{\cal K}(D(A)) & =  \Psi_{\cal K}({\overline h}(\Psi_{\cal K}(G)))  = {\overline h}(\Psi_{\cal K}(\Psi_{\cal K}(G))) & \!\!\!\!\!\!\!\!\!\!\!\!\!\!\!\!\!\!\!\!\!\!\!\!\!\!\!\!\!\!\!\!\!\!\!\!\text{ by property (iv) of } \Psi \\
& \subseteq {\overline h}(\Psi_{\cal K}(G)) = D(A) & \!\!\!\!\!\!\!\!\!\!\!\!\!\!\!\!\!\!\!\!\!\!\!\!\!\!\!\!\!\!\!\!\!\!\!\!\text{ by property (iii) of } \Psi\\

\end{array}$$
As $A \models_w {\mathcal K}$, 
$A$ weakly embeds into a total algebra $B$ satisfying 
${\mathcal T}_0 \cup {\mathcal K}$. 
But then $B \models G$, 
so $B \models {\mathcal T}_0 \cup {\mathcal K} \cup G$, 
which is a contradiction. \QED

\medskip
\noindent 
Analyzing the proof of Theorem~\ref{rel-loc-embedding} we notice that 
the $\Sigma_1$-linearity restriction can be relaxed. We can 
allow a variable $x$ to occur below two unary function symbols 
$g$ and $h$ in a clause $C$ if $\Psi_{\cal K}$ has the property that 
for every constant $c$, if 
$g(c) \in \Psi_{\cal K}(G)$ then $h(c) \in \Psi_{\cal K}(G)$
or vice versa. (In terms of partial models this means that 
we consider models $A$ with the property that if $g_A(a)$ is defined 
then $h_A(a)$ is defined or vice versa.)

The linearity condition can be similarly relaxed 
in the presence of $n$-ary functions, namely for groups of function symbols
$(g_1, \dots, g_n, h)$ -- which occur in axioms containing clauses in which 
the following sets of terms occur at the same time:
$$\{ g_i({\overline x}_i) \mid 1 \leq i \leq n \} \cup \{ h({\overline x}_1, \dots, {\overline x}_n)\},$$
where the sets of variables ${\overline x}_i$ and ${\overline x}_j$ are disjoint 
for $i \neq j$ --  
with the property that if 
($g_i(c^i_1, \dots, c^i_{n_i}) \in \Psi_{\cal K}(G)$ for all $i$) then  $h({\overline c^1}, \dots, {\overline c^n}) \in \Psi_{\cal K}(G)$ or vice versa.

\vspace{-2mm}
\section{Locality and complexity of $\mathcal{EL}^+$ and $\mathcal{EL}$ and extensions thereof}
\label{complexity}

We now show that the classes of algebraic models of 
$\mathcal{EL}^+$ and of $\mathcal{EL}$ (and of their extensions presented 
in Sections~\ref{alg-sem-el} and~\ref{extensions-of-el-n-ary}) 
have presentations which satisfy 
certain locality properties. This gives an alternative, 
algebraic explanation of the fact that CBox subsumption in these logics is 
decidable in PTIME, and makes generalizations possible.

\vspace{-3mm}
\subsection{Locality and $\mathcal{EL}$} 
\label{loc-el}
In \cite{Sofronie-amai-07} we proved that the algebraic counterpart of the 
description logic $\mathcal{EL}$ --  namely the class of semilattices 
with monotone operators --  has a local 
axiomatization -- ${\cal SL} \cup {\sf Mon}(\Sigma)$ --  
i.e.\ an axiomatization with the property 
that for every set $G$ of ground clauses 
\[ {\cal SL} \cup {\sf Mon}(\Sigma) \cup G  \models \perp \quad \mbox{  if and only if } \quad ({\cal SL} \cup {\sf Mon}(\Sigma))[G] \cup G \models \perp. \]
We denoted by ${\sf Mon}(\Sigma)$ the set 
$\{ {\sf Mon}(f) \mid f \in \Sigma \}$, where 
$${\sf Mon}(f)~~~~ \forall x, y (x \leq y \rightarrow f(x) \leq f(y)).$$
In \cite{Sofronie-cade-05} we  
showed that the extension $SLO_{\Sigma} = SL {\cup} {\sf Mon}(\Sigma)$ of the theory $SL$ of bounded semilattices with a family of monotone functions is local. 
\begin{theorem}[\cite{Sofronie-cade-05,sofronie-ihlemann-ismvl-07}]
\label{locality-of-el}
Let $G$ be a set of ground clauses. The following are equivalent:
\begin{itemize}
\item[(1)] $SL \cup {\sf Mon}(\Sigma)  \cup G  \models \perp$.
\item[(2)] $SL \cup {\sf Mon}(\Sigma)[G] \cup G$ has no 
partial model $A$ such that its $\{ \wedge, 0, 1 \}$-reduct is a 
(total) bounded semilattice, the functions in $\Sigma$ are partial
and all $\Sigma$-subterms of $G$ are defined.
\end{itemize}
\end{theorem}
Let 
${\sf Mon}(\Sigma)[G]_0 \cup G_0 \cup {\sf Def}$ be 
obtained from ${\sf Mon}(\Sigma)[G] \cup G$ by purification, 
i.e.\ by replacing, in a bottom-up manner, all subterms 
$f(g)$ with $f \in \Sigma$, with newly introduced constants $c_{f(g)}$ and  
adding the definitions $f(g) = c_t$ to the set ${\sf Def}$.

\begin{theorem}
The following are equivalent (and equivalent to (1) and (2) above):  
\begin{itemize}
\item[(3)] ${\sf Mon}(\Sigma)[G]_0 \cup G_0 \cup 
{\sf Def}$ has no 
partial model $A$ such that its $\{ \wedge, 0, 1 \}$-reduct is a 
(total) bounded semilattice, the functions in $\Sigma$ are partial
and all $\Sigma_1$-subterm of $G$ are defined.
\item[(4)] ${\sf Mon}(\Sigma)[G]_0 \cup G_0$ is unsatisfiable in $SL$. 

(Note that in the presence of ${\sf Mon}(\Sigma)$ the instances 
${\sf Con}[G]_0$ of the 
congruence axioms for the functions in $\Sigma$ are not necessary.)

\smallskip
${\sf Con}[G]_0  = \{ g {=} g' \rightarrow c_{f(g)} {=} c_{f(g')} \mid 
f(g) {=} c_{f(g)}, f(g') {=} c_{f(g')} \in {\sf Def} \}.$
\end{itemize}
\end{theorem}
This equivalence allows us to hierarchically reduce, in polynomial time, 
proof tasks in $SL \cup {\sf Mon}(\Sigma)$ to proof tasks in $SL$ (cf. e.g.\ 
\cite{sofronie-ihlemann-ismvl-07}) which can then be solved in 
polynomial time.

\begin{example}
We illustrate the method on an example first considered in \cite{Baader2003}. 
Consider the $\mathcal{EL}$ TBox ${\mathcal T}$ consisting of the following definitions: 
$$\begin{array}{lll}
A_1 & = & P_1 \sqcap A_2 \sqcap \exists r_1. \exists r_2. A_3 \\
A_2 & = &  P_2 \sqcap A_3 \sqcap \exists r_2. \exists r_1. A_1 \\
A_3 & = &  P_3 \sqcap A_2 \sqcap \exists r_1.(P_1 \sqcap P_2) \\
\end{array}$$
We want to prove that 
$P_3 \sqcap A_2 \sqcap \exists r_1.(A_1 \sqcap A_2) \sqsubseteq_{\mathcal T} A_3$.
We translate this subsumption problem to the following satisfiability problem: 
\begin{eqnarray*}
{\sf SL} \cup {\sf Mon}(f_1, f_2) & \cup & 
\{ \, a_1 =  (p_1 \wedge a_2 \wedge f_1(f_2(a_3))),  \\
& & ~~ a_2 =  (p_2 \wedge a_3 \wedge f_2(f_1(a_1))), \\
& & ~~ a_3 = (p_3 \wedge a_2 \wedge f_1(p_1 \wedge p_2)), \\
& & ~~\neg (p_3 \wedge a_2 \wedge f_1(a_1 \wedge a_2)  \leq a_3) \} \models \perp.
\end{eqnarray*}
We proceed as follows: We flatten and purify the set $G$ of ground clauses by 
introducing new names for the terms starting with the function symbols
$f_1$ or $f_2$.
Let ${\sf Def}$ be the corresponding set of definitions. 
We then take into account only those instances of the monotonicity and 
congruence axioms 
for $f_1$ and $f_2$ which correspond to the instances in ${\sf Def}$, 
and purify them as well, by replacing the terms themselves with the constants
which denote them. We obtain the following separated set of formulae:
$$\begin{array}{|l|ll|}
\hline 
~{\sf Def} & ~~~~~~~~~~~~~~~G_0 ~~~~~~~ \cup & ({\sf Mon}(f_1, f_2)[G])_0 \cup {\sf Con}[G]_0  \\
\hline 
\hline 
~f_2(a_3) = c_1~ & ~(a_1 =  p_1 \wedge a_2 \wedge c_2)~~~~~ & a_1 R c_1 \rightarrow c_3 R c_2, ~~R \in \{ \leq, \geq, = \} \\
~f_1(c_1) = c_2~ & ~(a_2 =  p_2 \wedge a_3 \wedge c_4) & a_3 R c_3 \rightarrow c_1 R c_4, ~~R \in \{ \leq, \geq, = \} \\
~f_1(a_1) = c_3~ & ~(a_3 = p_3 \wedge a_2 \wedge d_1)  & a_1 R e_1 \rightarrow c_3 R d_1, ~~R \in \{ \leq, \geq, = \} \\
~f_2(c_3) = c_4~ & ~(p_3 \wedge a_2 \wedge d_2 \not\leq a_3) & a_1 R e_2 \rightarrow c_3 R d_2, ~~R \in \{ \leq, \geq, = \} \\
~f_1(e_1) = d_1~ & ~p_1 \wedge p_2 = e_1 & c_1 R e_1 \rightarrow c_2 R d_1, ~~R \in \{ \leq, \geq, = \} \\
~f_1(e_2) = d_2~ & ~a_1 \wedge a_2 = e_2 & c_1 R e_2 \rightarrow c_2 R d_2, ~~R \in \{ \leq, \geq, = \} \\
& & e_1 R e_2 \rightarrow d_1 R d_2, ~~R \in \{ \leq, \geq, = \} \\
\hline 
\end{array}$$
The subsumption is true iff $G_0  \cup ({\sf Mon}(f_1, f_2)[G])_0 \cup {\sf Con}[G]_0$ is unsatisfiable in the 
theory of semilattices. We can see this as follows: note that $a_1 \wedge a_2 \leq p_1 \wedge p_2$, i.e. $e_2 \leq e_1$. 
Then (using an instance of monotonicity) $d_2 \leq d_1$, so $p_3 \wedge a_2 \wedge d_2 \leq p_3 \wedge a_2 \wedge d_1 = a_3$.

This can also be checked automatically in PTIME either by using the 
fact that there exists a local presentation of ${\sf SL}$ 
(cf.\ also Sect.~\ref{el-compl})
or using the fact that 
${\sf SL} = ISP(S_2)$ (i.e. every semilattice is isomorphic with a 
sublattice of a power of $S_2$), where $S_2$ is the 
semilattice with two elements, hence 
${\sf SL}$ and $S_2$ satisfy the same Horn clauses. Since 
the theory of semilattices is convex, satisfiability of 
ground clauses w.r.t. ${\sf SL}$  
can be reduced to SAT solving.
\end{example}
\vspace{-3mm}
\subsection{Locality and $\mathcal{EL}^+$} 
\label{loc-el+}
We prove that similar results hold for 
the class $SLO_{\Sigma}(RI)$ of 
semilattices with monotone operators in a set $\Sigma$ satisfying a 
family $RI$  axioms of the form: 
\begin{eqnarray*}
\forall x  ~ &  g(x) & \leq  h(x) \\
\forall x ~ &  f(g(x)) & \leq  x\\
\forall x ~ &  f(g(x)) & \leq  h(x) 
\end{eqnarray*}
Since the characterization of locality in Theorem~\ref{rel-loc-embedding} 
refers to 
sets of {\em flat} clauses, instead of $RI$ we consider  
the flat versions $RI^{\sf flat}$ of this family of axioms: 
\begin{eqnarray*}
\forall x ~ ~  ~ ~ &  & g(x)  \leq  h(x) \\
\forall x, y ~  & (y \leq g(x)  \rightarrow & f(y) \leq x)\\
\forall x, y ~  & (y \leq g(x)  \rightarrow & f(y) \leq h(x)) 
\end{eqnarray*}
\begin{theorem}
The extension $SL \cup {\sf Mon}(\Sigma) \cup RI^{\sf flat}(2)$ of the 
theory of semilattices 
with monotone functions $f, g$ satisfying axioms of the second type in  
$RI^{\sf flat}$ above is local.
\label{loc-2}  
\end{theorem}
\Proof We have to prove that every weak partial model of 
$SL \cup {\sf Mon}(\Sigma) \cup RI^{\sf flat}(2)$ weakly embeds into 
a total model of $SL \cup {\sf Mon}(\Sigma) \cup RI^{\sf flat}(2)$. 
This follows from Lemma~\ref{local-alg-el+}. \QED
\begin{theorem}
The extension $SL \cup {\sf Mon}(\Sigma) \cup RI^{\sf flat}(1, 3)$ 
of the theory of lattices 
with monotone functions satisfying axioms of the first or 
third type in $RI^{\sf flat}$ above is $\Psi$-local, where 
$\Psi(T) = \bigcup_{i \geq 1} \Psi_i(T)$, with 
$\Psi_0(T) = T$, and 
$$\begin{array}{rcl}
\Psi_{i+1}(T) & = & \{ h(c) \mid \forall x (g(x)  \rightarrow h(x)) \in RI^{\sf flat} \text{ and } g(c) \in T \} \cup \\
& &  \{ h(c) \mid \forall x (y \leq g(x) \rightarrow f(y) \leq h(x)) \in RI^{\sf flat} \text{ and } g(c) \in T \}.
\end{array}$$
\label{loc-1-3}
\end{theorem}
\Proof Note first that the clauses we consider (see below) are flat, 
but not linear. 
\begin{eqnarray*}
\forall x ~~~~ ~ &  & g(x)  \leq  h(x) \\
\forall x, y ~  & (y \leq g(x)  \rightarrow & f(y) \leq h(x)) 
\end{eqnarray*}
As mentioned before, a small change in the 
proof of Theorem~\ref{rel-loc-embedding} allows us to relax the linearity 
condition on the sets of clauses. By Theorem~\ref{rel-loc-embedding}, 
an extension of $SL$ with monotonicity axioms and 
clauses of the type above is $\Psi$-local provided that every 
partial model $S$ of $SL \cup {\sf Mon}(\Sigma) \cup RI(1, 3)$ with a total bounded semilattice 
reduct and with the property that 
if $g_S(a)$ is defined then $h_S(a)$ is defined (for all $g$ and $h$ occurring
at the positions they have in the axioms above) weakly embeds into a 
total model of $SL \cup{\sf Mon}(\Sigma) \cup  RI(1, 3)$. 
The proof of the fact that this embeddability result holds 
is a consequence of Lemma~\ref{local-alg-el+}. \QED
\begin{theorem}
Any extension of the theory $SL$ of semilattices 
with a set of monotone functions satisfying axioms of type $RI$ is $\Psi$-local, where 
$\Psi$ is defined as above. 
\label{locality-of-el+}
\end{theorem}
\Proof This is a consequence of Theorems~\ref{loc-2} and~\ref{loc-1-3} 
and of the fact that the same completion was used in all cases.
\QED
\begin{theorem}
Any theory of the form $SL \cup {\sf Mon}(\Sigma) \cup RI \cup GRI(c_1, \dots, c_n)$ -- where 
$GRI$ are guarded forms of axioms corresponding to role inclusions, as
discussed in Section~\ref{sect:slo-el+} 
-- is $\Psi$-local, where $\Psi(T)$ is as defined above.
\end{theorem} 
\Proof The proof is analogous to the proof of Theorems~\ref{loc-2} 
and~\ref{loc-1-3}.  
We illustrate, as an example, the completion process for the case of 
axioms of the type 
$$ \forall x (x \leq c \wedge y \leq g(x) \rightarrow f(y) \leq h(x)).$$
Let $S$ be a bounded 
semilattice with partial operators satisfying the axioms in 
${\sf Mon}(\Sigma) \cup RI \cup GRI(c_1, \dots, c_n)$. We extend the functions 
to ${\overline S}  = {\cal OI}(S)$ as explained in Lemma~\ref{local-alg-el+}.
Let $\eta : S \rightarrow {\cal OI}(S)$ defined by $\eta(x) = \downarrow x$. 
Then $i(c) = \downarrow c$. 
Let now $U, V \in {\overline S}$ be such that 
$V \subseteq \downarrow c$ and $U \subseteq {\overline g}(V)$.
Let $x \in {\overline f}(U)$, so there exist  $u \in U$ 
for which $f(u)$ is defined, and $v \in V$ with 
$g(v)$ defined such that $v \leq c$,  $x \leq f(u)$ and $u \leq g(v)$. 
By the $\Psi$-closure condition, $h(v)$ is defined as well. Thus, 
$x \leq f(u) \leq h(v)$, i.e.\ $x \in {\overline h}(V)$. 
The other guarded cases can be handled similarly. \QED

\medskip
\begin{example}
We illustrate the ideas on an example presented in 
\cite{Baader-2005} (here slightly simplified). 
Consider the CBox ${\mathcal C}$ consisting of the following $GCI$:

\vspace{-3mm}  
$$\begin{array}{@{}r@{}c@{}l} 
{\sf Endocard} & \,\sqsubseteq\, & {\sf Tissue} \sqcap \exists {\sf cont}\text{-}{\sf in}.{\sf HeartWall} \sqcap  
\exists {\sf cont}\text{-}{\sf in}.{\sf HeartValve}  \\
{\sf HeartWall} & \,\sqsubseteq\, & \exists {\sf part}\text{-}{\sf of}.{\sf Heart}  \\
{\sf HeartValve} & \,\sqsubseteq\, & \exists {\sf part}\text{-}{\sf of}.{\sf Heart}  \\
{\sf Endocarditis} & \,\sqsubseteq\, & {\sf Inflammation} \sqcap \exists {\sf has}\text{-}{\sf loc}.{\sf Endocard} \\ 
{\sf Inflammation} & \,\sqsubseteq\, & {\sf Disease} \\
{\sf Heartdisease} & = & {\sf Disease} \sqcap \exists {\sf has}\text{-}{\sf loc}.{\sf Heart} 
\end{array}$$

\vspace{-2mm}
\noindent and the following role inclusions $RI$: 

\vspace{-2mm}
$$\begin{array}{@{}r@{}c@{}l} 
{\sf part}\text{-}{\sf of} \circ {\sf part}\text{-}{\sf of} & \,\sqsubseteq\, & {\sf part}\text{-}{\sf of} \\
{\sf part}\text{-}{\sf of} & \,\sqsubseteq\, &  {\sf cont}\text{-}{\sf in}\\
{\sf has}\text{-}{\sf loc} \circ  {\sf cont}\text{-}{\sf in} & \,\sqsubseteq\, & {\sf has}\text{-}{\sf loc}
\end{array}$$

\vspace{-1mm}
\noindent We want to check whether
 ${\sf Endocarditis} \sqsubseteq_{\mathcal C} {\sf Heartdisease}$. 
This is the case iff (with some abbreviations -- e.g. 
$f_{\sf ci}$ stands for $f_{\exists {\sf cont}\text{-}{\sf in}}$ and 
$f_{\sf po}$ for $f_{\exists {\sf part}\text{-}{\sf of}}$, $h_w$ and 
$h_v$ for ${\sf HeartWall}$ resp. ${\sf HeartValve}$, 
$e$ for ${\sf  Endocard}$, $h$ for ${\sf Heart}$, etc.): 

\medskip
\noindent $\begin{array}{@{}lllll} 
SL & \cup & {\sf Mon}(f_{\sf ci}, f_{\sf hl}, f_{\sf po}) & \cup & \{ \forall x ~ y \leq f_{\sf ci}(x) \rightarrow f_{\sf ci}(y) {\leq}   f_{\sf ci}(x), \\
 && & &   ~\, 
\forall x ~ f_{\sf po}(x) {\leq} f_{\sf ci}(x), \\
 &&  & &   ~\, 
\forall x ~ y \leq f_{\sf ci}(x) \rightarrow f_{\sf hl}(y) {\leq}  f_{\sf hl}(x) \}  \\
\end{array}$ 

$\begin{array}{lll} 
~~~~   & \cup & \{ e \leq t \wedge f_{\sf ci}(h_w) \wedge  f_{\sf ci}(h_v),  h_w \leq f_{\sf po}(h), ~~h_v \leq  f_{\sf po}(h), \\
& & ~ {\sf Endocarditis} \leq i \wedge f_{\sf hl}(e), ~~ i \leq d, ~~ {\sf Heartdisease}= d \wedge f_{\sf hl}(h), \\
& & ~  {\sf Endocarditis} \not\leq  {\sf Heartdisease} \} ~~\models~~ \perp.
\end{array}$

\medskip
\noindent Then ${\sf st}({\mathcal K}, G) = \{ f_{\sf ci}(h_w),f_{\sf ci}(h_v), f_{\sf po}(h), f_{\sf hl}(e), f_{\sf hl}(h)\}$. It follows that 
$\Psi_{\mathcal K}(G)$ consists of the following terms: 
$\{ f_{\sf ci}(h_w),f_{\sf ci}(h_v), f_{\sf ci}(h), f_{\sf po}(h), f_{\sf hl}(e), f_{\sf hl}(h), f_{\sf hl}(h_w), f_{\sf hl}(h_v)\}$.
After computing $( RI_a \cup {\sf Mon}(f_{\sf ci}, f_{\sf hl}, f_{\sf po}) \cup {\sf Con}){[\Psi(G)]}$ we obtain:
$$\begin{array}{|l|l@{}l|}
\hline 
~G & (RI_a \cup {\sf Mon} \cup  {\sf Con})[\Psi(G)] & \\
\hline 
\hline 
~e \leq t \wedge f_{\sf ci}(h_w) \wedge f_{\sf ci}(h_v) & ~y \leq f_{\sf ci}(x) \rightarrow f_{\sf ci}(y) \leq   f_{\sf ci}(x)~~~ & \text{ for  } x, y \in \{ h_v, h_w, h \}, x \neq y \\
~h_w \leq f_{\sf po}(h) & ~f_{\sf po}(h) \leq f_{\sf ci}(h) & \\ 
~h_v \leq  f_{\sf po}(h) & ~y \leq f_{\sf ci}(x) \rightarrow f_{\sf hl}(y) \leq  f_{\sf hl}(x) & \text{ for  } x \in \{ h_v, h_w, h \}\\
& & ~~~~~~ y \in \{ e, h, h_w, h_v \}, x \neq y \\
~{\sf Endocarditis} \leq i \wedge f_{\sf hl}(e) & ~~~ &  \\
~i \leq d &  ~x R y \rightarrow f_{\sf ci}(x) R f_{\sf ci}(y) & \text{ for  } x, y \in \{ h_w, h_v, h \}, x \neq y \\
~ {\sf Heartdisease}= d \wedge f_{\sf hl}(h) & ~ x R y \rightarrow f_{\sf hl}(x) R f_{\sf hl}(y) & \text{ for  } x, y \in \{ e, h, h_w, h_v \} \\
~{\sf Endocarditis} \not\leq  {\sf Heartdisease}~ &  R \in \{ \leq, \geq \} & \\[1ex]
\hline 
\end{array}$$
We can simplify the problem even further by replacing the ground terms in 
$\Psi(G)$ with new constants, and taking into account the corresponding 
definitions $c_t = t$. Let 
$(RI_a \cup {\sf Mon} \cup  {\sf Con}){[\Psi(G)]}_0$ be the set of 
clauses obtained this way. 
$$\begin{array}{|l|l@{}l|}
\hline 
~G_0 & (RI_a \cup {\sf Mon} \cup  {\sf Con})[\Psi(G)]_0 & \\
\hline 
\hline 
~e \leq t \wedge c_{f_{\sf ci}(h_w)} \wedge c_{f_{\sf ci}(h_v)} & ~y \leq c_{f_{\sf ci}(x)} \rightarrow c_{f_{\sf ci}(y)} \leq   c_{f_{\sf ci}(x)} & \text{ for  } x, y \in \{ h_v, h_w, h \}, x \neq y \\
~h_w \leq c_{f_{\sf po}(h)} & ~c_{f_{\sf po}(h)} \leq c_{f_{\sf ci}(h)} & \\ 
~h_v \leq  c_{f_{\sf po}(h)} & ~y \leq c_{f_{\sf ci}(x)} \rightarrow c_{f_{\sf hl}(y)} \leq  c_{f_{\sf hl}(x)}~~~ & \text{ for  } x \in \{ h_v, h_w, h \}\\
& & ~~~~~~ y \in \{ e, h, h_w, h_v \}, x \neq y~~\\
~{\sf Endocarditis} \leq i \wedge c_{f_{\sf hl}(e)} & ~~~ &  \\
~i \leq d &  ~x R y \rightarrow c_{f_{\sf ci}(x)} R c_{f_{\sf ci}(y)} & \text{ for  } x, y \in \{ h_w, h_v, h \}, x \neq y \\
~ {\sf Heartdisease}= d \wedge c_{f_{\sf hl}(h)} & ~ x R y \rightarrow c_{f_{\sf hl}(x)} R c_{f_{\sf hl}(y)} & \text{ for  } x, y \in \{ e, h, h_w, h_v \} \\
~{\sf Endocarditis} \not\leq  {\sf Heartdisease}~ &  R \in \{ \leq, \geq \} & \\[1ex]
\hline 
\end{array}$$
With the notation in the previous table, by Corollary~\ref{cor-stable-loc}, 
${\sf Endocarditis} \sqsubseteq_{\mathcal C} {\sf Heartdisease}$ iff 
$G_0 \cup (RI_a \cup {\sf Mon} \cup  {\sf Con}){[\Psi(G)]}_0 \models_{SL} \perp$
(i.e.\ it is unsatisfiable w.r.t. the theory of semilattices with 0 and 1). 
The satisfiability of $\phi$ 
can therefore be checked automatically 
in polynomial time in the size of $\phi$ which in its turn 
is polynomial in the size of ${\Psi}_{\mathcal K}(G)$. Hence, in this case, 
the size of $\phi$ is polynomial in the size of $G$. 

Unsatisfiability can also be proved directly: 
$G$ entails the inequalities:
$$\begin{array}{lllll}
(1) & {\sf Endocarditis} \leq (d \wedge f_{\sf hl}(e)); & ~~~~&  (2) &  
e \leq (f_{\sf ci}(h_w) \wedge f_{ci}(h_v)); \\
(3) & (h_w \leq f_{\sf po}(h));~~~~~~~~~~~~~~~~ & & (4) &  (h_v \leq f_{\sf po}(h)). 
\end{array}$$
Hence $G \wedge (RI_a \wedge {\sf Mon} \wedge  {\sf Con})^{[\Psi(G)]} \models e \leq f_{\sf ci}(f_{\sf po}(h)) \leq f_{\sf ci}(f_{\sf ci}(h)) \leq  f_{\sf ci}(h)$.
Thus, $G \wedge (RI_a \wedge {\sf Mon} \wedge  {\sf Con})^{[\Psi(G)]} 
\models f_{\sf hl}(e) \leq f_{\sf hl}(f_{\sf ci}(h)) \leq f_{\sf hl}(h)$, 
so $G \wedge (RI_a \wedge {\sf Mon} \wedge  {\sf Con})^{[\Psi(G)]} 
\models {\sf Endocarditis} \leq d \wedge f_{\sf hl}(h)$, 
which together with $d \wedge f_{\sf hl}(h)  = {\sf Heartdisease}$ and 
${\sf Endocarditis} \not\leq  {\sf Heartdisease}$ leads 
to a contradiction.
\end{example}

\subsection{Complexity} 
\label{el-compl}
We now analyze the complexity of the problem of checking 
CBox subsumption in the extensions of ${\cal EL}^+$ considered in this paper. 
Note that by Theorems~\ref{locality-of-el} and~\ref{locality-of-el+}, 
in all cases considered in Section~\ref{loc-el} and \ref{loc-el+} 
we can reduce CBox subsumption to the task of 
checking the satisfiability of a set of constraints of the form 
$$RI[\Psi(G)]_0 \cup {\sf Mon}(\Sigma)[\Psi(G)]_0 \cup G_0$$ 
w.r.t.\ the theory of bounded semilattices.

\begin{lemma} 
For the specific closure operator $\Psi$ we consider, the following hold:
\begin{itemize}
\item The size of $\Psi(G)$ is linear in the size of $|st(G)|$, where 
$|st(G)|$ is the number of subterms of $G$ which start with a function 
symbol in $\Sigma$.  
\item The size of ${\sf Mon}(\Sigma)[\Psi(G)]$ (and hence also 
the size of ${\sf Mon}(\Sigma)[\Psi(G)]$) is $2 |\Psi(G)|^2$, hence it is 
quadratic in the size of  $|st(G)|$.  
\item The size of 
$RI[\Psi(G)]$ (hence also the size of $RI[\Psi(G)]_0$) is quadratic in the size of $\Psi(G)$, hence also in the size of  $|st(G)|$. 
\end{itemize}
\end{lemma}
We reduced the initial problem to the 
problem of checking satisfiability 
w.r.t.\ the theory of bounded semilattices of a conjunction between a 
set $G_0$ of ground unit clauses of the form 
$$c_1 \wedge c_2 \leq c, \quad c_1 \leq c_2 \quad d_1 \not\leq d_2$$ 
of size linear in  $|st(G)|$ and a set 
of Horn clauses of length at most $n+1$, where $n$ is the maximal arity 
of a function symbol in $\Sigma$ of the form 
$$ c_1 \leq c'_1 \wedge \dots \wedge c_n \leq c'_n \rightarrow c \leq c'.$$
It is easy to see (cf.\ also \cite{Sofronie-ijcar-06,Sofronie-lmcs-08}) 
that one can give a polynomial decision procedure for checking the 
satisfiability of such sets of clauses, by noticing that if the set of 
clauses is unsatisfiable then there exists an instance of monotonicity 
with all premises entailed by the unit clauses from $G_0$. We can add the 
conclusion to $G_0$ and recursively repeat the argument. 

\medskip
\noindent In order to obtain an even more efficient method for checking TBox 
subsumption we use a reduction to reachability in the theory of posets.
It is known that the theory of semilattices allows a local Horn axiomatization 
(cf.\ e.g.\ \cite{Skolem20,Burris95}), by means of the following axioms:
$$\begin{array}{lll}
(S1)~~~~~ ~~~~~ & \forall x, y, z~~~~~~~~~ &   (x \leq y \wedge y \leq z ~
  \rightarrow ~ z \leq z)\\
(S2)~~~~~ & \forall x & (0 \leq x  \wedge x \leq 1)  \\
(S3) & \forall x, y  & (x \wedge y \leq x  \wedge  x \wedge y \leq y) \\
(S4)& \forall x, y, z & (z \leq x \wedge z \leq y ~\rightarrow~ z \leq x \wedge y) 
\end{array}$$ 
We denote by ${\cal SL}$ this set of axioms 
for the theory of bounded semilattices. 

\begin{theorem}
The set of Horn clauses ${\cal SL}$ define a local extension of the pure 
theory of bounded partial orders, i.e.\ for every 
set $G$ of ground clauses in the signature of bounded semilattices, 
${\cal SL} \cup G \models \perp$ iff ${\cal SL}[G] \cup G \models \perp$.
\end{theorem}
\Proof Let $(P, \leq, \wedge, 0, 1)$ be a weak partial model of ${\cal SL}$.
Then $(P, \leq, 0, 1)$ is a poset with first and last element.
Let 
${\cal OI}(P) = ({\cal OI}(P, \leq), \cap, \{ 0 \}, P)$ be the semilattice of 
all order ideals of $P$. 
We show that the map 
$i : P \rightarrow {\cal OI}(P)$ defined by $i(x) = {\downarrow} x$ 
is a weak embedding: $i$ is obviously injective and an order embedding. 
Clearly, $i(0) = {\downarrow} 0 = \{0 \}$ and 
$i(1) = P$. 
Assume that $x \wedge y$ is defined in $P$. Then 
$i(x \wedge y) = {\downarrow} (x \wedge y)$. 
If $x \wedge y$ is defined in $P$, since $P$ weakly satisfies (S3), 
$x \wedge y \leq x$ and $x \wedge y \leq y$, so 
$x \wedge y \in {\downarrow} x \cap {\downarrow} y$. Hence, 
${\downarrow}(x \wedge y) \subseteq {\downarrow} x \cap {\downarrow} y$.
Conversely, let $z \in {\downarrow} x \cap {\downarrow} y$. Then 
$z \leq x$ and $z \leq y$ and as $x \wedge y$ is defined and $P$ weakly satisfies (S4), $z \leq x \wedge y$.
It follows that 
$i(x \wedge y) = {\downarrow} (x \wedge y) = {\downarrow} x \cap {\downarrow} y = i(x) \cap i(y). $ \QED
\begin{corollary}
The following are equivalent: 
\begin{itemize}
\item[(1)] $SL \cup {\sf Mon}(\Sigma) \cup RI \models \forall {\overline x} 
\bigwedge_{i = 1}^n s_i({\overline x}) \leq s'_i({\overline x}) \rightarrow s({\overline x}) \leq s'({\overline x})$.

\item[(2)] $SL \cup {\sf Mon}(\Sigma) \cup RI {\cup} G  {\models} \perp$, where $G = \bigwedge_{i = 1}^n s_i({\overline c}) {\leq} s'_i({\overline c}) {\wedge}  s({\overline c}) {\not\leq} s'({\overline c})$.  

\item[(3)] $SL \cup ({\sf Mon}(\Sigma) \cup RI)[\Psi(G)] {\cup} G  {\models} \perp$, where $G = \bigwedge_{i = 1}^n s_i({\overline c}) {\leq} s'_i({\overline c}) {\wedge}  s({\overline c}) {\not\leq} s'({\overline c})$, 
and $\Psi$ is defined as in Theorem~\ref{loc-1-3}. 

\item[(4)] ${\cal SL} \cup ({\sf Mon}(\Sigma) \cup RI)[\Psi(G)]_0 {\cup} G_0  {\models} \perp$, for the purified semilattice part of the problem.

\item[(5)] ${\cal SL}[G'] \cup G' \models \perp,$ where 
$G' = ({\sf Mon}(\Sigma) \cup RI){[{\Psi}(G)]}_0 \cup G_0$.
\item[(6)] ${\cal SL}[G']_0 \cup G'_0 \cup {\sf Con}(\wedge)[G'] 
\models \perp$. 
\end{itemize}
\label{cor-stable-loc}
\end{corollary}
\begin{theorem}
CBox subsumption can be checked in cubic time in the size of the 
original CBox for all CBoxes in the language of the extension of 
${\cal EL}^+$ considered in this paper.
\end{theorem}
\Proof 
We analyze the complexity of the problem in item (6) of 
Corollary~\ref{cor-stable-loc}, as a function of the 
size of the input CBox, i.e.\ as a function of the size of $RI$ and $G$. 
We first estimate the size of $G'$. 
Note that $\Psi(G)$ can have at most $|{\sf st}(G)| \cdot |N_R|$ 
elements. Thus, its size is linear in the size of $G$ if $N_R$ is fixed. 
The number of clauses in 
$({\sf Mon}(\Sigma) \cup RI){[{\Psi}(G)]}$
is quadratic in $|\Psi(G)|$. By purification, the size grows 
linearly. Thus: 
\begin{itemize}
\item The size of $G'$ is quadratic in the number of subterms of $G$.
\item $G'$ contains a set of ground unit clauses (of size linear in the size 
of $G$) and a set of ground Horn clauses (of size quadratic in in the size 
of $G$).
\item The number of subterms in $G'$ is linear in the number of subterms of $G$.\end{itemize}
If we consider the form of the clauses in ${\cal SL}$ we note that 
the number of clauses in ${\cal SL}[G'] \cup {\sf Con}(\wedge)[G']$ 
is at most cubic in the number of subterms 
in $G'$, i.e.\ cubic in the number of subterms of $G$. 
The conclusion of the theorem now follows easily if we 
note that 
\begin{itemize}
\item ${\cal SL}[G']_0 \cup G'_0 \cup {\sf Con}(\wedge)[G']$ is a set 
of ground Horn clauses, and 
\item in order to check the satisfiability of any set of $N$ 
ground clauses w.r.t. the theory of posets we only need to take 
into account those instances of the poset axioms in which the variables 
are instantiated with the (ground) terms occurring in $N$. 
\end{itemize}
We can thus reduce the verification problem to the problem of checking the 
satisfiability of a set of Horn clauses of size at most cubic in 
the number of subterms of $G$.  
Since the satisfiability of Horn clauses can be tested in linear time 
\cite{DowlingGallier}, 
this shows that the uniform word problem for the class 
${\sf SLO}_{\Sigma}(RI)$ 
(and thus  for ${\sf SLO}^{\exists}_{NR}(RI)$)
is decidable in cubic time.
\QED




\subsection{Extensions of $\mathcal{EL}$ with $n$-ary roles and concrete domains}
\label{sect-extensions}

The previous results can easily be generalized to 
semilattices with $n$-ary monotone functions satisfying composition axioms.

\subsubsection{Extensions of $\mathcal{EL}$ with $n$-ary roles}

We now consider the extensions of $\mathcal{EL}$ with 
$n$-ary roles 
introduced in 
Section~\ref{ext1-n-ary}. 
The semantics is defined in terms of interpretations
${\mathcal I} = (D^{\mathcal I}, \cdot^{\mathcal I})$, where $D^{\mathcal I}$  
is a non-empty set, concepts are interpreted as usual, and each $n$-ary 
role $R \in N_R$ is interpreted as an $n$-ary relation
$R^{\mathcal I} \subseteq (D^{\mathcal I})^n$. 
All results in the previous section extend in a natural way to this case, 
because, independently of the arities of the functions, 
the extension of the theory of bounded semilattices with monotone 
functions is local 
and the number of instances of the monotonicity axioms 
in ${\sf Mon}[\Psi(G)]$ is quadratic in the size of $\Psi(G)$.

\subsubsection{Extensions of $\mathcal{EL}^+$ with $n$-ary roles}
In this case we need to take into account role inclusions of type: 

\vspace{-7mm}
\begin{eqnarray}
r_1 & \sqsubseteq & r_2 \label{n-ary-incl-1} \\
r_1 \circ (s_1, \dots, s_n) & \sqsubseteq & r_2 \label{n-ary-1} \\
r_1 \circ (s_1, \dots, s_n) & \sqsubseteq & id ~~~ \text{ for binary relations } s_i \label{n-ary-inv-1} 
\end{eqnarray}
We proved that TBox subsumption 
can be expressed as a uniform word problem w.r.t.\ the class of 
semilattices with monotone operators associated with the roles, 
satisfying axioms $RI_a$ 
corresponding in a natural way to the role inclusion laws above. 
Below we write the flat form of those axioms $RI^{\sf flat}$: 

\vspace{-7mm}
\begin{eqnarray*}
\forall x_1, \dots, x_n & & f_{\exists r_1}(x_1, \dots, x_n)  \leq  f_{\exists r_2}(x_1, \dots, x_n) \\
\forall {\overline y^k_j} z_1,\dots,z_n~~ & \bigwedge_i z_i {\leq} f_{\exists s_i}(y^i_1, \dots, y^i_{m_i})  \rightarrow &  
f_{\exists r_1}(z_1, \dots, z_n)  \leq  f_{\exists r_2}(y^1_1, \dots, y^1_{m_1},  \dots, y^n_1, \dots, y^n_{m_n}) \\
\forall x, z_1,\dots,z_n~~ & \bigwedge_i z_i {\leq} f_{\exists s_i}(x)  \rightarrow  &  f_{\exists r_1}(z_1, \dots, z_n) \leq x\end{eqnarray*}
\begin{theorem}
Any extension of the theory $SL$ of lattices 
with a set of monotone functions satisfying any combination of 
axioms containing axioms of type $RI^{\sf flat}$ is $\Psi$-local, where 
$\Psi(T) = \bigcup_{i \geq 1} \Psi^i(T)$, with 
$\Psi_0(T) = T$, and 
$$\begin{array}{rcl}
\Psi_{i+1}(T) & = & \{ h({\overline c}) \mid \exists \forall {\overline x} (\bigwedge_i g_i({\overline x})  \rightarrow h({\overline x})) \in RI^{\sf flat} \text{ and } g({\overline c}) \in T \} \cup \\
& &  \{ h(c) \mid \exists \forall x,{\overline y} (\bigwedge y_i \leq g_i(x) \rightarrow f(y_1, \dots, y_n) \leq h(x)) \in RI^{\sf flat} \text{ and } \forall i g_i(c) \in T  \} \cup \\
& & \{ h({\overline c_1}, \dots, {\overline c}_n) \mid \exists \forall {\overline x}_i,{\overline y} (\bigwedge y_i \leq g_i({\overline x}_i) \rightarrow f(y_1, \dots, y_n) \leq h({\overline x}_1, \dots, {\overline x}_n)) \in RI^{\sf flat}  \\
& & ~~~~~~~~~~~~~~~~~~~~~~~~\text{ and } g_i({\overline c}_i) \in T \text{ for all } i \}.
\end{array}$$ 
\label{local-el+-n-ary-psi}
\end{theorem}
\Proof The proof is analogous to the proof of Theorem~\ref{locality-of-el+}. 
We illustrate as an example the fact that any axiom in $RI^{\sf flat}$ 
of the second type is $\Psi$-local. Consider an axiom of this type
$$ \forall {\overline y^k_j} z_1,\dots,z_n~~ \bigwedge_i z_i {\leq} f_{\exists s_i}(y^i_1, \dots, y^i_{m_i})  \rightarrow   
f_{\exists r_1}(z_1, \dots, z_n)  \leq  f_{\exists r_2}(y^1_1, \dots, y^1_{m_1},  \dots, y^n_1, \dots, y^n_{m_n})$$
Let $U^k_j, V_1, \dots, V_n \in {\cal OI}(S)$ be such that 
$V_i \subseteq {\overline g}_i(U^i_1, \dots, U^i_{m_i})$.
Let $x \in {\overline f}(V_1, \dots, V_n)$. Then there exist $v_i \in V_i$ 
such that $f(v_1, \dots, v_n)$ is defined and $x \leq f(v_1, \dots, v_n)$.
Since $v_i \in V_i \subseteq {\overline g}_i(U^i_1, \dots, U^i_{m_i})$, 
there exist $u^i_j \in U^i_j$ with $g_i({\overline u}^i)$ defined and 
such that $v_i \leq g_i({\overline u}^i)$. By the $\Psi$-closure properties 
of the models we consider it follows that $h({\overline u}^1, \dots, {\overline u}^n)$ is also defined and since $S$ weakly satisfies the corresponding 
axiom, it follows that $x \leq f(v_1, \dots, v_n) \leq h({\overline u}^1, \dots, {\overline u}^n)$. Thus, $x \in {\overline h}({\overline U}^1, \dots, {\overline U}^n)$. \QED

\medskip
\noindent The extension to guarded role inclusions follows exactly 
as in the case of binary relations. Because of the flatness restriction 
in the definition of locality we need to consider flat versions of 
$GRI_a$ axioms, $GRI^{\sf flat}$ which are defined analogously to 
$RI^{\sf flat}$.  

\subsubsection{Extensions with existential role restrictions}
In the presence of existential role restrictions we can prove the following 
result. 

\begin{theorem}
Any extension of the theory $SL$ of lattices 
with a set of monotone functions satisfying any combination of 
axioms containing axioms of type $RI^{\sf flat}$, $GRI^{\sf flat}$
and existential restrictions $ER$ of the form:
$$ \forall x_1, \dots, x_{i-1}, x_{i+1}, \dots, x_n~~ g(x_1, \dots, x_{i-1}, x_{i+1}, \dots, x_n) = h(x_1, \dots, x_{i-1}, c, x_{i+1}, \dots, x_n),$$ 
is $\Psi$-local, where 
$\Psi(T) = \bigcup_{i \geq 1} \Psi^i(T)$, with 
$\Psi_0(T) = T$, 
$$\begin{array}{rcl}
\Psi_{i+1}(T) & = & \{ h({\overline c}) \mid \exists \forall {\overline x} (g \wedge \bigwedge_i g_i({\overline x})  \rightarrow h({\overline x})) \in (G)RI^{\sf flat} \text{ and } g({\overline c}) \in T \} \cup  \\
& &  \{ h(c) \mid \exists \forall x,{\overline y} (g \wedge \bigwedge y_i \leq g_i(x) \rightarrow f(y_1, \dots, y_n) \leq h(x)) \in (G)RI^{\sf flat} \\
& & ~~~~~~~~~~~~~\text{ and } g_i(c) \in T \text{ for all } i \} \cup \\
& & \{ h({\overline c_1}, \dots, {\overline c}_n) \mid \exists \forall {\overline x}_i,{\overline y} (g \wedge \bigwedge y_i \leq g_i({\overline x}_i) \rightarrow f(y_1, \dots, y_n) \leq h({\overline x}_1, \dots, {\overline x}_n)) \in (G)RI^{\sf flat} \\
& & ~~~~~~~~~~~~~~~~~~~~~~~~~ \text{ and } g_i({\overline c}_i) \in T \text{ for all } i \},
\end{array}$$ 
where $g$ are either $true$ or a suitable conjunction of guards of the form 
$x_i \leq d_i$. 
\end{theorem}
\Proof The only issue to be clarified is the locality of the extension 
with axioms in $ER$. 
The axioms in $ER$ are extensions by definitions like the ones considered 
in \cite{sofronie-ihlemann-ismvl-07}. Due to arity reasons, they are acyclic.
Thus, we have the following chain of extensions: 
$SLO_{\Sigma} \subseteq SLO_{\Sigma}(ER) \subseteq SLO_{\Sigma}(RI \cup ER)$. 
 \QED

\subsubsection{Extensions with $n$-ary roles and concrete domains}
We now consider the extension with concrete domains studied in 
Section~\ref{ext1-n-ary-concrete}. We showed that an algebraic semantics 
can be given in terms of the class 
$SL_S$ of all structures 
 ${\mathcal A} = (A, {\mathcal P}(A_1), \dots, {\mathcal P}(A_n))$, 
with signature 
$\Pi = (S, \{ \wedge \} {\cup} \Sigma, {\sf Pred})$ with 
$S {=} \{ {\sf concept}, {\sf s_1}, \dots, {\sf s_n} \}$, 
${\sf Pred} {=} \{ \leq \} {\cup} \{ \subseteq_i \mid 1 \leq i \leq n \}$, 
where $A \in SL$, the support of sort ${\sf concept}$ of ${\mathcal A}$ 
 is $A$, and for all $i$ 
the support sort $s_i$ of ${\mathcal A}$  is ${\mathcal P}(A_i)$. 
\begin{theorem}[\cite{sofronie-ihlemann-ismvl-07}]
Every structure 
$(A, {\mathcal P}(A_1), \dots, {\mathcal P}(A_n), \{ f_A \}_{f \in \Sigma})$, 
where 
\begin{itemize}
\item[(i)] $(A, {\mathcal P}(A_1), \dots, {\mathcal P}(A_n)) \in SL_S$, and 
\item[(ii)] for every 
$f {\in} \Sigma$ of arity $s'_1 {\dots} s'_n {\rightarrow} s$, with $s'_1,
  \dots, s'_n, s \in S$, $f_A$ is a partial function from 
$\prod_{i = 1}^n U_{s'_i}$ to $U_s$ which is monotone on its domain of
definition (here $U_{\sf concept} = A$ and $U_{s_i} =
  {\mathcal P}(A_i)$ are the universes of the many-sorted structure in (i)).  

\end{itemize}
weakly embeds into a total model of $SL_S {\cup} {\sf Mon}(\Sigma)$.  
\end{theorem}
\begin{corollary}
\label{cor-red-n-ary}
Let $G = \bigwedge_{i = 1}^n s_i({\overline c}) {\leq} s'_i({\overline c}) \wedge  s({\overline c}) {\not\leq} s'({\overline c})$ be a set of ground unit 
clauses in the extension $\Pi^c$ of $\Pi$ with new 
constants $\Sigma_c$.  The following are 
equivalent:
\begin{itemize}
\item[(1)] $SL_S \cup {\sf Mon}(\Sigma)  \cup G  \models \perp$. 
\item[(2)] $SL_S \cup {\sf Mon}(\Sigma)[G] \cup G$ has no partial model with a total $\{ \wedge_{SL} \}$-reduct in which all terms in $G$ are defined.
\end{itemize}
\end{corollary}
A hierarchical reduction to the problem of checking satisfiability of 
constraints in the disjoint combinations of the theory of 
semilattices and the theories ${\cal P}(A_i)$ follows immediately 
from this locality result. 
Let $\bigcup_{i = 0}^n {\sf Mon}(\Sigma)[G]_i \cup G_i \cup {\sf Def}$ be 
obtained from ${\sf Mon}(\Sigma)[G] \cup G$ by purification, 
i.e. by replacing, in a bottom-up manner, all subterms $f(g)$ of sort $s$
with $f \in \Sigma$, with newly introduced constants $c_{f(g)}$ of 
sort $s$ and  adding the definitions $f(g) = c_t$ to the set ${\sf Def}$.
We thus separate ${\sf Mon}(\Sigma)[G] \cup G$ into a conjunction 
of constraints  $\Gamma_i = {\sf Mon}(\Sigma)[G]_i \cup G_i$, 
where $\Gamma_0$ is a constraint of sort ${\sf semilattice}$ and 
for $1 \leq i \leq n$, $\Gamma_i$ is a set of constraints over terms 
of sort $i$ ($i$ being the concrete sort with fixed support ${\mathcal P}(A_i)$).
\begin{corollary}
The following are equivalent (and are also equivalent to (1) and (2)):
\begin{itemize}
\item[(3)] $\bigcup_{i = 0}^n {\sf Mon}(\Sigma)[G]_i \cup G_i \cup {\sf Def}$
has no partial model with a total $\{ \wedge, 0, 1 \}$-reduct in which all 
terms in ${\sf Def}$ are defined.
\item[(4)] $\bigcup_{i = 0}^n {\sf Mon}(\Sigma)[G]_i \cup G_i$ 
is unsatisfiable in the many-sorted disjoint combination of $SL$ 
and the concrete theories of ${\mathcal P}(A_i)$, $1 \leq i \leq n$. 
\end{itemize}
\end{corollary}
The complexity of the uniform word problem of 
$SL_S \cup {\sf Mon}(\Sigma)$ 
depends on the complexity of the problem of testing the 
satisfiability --- in  the many-sorted 
disjoint combination of $SL$ 
with the concrete theories of ${\mathcal P}(A_i)$, $1 \leq i \leq n$ --- 
of sets of clauses 
$C_{\sf concept} \cup \bigcup_{i = 1}^n C_i \cup {\sf Mon}$, 
where $C_{\sf concept}$ and $C_i$ are unit clauses of sort 
${\sf concept}$ resp. $s_i$, and ${\sf Mon}$ consists of possibly mixed
ground Horn clauses.  

\smallskip Specific extensions of the logic $\mathcal{EL}$ can be obtained 
by imposing additional restrictions on the interpretation of the 
``concrete''-type concepts within ${\mathcal P}(A_i)$. For instance, 
we can require that numerical concepts are always interpreted as 
intervals, as in Example~\ref{ex1}.

\begin{theorem} 
Consider the 
extension of $\mathcal{EL}$ with two sorts, ${\sf concept}$ 
and ${\sf num}$, where the semantics of classical concepts is the usual 
one, and the concepts of sort ${\sf num}$ are interpreted 
as elements in the ORD-Horn, 
convex fragment of Allen's interval algebra \cite{Nebel-Buerkert95}, 
where any CBox can contain many-sorted GCI's over concepts, as well as 
constraints over the numerical data expressible in the ORD-Horn fragment.

In this extension, CBox subsumption is decidable in PTIME. 
\label{el-many-sorted}
\end{theorem} 
\Proof 
The assumption on the semantics of the extension of $\mathcal{EL}$ we made 
ensures that all algebraic models are two-sorted structures 
of the form 
${\mathcal A} = ((A, \wedge), 
{\sf Int}({\mathbb R}, O), \{ f_{\mathcal A} \}_{f \in \Sigma})$, 
with sorts $\{{\sf concept}, {\sf num} \}$, such that  
$(A, \wedge)$ is a semilattice, ${\sf Int}({\mathbb R}, O)$ is an interval 
algebra in the Ord-Horn fragment of Allen's interval arithmetic 
\cite{Nebel-Buerkert95}, 
and for all $f \in \Sigma$, $f_A$ is a monotone (many-sorted) function.
We will denote the class of all these structures by $SL_{\sf OrdHorn}$. 

Note that the 
Ord-Horn fragment of Allen's interval arithmetic  has the property 
that all operations and relations between intervals can be represented 
by Ord-Horn clauses, i.e.\ clauses over atoms $x \leq y, x = y$, 
containing at most one positive literal ($x \leq y$ or $x = y$) and 
arbitrarily many negative literals (of the form $x \neq y$). Nebel and 
B{\"u}rckert \cite{Nebel-Buerkert95} proved that  a finite set of 
Ord-Horn clauses is satisfiable over the real numbers iff it is satisfiable 
over posets. As the theory of partial orders is convex, 
this means that although the theory of reals is not convex w.r.t. $\leq$, 
we can always assume that the theory of Ord-Horn clauses is convex. 
The main result in Corollary~\ref{cor-red-n-ary} can be adapted without 
problems to show that if  
$G = \bigwedge_{i = 1}^n s_i({\overline c}) {\leq} s'_i({\overline c}) 
\wedge  s({\overline c}) {\not\leq} s'({\overline c})$
is a set of  ground unit clauses  in the extension $\Pi^c$ of $\Pi$ with new 
constants $\Sigma_c$, and  if 
${\sf Mon}(\Sigma)[G]_{\sf c} \cup 
{\sf Mon}(\Sigma)[G]_{\sf num} \cup 
G_{\sf c} \cup G_{\sf num} \cup {\sf Def}$ are 
obtained from ${\sf Mon}(\Sigma)[G] \cup G$ by purification, 
the following are equivalent:
\begin{itemize}
\item $SL_{\sf OrdHorn} \cup {\sf Mon}(\Sigma)  \cup G  \models \perp$;
\item ${\sf Mon}(\Sigma)[G]_0 \cup G_0 \cup {\sf Con}[\sf Def]_0$ 
is unsatisfiable in the combination of $SL$ 
and the Ord-Horn fragment of Allen's interval arithmetic. 
\end{itemize}
In order to test the unsatisfiability of the latter problem we proceed 
as follows. We first note that, due to the convexity of the theories 
involved and to the fact that all constraints in 
$G_0 \cup {\sf Mon}(\Sigma)[G]_0 \cup {\sf Con}[\sf Def]_0$ 
are separated (in the sense that there are no mixed atoms) if 

\begin{itemize}
\item[(1)] $G_0 \cup {\sf Mon}(\Sigma)[G]_0 \cup {\sf Con}[\sf Def]_0 \models \perp$, then: 

\item[(2)] there exists a clause $C = (\bigwedge c_i = d_i \rightarrow c = d)$ in 
${\sf Mon}(\Sigma)[G]_0 \cup {\sf Con}[\sf Def]_0$ such that 
$G_0 \models \bigwedge c_i = d_i$ and 
$G_0 \cup \{ c = d \} \cup ({\sf Mon}(\Sigma)[G]_0 
\cup {\sf Con}[\sf Def]_0) \backslash \{C \} \models \perp$.
\end{itemize}
In order to prove this, let 
${\mathcal D}$ be the set of all atoms $c_i R_i d_i$
occurring in premises of clauses in 
${\sf Mon}(\Sigma)[G]_0 \cup {\sf Con}[\sf Def]_0$. 
As every model of 
$G_0 \wedge \bigwedge_{(c R d) \in {\mathcal D}} \neg (c R d)$
is also a model of $G_0 \cup {\sf Mon}(\Sigma)[G]_0 \cup {\sf Con}[\sf Def]_0$, and the last formula is by (1) unsatisfiable, it follows that 
$G_0 \wedge
\bigwedge_{(c R d) \in {\mathcal D}} \neg (c R d) 
\models \perp$ in the combination of the Ord-Horn fragment over posets with 
the theory of semilattices. 
Let $G_0^+$ be the conjunction of all atoms in $G_0$, and $G_0^-$ be the 
set of all negative literals in $G_0$.
 Then 
$G_0^+ \models \bigvee _{(c R d) \in {\mathcal D}} (c R d) \vee \bigvee_{\neg L \in G_0^-} L.$ Since the constraints are sort-separated and both theories 
involved are convex, it follows that either $G_0 \models \perp$ or else 
$G_0 \models c R d$ for some $(c R d) \in {\mathcal D}$. 
We can repeat the process until all the premises of some clause in 
${\sf Mon}(\Sigma)[G]_0 \cup {\sf Con}[\sf Def]_0$ are proved to be entailed 
by $G_0$. Thus, (2) holds.

By iterating the argument above we can always  -- if (1) holds -- 
successively entail sufficiently many premises of monotonicity and 
congruence axioms in order to ensure that, in the end, 
\begin{itemize}
\item[(3)] there exists a set $\{ C_1, \dots, C_n \}$  of clauses in 
${\sf Mon}(\Sigma)[G]_0 \cup {\sf Con}[\sf Def]_0$ with 
$C_j = (\bigwedge c^j_i = d^j_i \rightarrow c^j = d^j)$,
such that for all $k \in \{0, \dots, n-1 \}$,  
$$G_0 \wedge \bigwedge_{j = 1}^k  (c^j = d^j) \models 
\bigwedge c^{k+1}_i = d^{k+1}_i \text{ and } G_0 \wedge 
\bigwedge_{j = 1}^n  (c^j = d^j) 
\models \perp.$$ 
\end{itemize}
Note that (3) implies (1), since the conditions in (3) imply that 
$G_0 \wedge \bigwedge_{j = 1}^n  (c^j = d^j)$ is logically equivalent 
with $G_0 \wedge C_1 \wedge \dots C_n$, which (as set of clauses) 
is contained in the set of clauses 
$G_0 \cup {\sf Mon}(\Sigma)[G]_0 \cup {\sf Con}[\sf Def]_0$.

This means that in order to 
test satisfiability of 
$G_0 \cup {\sf Mon}(\Sigma)[G]_0 \cup {\sf Con}[\sf Def]_0$
we need to test entailment of the premises of 
${\sf Mon}(\Sigma)[G]_0 \cup {\sf Con}[\sf Def]_0$ from $G_0$; when all 
premises of some clause are provably true we delete the clause and add its 
conclusion to $G_0$. The PTIME assumptions for concept subsumption and for 
the Ord-Horn fragment ensure that this process terminates in PTIME.
\QED

\medskip
\begin{example}
Consider the special case described in Example~\ref{ex1}. 
Assume that the concepts of sort 
${\sf num}$ used in any TBox are of the form ${\uparrow}n, {\downarrow}m$ 
and $[n, m]$.  Consider the TBox ${\mathcal T}$ consisting of the following GCIs: 
$$\begin{array}{@{}l@{}l}
\{ & \exists {\sf price}({\downarrow}n_1) \sqsubseteq {\sf affordable}, ~~\exists {\sf weight}({\uparrow}m_1) \sqcap {\sf car} \sqsubseteq {\sf truck}, \\ 
   & \mbox{ {\sf has-weight-price}}({\uparrow}m, {\downarrow}n) \sqsubseteq \exists {\sf price}({\downarrow}n) \sqcap \exists {\sf weight}({\uparrow}m), \\
& {\downarrow}n \sqsubseteq {\downarrow}n_1, ~~{\uparrow}m \sqsubseteq {\uparrow}m_1,~~ C \sqsubseteq {\sf car}, ~~ C \sqsubseteq \exists \mbox{ {\sf has-weight-price}}({\uparrow}m, {\downarrow}n) ~~\} 
\end{array}$$
In order to prove that $C \sqsubseteq_{\mathcal T} {\sf affordable} \sqcap {\sf truck}$ we proceed as  follows. We refute $\bigwedge_{D \sqsubseteq D' \in {\mathcal T}} {\overline D} \leq {\overline D}' \wedge {\overline C} \not\leq {\sf affordable} \wedge {\sf truck}$. We purify the problem introducing definitions for the 
terms starting with existential restrictions, and express the interval constraints using constraints over ${\mathbb Q}$ and obtain the following set 
of constraints: 

\bigskip
\noindent 
$\begin{array}{|l||l|l|l|}
\hline 
{\sf Def} & C_{\sf num} & C_{\sf concept} & {\sf Mon} \\
\hline 
\hline 
f_{\sf price}({\downarrow}n_1) = c_1 & n \leq n_1 & c_1 \leq {\sf affordable} & n_1 \leq n \rightarrow c_1 \leq c \\
f_{\sf price}({\downarrow}n) = c & m \geq m_1 & d_1 \wedge {\sf car} \leq {\sf truck} & n_1 \geq n \rightarrow c_1 \geq c \\
f_{\sf weight}({\uparrow}m_1) = d_1 & & e \leq c \wedge d & m_1 \geq m \rightarrow d_1 \leq d \\
f_{\sf weight}({\uparrow}m) = d  & & C \leq {\sf car} &  m_1 \leq m \rightarrow d_1 \geq d \\
f_{\sf \text{h-w-p}}({\uparrow}m, {\downarrow}n) = e & & C \leq e & \\
            & & C \not\leq {\sf affordable} \wedge {\sf truck} & \\ 
\hline 
\end{array}$

\bigskip
\noindent The task of proving 
$C \sqsubseteq_{\mathcal T} {\sf affordable} \sqcap {\sf truck}$
can therefore be reduced to checking whether 
$C_{\sf num} \wedge C_{\sf concept} 
\wedge {\sf Mon}$ is satisfiable w.r.t.\ the combination of $SL$ (sort {\sf concept}) with $LI({\mathbb Q})$ (sort ${\sf num}$). For this, we note 
that $C_{\sf num}$ entails the premises of the first, second, and fourth 
monotonicity rules. Thus, we can add $c \leq c_1$ and $d \leq d_1$ to 
$C_{\sf concept}$. Thus, we deduce that 
$C \leq e \wedge {\sf car} \leq (c \wedge d) \wedge {\sf car} 
\leq c_1 \wedge (d_1 \wedge {\sf car}) \leq {\sf affordable} \wedge {\sf truck}$, which contradicts the last clause in $C_{\sf concept}$. 

\smallskip
\noindent 
A similar procedure can be used in general for testing (in PTIME) 
the satisfiability of mixed constraints in the many-sorted combination 
of $SL$ with concrete domains of sort ${\sf num}$, assuming that all 
concepts of sort ${\sf num}$ are interpreted as intervals and 
the constraints $C_{\sf num}$ are expressible in a PTIME, convex fragment 
of Allen's interval algebra. 

\end{example}
These results lift in a natural way to $n$-ary roles satisfying 
(guarded) role inclusion axioms. 

\vspace{-2mm}
\section{Interpolation in semilattices with operators and applications}
\label{interpolation}
Interpolation theorems are important in the study of distributed or 
evolving ontologies. 

A theory ${\mathcal T}$ has interpolation if, for all
formulae  $\phi$ and $\psi$ in the signature of ${\mathcal T}$, 
if $\phi \models_{\mathcal T} \psi$ then there exists a formula 
$I$ containing only symbols which occur in both $\phi$ and 
$\psi$ such that  $\phi \models_{\mathcal T} I$ and 
$I  \models_{\mathcal T} \psi$.  
First order logic has interpolation but -- for an arbitrary theory 
${\mathcal T}$ -- even if 
$\phi$ and $\psi$ are e.g. conjunctions of ground literals,    
$I$ may still be an arbitrary formula, containing alternations of quantifiers.
It is often important to identify situations in which  
{\em ground clauses} have {\em ground interpolants}. 
In recent literature, when defining ground interpolation, instead of
considering formulae  $\phi$ and $\psi$ such that $\phi \models_{\mathcal T}
\psi$, formulae $A$ and $B$ are considered such that 
$A \wedge B \models_{\mathcal T} \perp$. The two formulations are clearly 
equivalent. In what follows we will use the second one. 
\begin{definition}[Ground interpolation]
We say that a theory ${\mathcal T}$ has the {\em ground interpolation 
property} (or, shorter, that ${\mathcal T}$ has {\em ground interpolation}) 
if for all ground 
clauses $A({\overline c}, {\overline d})$ and  
$B({\overline c}, {\overline e})$,  if 
$A({\overline c}, {\overline d}) \wedge B({\overline c}, {\overline e}) \models_{\mathcal T} \perp$ then 
there exists a ground formula $I({\overline c})$, containing only 
the constants ${\overline c}$ occurring both in $A$ and $B$ (and, ideally, 
only function symbols shared by $A$ and $B$), such that 
$A({\overline c}, {\overline d}) \models_{\mathcal T} I({\overline c}) 
\text{ and } 
B({\overline c}, {\overline e}) \wedge 
I({\overline c}) \models_{\mathcal T} \perp.$
\end{definition} 
\begin{definition}[Equational interpolation property]
An equational theory ${\mathcal T}$ 
(in signature $\Pi = (\Sigma, {\sf Pred})$ where 
${\sf Pred} = \{ \approx \}$) has 
the {\em equational interpolation property} if
whenever 
$$\bigwedge_i A_i({\overline a}, {\overline c}) \wedge 
\bigwedge_j B_j({\overline c}, {\overline b}) \wedge 
\neg B({\overline c}, {\overline b}) \models_{\mathcal T} \perp,$$ where 
$A_i$, $B_j$ and $B$ are ground atoms,  
there exists a conjunction $I({\overline c})$ of ground atoms
containing only 
the constants ${\overline c}$ occurring both in 
$\bigwedge_i A_i({\overline a}, {\overline c})$ and $\bigwedge_j B_j({\overline c}, {\overline b}) \wedge \neg B({\overline c}, {\overline b})$, 
such that  
$\bigwedge_i A_i({\overline a}, {\overline c})  \models_{\mathcal T}
I({\overline c}) \text{ and } I({\overline c}) \wedge\bigwedge_j   B_j  \models_{\mathcal T} B.$
\label{equational-interpolation}
\end{definition}
There exist results which relate ground interpolation to 
amalgamation or the injection transfer property 
\cite{Jonsson65,Bacsich75,Wronski86} 
and thus allow us to recognize 
many theories with ground interpolation. However, 
just knowing that ground 
interpolants exist is usually not sufficient: we would like to 
construct the interpolants fast. 
In \cite{Sofronie-ijcar-06,Sofronie-lmcs-08} a class of theory extensions 
was identified which have ground interpolation, and for which 
hierarchical methods for computing the interpolants exist. 
We present the results below.
The theories we consider are theory extensions ${\cal T}_0 \subseteq {\cal T}_1 = {\cal T}_0 \cup {\cal K}$ which satisfy the following assumptions:

\medskip
\noindent ${\mathcal T}_0$ is a theory with the following properties: 
\begin{description}
\item[Assumption 1:] ${\mathcal T}_0$ is {\em convex}\/ w.r.t.\  the set 
${\sf Pred}$ (including equality $\approx$), i.e., for all conjunctions $\Gamma$ of ground 
atoms, relations $R_1, \dots, R_m \in {\sf Pred}$ and ground tuples of 
corresponding arity 
${\overline t}_1, \dots, {\overline t}_n$, if $\Gamma
  \models_{{\mathcal T}_0} \bigvee_{i = 1}^m R_i({\overline t}_i)$ 
then there exists $j \in
  \{ 1, \dots, m \}$ such that 
$\Gamma \models_{{\mathcal T}_0} R_j({\overline t}_j)$.
\item[Assumption 2:]  ${\mathcal T}_0$ is {\em $P$-interpolating} w.r.t.\  a subset $P \subseteq {\sf Pred}$
and the separating terms $t_i$ can be effectively computed, i.e.\ for all conjunctions $A$ and $B$ of ground literals, all binary  
  predicates $R \in P$ and all
  constants $a$ and $b$ such that $a$ occurs in $A$ 
  and $b$ occurs in $B$ 
  (or vice versa), if 
$A  \wedge B \models_{{\mathcal T}_0} a R b$ then there exists a term $t$
  containing only constants common to $A$ and $B$ with 
$A \wedge B \models_{{\mathcal T}_0} a R t \wedge t R b$. 
(If we can always find a term $t$ containing only
constants common to $A$ and $B$ with 
$A \models_{{\mathcal T}_0} a R t$ and $B \models_{{\mathcal T}_0} t R b$ 
we say that ${\mathcal T}_0$ is
{\em strongly $P$-interpolating}.).
\item[Assumption 3:] ${\mathcal T}_0$ has ground interpolation. 
\end{description}
The extension 
${\mathcal T}_1 = {\mathcal T}_0 \cup {\mathcal K}$ of 
${\mathcal T}_0$ has the following properties: 
\begin{description}
\item[Assumption 4:] ${\mathcal T}_1$ is a local extension of ${\mathcal T}_0$; and 
\item[Assumption 5:] ${\mathcal K}$ consists of the following type of 
combinations of clauses:
\begin{eqnarray*}
\left\{ \begin{array}{l} x_1 \, R_1 \, s_1 \wedge \dots 
\wedge x_n \, R_n \, s_n \rightarrow 
f(x_1, \dots, x_n) \, R \, g(y_1, \dots, y_n) \\ 
x_1 \, R_1 \, y_1 \wedge \dots 
\wedge x_n \, R_n \, y_n \rightarrow 
f(x_1, \dots, x_n) \, R \, f(y_1, \dots, y_n) \end{array} \right.
\end{eqnarray*}
where $n \geq 1$, $x_1, \dots, x_n$ are variables, $R_1, \dots, R_n, R$ 
are binary relations, $R_1, \dots, R_n \in P$, $R$ is transitive, 
and each $s_i$ is either a variable 
among the arguments of $g$, or a term of the form $f_i(z_1, \dots, z_k)$, 
where $f_i \in \Sigma_1$ and all the arguments of $f_i$ are 
variables occurring  among the arguments of $g$.  
\end{description}
Because of the presence of several function symbols in the axioms in 
${\cal K}$ we need to define a more general 
notion of ``shared function symbols''. 
\begin{definition}[Shared function symbols] 
We define a relation $\sim$ between extension functions, where 
$f \sim g$ if $f$ and $g$ occur in the same clause in ${\mathcal K}$. 
We henceforth consider that a function $f \in \Sigma_1$ is 
common to  $A$ and $B$ if there exist $g, h \in \Sigma_1$ 
such that $f \sim g$, $f \sim h$, $g$ occurs in $A$ and $h$ occurs in $B$.
\label{shared}
\end{definition}
\begin{theorem}
Assume that the theories ${\cal T}_0$ and ${\cal T}_0 \cup {\cal K}$ 
satisfy Assumptions 1--5. 

For every conjunction $A \wedge B$ of ground unit clauses in the signature 
$\Pi^c$ of ${\mathcal T}_1$ (possibly containing additional constants)
with $A \wedge B \models_{{\mathcal T}_1} \perp$ a ground interpolant $I$ 
for $A \wedge B$ exists. 
In \cite{Sofronie-ijcar-06,Sofronie-lmcs-08} a procedure for 
hierarchically computing interpolants is given.

\medskip
\noindent If in addition ${\cal T}_0$ is strongly $P$-interpolating and the 
interpolants for conjunctions of ground literals are 
again conjunctions of ground literals, the same is true in the extension. 
\end{theorem}
The theory ${\cal T}_0$ of bounded semilattices 
has the following properties (cf.\ \cite{Sofronie-ijcar-06,Sofronie-lmcs-08}):
\begin{itemize}
\item it is convex w.r.t. $\approx$ and $\leq$; 
\item it is strongly $P$-interpolating w.r.t.\ $\leq$ and separating 
terms can be effectively computed; 
\item it has ground interpolation (in fact, the equational interpolation property (cf.\ \cite{Sofronie-lmcs-08})). 
\end{itemize}
Thus, Assumptions 1, 2 and 3 above are fulfilled.
The class $SLO_{\Sigma}(RI)$ of all semilattices with monotone 
operators which satisfy a set $RI$ of axioms satisfies also Assumptions 
4 and 5 provided that  $RI$ contains (flat) axioms of the following types: 
$$\begin{array}{lrrcl}
\forall x ~~~~~~~~~~~~& & f(x) & \leq & g(x) \\
\forall x, y & x \leq g(y)  \rightarrow & f(x) & \leq & h(y) \\
\forall x, y & x \leq g(y) \rightarrow & f(x) & \leq & y \\
\end{array}$$
as well as of the more general type:
$$\begin{array}{lrrcl}
\forall x_1, \dots, x_n & &  f(x_1, \dots, x_n) & \leq & g(x_1, \dots, x_n) \\
\forall x_1, \dots, x_n, y^k_1, \dots, y^k_n ~~~~~~& \displaystyle{\bigwedge_k} x_k \leq g_k(y^k_1, \dots y^k_{m_k}) \rightarrow & f(x_1, \dots, x_n) & \leq & g({\overline y}^1, \dots, {\overline y}^n) \\
\forall x_1, \dots, x_n, y^k_1, \dots, y^k_n & \displaystyle{\bigwedge_k} x_k \leq g_k(y) \rightarrow & f(x_1, \dots, x_n) & \leq & y \\
\end{array}$$
\begin{corollary}
The class $SLO_{\Sigma}(RI)$ has ground interpolation (in fact the equational interpolation property) and interpolants can 
be computed in a hierarchical manner.
\label{int-slo}
\end{corollary}
\begin{example}[cf.\ also \cite{Sofronie-lmcs-08}]
Let ${\mathcal T}_1 = SL \cup {\sf SGc}(f, g) \cup {\sf Mon}(f, g)$
be the extension of the theory of semilattices with two monotone functions
$f, g$ satisfying the semi-Galois condition 
$${\sf SGc}(f, g)~~~~~~~~~~~~~~~~ \forall x, y ~~~ x \leq g(y) \rightarrow f(x) \leq y.$$
Consider the following ground formulae $A$, $B$ in the signature of 
${\mathcal T}_1$: 

\bigskip
\noindent 
~~~~~~~~~~~~~$A:~~ d \leq g(a) ~\wedge~ a \leq c \quad \quad 
B:~~  b \leq d ~\wedge~ f(b) \not\leq c.$

\bigskip
\noindent 
where $c$ and $d$ are shared constants. We proved that 
${\mathcal T}_1$ is a local extension of the theory of (bounded) semilattices. 
To prove that 
$A \wedge B \models_{{\mathcal T}_1} \perp$ we proceed as follows:

\bigskip
\noindent
{\bf Step 1:} {\em Use locality.} By the locality condition, 
$A \wedge B$ is unsatisfiable w.r.t.\  
$SL \wedge {\sf SGc}(f, g) \wedge {\sf Mon}(f, g)$ iff 
$SL \wedge {\sf SGc}(f, g)[A \wedge B] \wedge 
{\sf Mon}(f, g)[A \wedge B] 
\wedge A \wedge B$ has no weak partial model in which all terms in $A$ and $B$
are defined. The extension terms occurring in $A \wedge B$ are $f(b)$ and 
$g(a)$, hence:
\begin{eqnarray*}
{\sf Mon}(f, g)[A \wedge B] & = & \{ a \leq a \rightarrow g(a) \leq g(a),~~ b \leq b \rightarrow f(b) \leq f(b) \} \\
{\sf SGc}(f, g)[A \wedge B] & = & \{ b \leq g(a) \rightarrow f(b) \leq a \}  
\end{eqnarray*}

\bigskip
\noindent 
{\bf Step 2:} {\em Flattening and purification.}
We purify and flatten the formula ${\sf SGc}(f, g) \wedge {\sf Mon}(f, g)$ by 
replacing the ground terms starting with $f$ and $g$ with new constants. 
The clauses are separated into a part containing definitions 
for terms starting with extension functions, $D_A \wedge D_B$, and a 
conjunction of formulae in the base signature, $A_0 \wedge B_0 \wedge {\sf SGc}_0  \wedge {\sf Mon}_0$.

\bigskip
\noindent {\bf Step 3:} 
{\em  Reduction to testing satisfiability in ${\mathcal T}_0$.}
As the extension $SL \subseteq {\mathcal T}_1$ is local, we have:
$$A \wedge B \models_{{\mathcal T}_1} \perp \quad \text{ iff } \quad
A_0  \wedge B_0 \wedge  {\sf SGc}_0   \wedge {\sf Mon}_0 \wedge {\sf Con}_0
\text{ is unsatisfiable w.r.t.\  } SL,$$ where
${\sf Con}_0 = {\sf Con}[A \wedge B]_0$ consists of the flattened form of 
those instances 
of the congruence axioms containing only $f$- and $g$-terms which 
occur in $D_A$ or $D_B$, and 
${\sf SGc}_0  \wedge {\sf Mon}_0$ consists of those instances 
of axioms in ${\sf SGc}(f, g) \wedge {\sf Mon}(f, g)$ 
containing only $f$- and $g$-terms which occur in 
$D_A$ or  $D_B$.

$$\begin{array}{l|ll}
\hline 
{\sf Extension} & ~~~~~~{\sf Base} \\
D_A \wedge D_B & ~A_0 \wedge B_0  \wedge {\sf SGc}_0 \wedge {\sf Mon}_0 \wedge {\sf Con}_0 & ~~~~~~~~~~~~~~~ \\
\hline 
a_1 \approx g(a) ~   & ~A_0 = d \leq a_1  \wedge a \leq c & {\sf SGc}_0   = b \leq a_1 \rightarrow b_1 \leq a  \\
b_1 \approx f(b)  & ~B_0 = b \leq d \wedge  b_1 \not\leq c & {\sf Con}_A \wedge {\sf Mon}_A = a \lhd a \rightarrow a_1 \lhd a_1, \lhd \in \{ \approx, \leq \}\\
& &  {\sf Con}_B \wedge {\sf Mon}_B =  b \lhd b \rightarrow b_1 \lhd b_1,~ \lhd \in \{ \approx, \leq \} \\
\hline 
\end{array}$$

\noindent 
It is easy to see that $A_0 \wedge B_0 \wedge {\sf SGc}_0 \wedge {\sf Mon}_0 \wedge {\sf Con}_0 $  
is unsatisfiable w.r.t.\  ${\mathcal T}_0$: 
$A_0 \wedge B_0$ entails $b \leq a_1$; together with ${\sf SGc}_0$ this 
yields $b_1 \leq a$, which together with $a \leq c$ and $b_1 \not\leq c$ 
leads to a contradiction.

In order to compute an interpolant we proceed as follows: 
Consider the conjunction $A_0 \wedge D_A \wedge B_0 \wedge D_B \wedge 
{\sf Con}[D_A \wedge D_B]_0 \wedge {\sf Mon}_0 \wedge {\sf SGc}_0$. 
The $A$ and $B$-part share the constants 
$c$ and $d$, and no function symbols. However, as $f$ and $g$ occur together 
in ${\sf SGc}$, $f \sim g$, 
so they are considered to be all shared. (Thus, the interpolant 
is allowed to contain both $f$ and $g$.)  
We obtain a separation for the clause 
$b \leq a_1 \rightarrow b_1 \leq a$ of ${\sf SGc}_0$ as follows:
\begin{itemize}
\item[(i)] We note that $A_0 \wedge B_0 \models b \leq a_1$. 
\item[(ii)] We can find an $SL$-term $t$ containing only shared 
constants of 
$A_0$ and $B_0$ such that $A_0 \wedge B_0 \models b \leq t \wedge t \leq a_1$.
(Indeed, such a term is $t = d$.) 
\item[(iii)] We show that, instead of the axiom 
$b \leq g(a) \rightarrow f(b) \leq a$, 
whose flattened form is in ${\sf SGc}_0$, we can use, without loss 
of unsatisfiability:
\begin{quote}
\begin{itemize}
\item[(1)] an instance of the monotonicity axiom for $f$: 
$b \leq d \rightarrow f(b) \leq f(d)$, 

\item[(2)] another instance of ${\sf SGc}$, namely: 
$d \leq g(a) \rightarrow f(d) \leq a$. 
\end{itemize}
\end{quote}
For this, we introduce a new constant $c_{f(d)}$ for $f(d)$
(its definition, $c_{f(d)} \approx f(d)$, is stored in a set $D_T$), 
and 
the corresponding instances ${\mathcal H}_{\sf sep} = {\mathcal H}^{A}_{\sf sep} 
\wedge {\mathcal H}^{B}_{\sf sep}$ 
of the congru\-ence, monotonicity and 
${\sf SGc}(f, g)$-axioms, which are now  
separated into an $A$-part 
(${\mathcal H}^{A}_{\sf sep}: d \leq a_1 \rightarrow c_{f(d)} \leq a$) and a 
$B$-part (${\mathcal H}^{B}_{\sf sep}: b \leq d \rightarrow b_1 \leq c_{f(d)}$).
We thus obtain a separated conjunction
${\overline A}_0 \wedge {\overline B}_0$  (where 
${\overline A}_0 =  {\mathcal H}^{A}_{\sf sep} \wedge A_0$ and  
${\overline B}_0 = {\mathcal H}^{B}_{\sf sep} \wedge B_0$),  
which can be proved to be unsatisfiable in 
${\mathcal T}_0 = SL$. 
\item[(iv)] To compute an interpolant in $SL$ for 
${\overline A}_0 \wedge {\overline B}_0$ 
note that 
${\overline A}_0$ is logically equivalent to the conjunction of unit 
literals 
\/ $d \leq a_1 ~\wedge~ a \leq c ~\wedge~ c_{f(d)} \leq a$
and ${\overline B}_0$ is logically equivalent to 
\/ $b \leq d ~\wedge~ b_1 {\not\leq} c ~\wedge~ b_1 \leq c_{f(d)}$. 
An interpolant  is 
$I_0 = c_{f(d)} \leq c$. 
\item[(v)] By replacing the new constants with the 
terms they denote we obtain the interpolant 
$I = f(d) \leq c$ for $A \wedge B$. 
\end{itemize}
\end{example}
An immediate consequence of Corollary~\ref{int-slo} is interpolation in 
${\cal EL}, {\cal EL}^+$ and their extensions considered in this paper.
A variant of the result for the case of ${\cal EL}$ occurs in \cite{wolter-ijcar-08}.
\begin{theorem}
${\cal EL}^+$ 
has the interpolation property, i.e.\ if 
${\cal T} \cup RI \models C \subseteq D$ then there exists a finite set 
${\cal T}_I$ of general concept inclusions containing only concept names and 
role names common\footnote{In the case of roles, by ``common'' we mean common
  or ``shared'' according to Definition~\ref{shared}.}  to 
${\cal T}$ and  $C \subseteq D$ such that ${\cal T} \cup RI \models {\cal T}_I$ and ${\cal T}_I \cup RI \models C \subseteq D$.

\noindent The same holds also for the generalization of ${\cal EL}^+$ with $n$-ary roles.
 \end{theorem}
\Proof Assume that ${\cal T} \cup RI \models C \subseteq D$.  Then  
$SLO^{\exists}_{N_R}(RI) \wedge A \wedge B \models \perp$, where 
$A = \bigwedge_{C_1 \sqsubseteq C_2} {\overline C_1} \leq {\overline C_2}$ and 
$B = {\overline C} \not\leq {\overline C}$. 
By Corollary~\ref{int-slo}, there exists a formula $I$ containing only constant names 
and role names common to $A$ and $B$ such that 
$SLO^{\exists}_{N_R}(RI) \wedge A \models I$ and 
$SLO^{\exists}_{N_R}(RI) \wedge I \wedge B \models \perp$. 
We actually showed that ${\sf SLO}_{\Sigma}(RI)$ has the equational interpolation property, so we can find an interpolant $I$ which is a conjunction 
of (positive) literals. Then ${\cal T}_I$ is this interpolant.
\QED

\section{$\mathcal{EL}^{++}$ constructors}
\label{el++}
In the definition of $\mathcal{EL}^{++}$ the following concept constructors 
are considered:
$${\sf ConcDom} ~~~~ p(f_1, \dots, f_n) = \{ x \mid \exists y_1, \dots, y_n: f_i(x) = y_i \text{ and }  p(y_1, \dots, y_n) \}.$$
Here, we show how to approach this type of problems, as well as the related concept constructions of the following type\footnote{These constructors are allowed if we allow concept construction also on the concrete domains.} (where $D_1, \dots, D_n$ are concepts terms in the concrete domains): 
$${\sf ConcDom} ~~~~ p(f_1, \dots, f_n)(D_1, \dots, D_n) = \{ x \mid \exists y_1 \in D_1, \dots, y_n \in D_n: f_i(x) = y_i \text{ and }  p(y_1, \dots, y_n) \}$$ 
within the framework of 
locality. Note that the following transfer of locality results holds:
\begin{theorem}
Let ${\cal T}_0$ be a theory and let ${\cal T}_0'$ be another theory, in the 
same signature $(\Sigma_0, {\sf Pred})$, 
with the property that every model of ${\cal T}_0'$ is a model 
of ${\cal T}_0$. Let $\Sigma_1$ be an additional set of function symbols, 
not contained in the signature  of  ${\cal T}_0$, and 
let ${\cal K}$ be a set of clauses over the signature 
$(\Sigma_0 \cup \Sigma_1, {\sf Pred})$. If the extension ${\cal T}_0 \subseteq 
{\cal T}_0 \cup {\cal K}$ has the property that every model in ${\sf PMod}_w(\Sigma_1, {\cal T}_0 \cup {\cal K})$ weakly embeds into a total model of 
${\cal T}_0' \cup {\cal K}$ then 
every model in ${\sf PMod}_w(\Sigma_1, {\cal T}_0' \cup {\cal K})$ weakly embeds into a 
total model of ${\cal T}_0' \cup {\cal K}$. 
\label{transfer}
\end{theorem}


\begin{theorem}
Assume that the only concept constructors are 
intersection, existential restriction, and ${\sf ConcDom}$.
Let 
${\mathcal C} {=} GCI {\cup} CD {\cup} RI$  be a  CBox containing a set 
$GCI$ of general concept inclusions, a set $CD$ of definitions of domains 
$\{ c_1, \dots, c_k \}$ using rules in ${\sf ConcDom}$:
$$ C_k = p_k(f^k_1, \dots, f^k_{n_k})$$
and a set $RI$ of (guarded) 
role inclusions. Assume that the only concepts names that appear are
$N_C {=} \{ C_1, {\dots}, C_n \}$.
Then for all concept descriptions $D_1, D_2$ the following are equivalent:
\begin{itemize}
\item[(1)] $D_1 {\sqsubseteq}_{\mathcal C} D_2$. 
\item[(2)] $\left( \bigwedge_{C {\sqsubseteq} D \in GCI} 
\overline{C} {\leq} \overline{D} \right) \wedge 
\overline{D_1} {\not\leq} \overline{D_2}$ is unsatisfiable w.r.t.\ 
the class $SetBAO(c_1, \dots, c_n)(RI)$ of all Boolean algebras of sets  
with monotone operators satisfying $RI_a$ (of the form 
${\cal P}({\bf D}) = ({\cal P}(D), \cap, \cup, \neg, \emptyset, 
D, \{ f_r \}_{r \in N_R}, c_1, \dots, c_k)$). 
\item[(3)] $\left( \bigwedge_{C {\sqsubseteq} D \in GCI} 
\overline{C} {\leq} \overline{D} \right) \wedge 
\overline{D_1} {\not\leq} \overline{D_2}$ is unsatisfiable w.r.t.\ 
the class $SetSL(c_1, \dots, c_n)(RI)$ of all semilattices of sets 
with monotone 
operators (i.e.\ semilattices of the form 
${\cal P}({\bf D}) = ({\cal P}(D), \cap, \emptyset, D, \{ f_r \}_{r \in N_R}, c_1, \dots, c_k)$) which satisfy $RI_a$.  
\end{itemize}
\label{set-sem}
\end{theorem}
\Proof (2) $\Rightarrow$ (1) follows from 
the definition of $D_1 {\sqsubseteq}_{\mathcal C} D_2$, and 
(3) $\Rightarrow$ (2) is immediate.
To prove that (1) $\Rightarrow$ (3), assume that (1) holds and (3) does not. 
Then there would exist a model  
${\cal P}({\bf D}) = ({\cal P}(D), \cap, \emptyset, D, \{ f_r \}_{r \in N_R}, c_1, \dots, c_k) \in SetSL(c_1, \dots, c_n)(RI)$ of 
$$G = \left( \bigwedge_{C {\sqsubseteq} D \in GCI} 
\overline{C} {\leq} \overline{D} \right) \wedge 
\overline{D_1} {\not\leq} \overline{D_2}.$$
Then  
${\cal P}({\bf D}) = ({\cal P}(D), \cap, \cup, \emptyset, D, \{ f_r \}_{r \in N_R}, c_1, \dots, c_k) \in SetSL(c_1, \dots, c_n)(RI)$ is a model of $G$. 
As the set of maximal filters of ${\cal P}({\bf D})$ is in bijective 
correspondence with $D$, the canonical definition of 
relations associated with the monotone functions $f_r$ on the Stone dual 
of  ${\cal P}({\bf D})$ induces a model ${\cal I} = (D, \cdot^{\cal I})$
which satisfies $G$, $RI$ and also $CD$. This contradicts (1). \QED

\medskip
\noindent 
We now show that $SetSL(c_1, \dots, c_n)(RI)$ is a local extension of 
$SetSL(c_1, \dots, c_n)$. We use the criterion in Theorem~\ref{transfer}.
\begin{lemma}
Let ${\bf S} = (S, \wedge, 0, 1, \{ f_r \}_{r \in \Sigma})$ be a bounded 
semilattice with partial unary functions $f_r$ weakly satisfying 
the monotonicity axioms and the $RI$ axioms. Then 
${\bf S}$ weakly embeds into a total semilattice of sets with 
monotone operators satisfying the axioms $RI_a$. 
\label{emb-sets}
\end{lemma}
\Proof By the proof of Theorem~\ref{local-alg-el+}, 
${\bf S}$ weakly embeds into the total semilattice reduct 
(in $SLO_{\Sigma}$) of the distributive lattice  
$L = {\cal OI}({\bf S}) \in DLO^{\exists}_{N_R}(RI)$. 
We can now use the proof of the last part in 
Lemma~\ref{embeddings-slo-dlo-bao} 
to show that if ${\cal F}_p$ is the set of prime filters 
of $L$ then the Boolean algebra of sets $B(L) = ({\cal P}({\cal F}_p),
\cap, \cup, \emptyset, {\cal F}_p, \{ \overline{f}_{\exists r} \}_{r \in N_R})$
(defined in Lemma~\ref{embeddings-slo-dlo-bao}) is a Boolean algebra in $BAO_{N_R}^{\exists}(RI)$. \QED

\medskip
\noindent 
We therefore can hierarchically reduce the problem of checking  if 
$D_1 {\sqsubseteq}_{\mathcal C} D_2$ as follows:
\begin{corollary}
Assume that the only concept constructors are 
intersection, existential restriction, and ${\sf ConcDom}$.
Let 
${\mathcal C} {=} GCI {\cup} CD {\cup} RI$  be a  CBox containing a set 
$GCI$ of general concept inclusions, a set $CD$ of definitions of domains 
$\{ c_1, \dots, c_k \}$ using rules in ${\sf ConcDom}$, as:
$$ c_k = p_k(f^k_1, \dots, f^k_{n_k})$$
 and sets $RI$, $GRI$ of (guarded) 
role inclusions. Assume that the concepts names that appear are 
$N_C {=} \{ C_1, \dots, C_n \}$.
 Then for all concept descriptions $D_1, D_2$ the following are equivalent:
\begin{itemize}
\item[(1)] $D_1 {\sqsubseteq}_{\mathcal C} D_2$. 
\item[(2)] $CD \wedge G$ --- where   
$ G = \left( \bigwedge_{C {\sqsubseteq} D \in GCI} 
\overline{C} {\leq} \overline{D} \right) \wedge 
\overline{D_1} {\not\leq} \overline{D_2}$ --- is unsatisfiable w.r.t.\ 
the class $SetSL(c_1, \dots, c_n)(RI)$ of all semilattices of sets 
with monotone 
operators satisfying $RI_a$, of the form 
${\cal P}({\bf D}) = ({\cal P}(D), \cap, \emptyset, D, \{ f_r \}_{r \in N_R}, c_1, \dots, c_k)$. 
\item[(3)] $CD \wedge G_0 \wedge RI[G]_0 \wedge {\sf Con}_0 \wedge {\sf Def}$
 is unsatisfiable w.r.t.\ 
the class $SetSL(c_1, \dots, c_n)(RI)$ of all semilattices of sets 
with monotone 
operators satisfying $RI_a$, of the form 
${\cal P}({\bf D}) = ({\cal P}(D), \cap, \emptyset, D, \{ f_r \}_{r \in N_R}, c_1, \dots, c_k)$. 
\item[(4)] $CD_0 \wedge G_0 \wedge RI[G]_0 \wedge {\sf Mon}_0$ is unsatisfiable
w.r.t.\ the extensions with free function symbols $\{ f_1, \dots, f_n \}$ 
of the many-sorted disjoint combination $(SetSL, {\sf Dom})$ 
of the theory $SetSL$ of sets with intersection and 
the theory ${\sf Dom}$ of the concrete domains. 
\end{itemize}
\label{set-red}
\end{corollary}  
\Proof (1) and (2) are equivalent by Theorem~\ref{set-sem}. 
It is obvious that (3) implies (2). 
We show that (2) implies (3). Assume that  
$CD \wedge G_0 \wedge RI[G]_0 \wedge {\sf Con}_0 \wedge {\sf Def}$
has a (partial) model 
${\bf S} = {\cal P}({\bf D}) = ({\cal P}(D), \cap, \emptyset, D, 
\{ f_r \}_{r \in N_R}, c_1, \dots, c_k)$.
By Theorem~\ref{emb-sets}, 
${\bf S}$ weakly embeds into a semilattice with operators ${\bf S'} = {\cal P}({\bf D'}) = ({\cal P}(D'), \cap, \emptyset, D, 
\{ f_r \}_{r \in N_R})$ which satisfies $RI \cup GRI$ 
(the interpretation of the constants is 
translated too). Then ${\bf S'}$ is also a model of $G_0, CD_0$ and 
${\sf Def}$, hence of $G \wedge CD$. Contradiction. 
The equivalence of (3) and (4) follows as a special case of 
Theorem~\ref{lemma-rel-transl}. \QED


\section{Conclusions}

In this paper we have shown that subsumption problems in $\mathcal{EL}$ can be 
expressed as uniform word problems in classes of semilattices with 
monotone operators, and that subsumption problems in $\mathcal{EL}^+$ can be 
expressed as uniform word problems in classes of semilattices with 
monotone operators satisfying certain composition laws. 
This allowed us to obtain, in a uniform way, 
PTIME decision procedures for  $\mathcal{EL}$, $\mathcal{EL}^+$,
and  extensions thereof. The use of the notion of local theory extensions 
allowed us to present a new family of PTIME (many-sorted) logics which 
extend $\mathcal{EL}$ with $n$-ary roles, (guarded) role inclusions, 
existential role restrictions and/or with numerical domains. 
These extensions are different  from other types of extensions studied in 
the description logic literature such as extensions with $n$-ary existential 
quantifiers (cf.\ e.g.\ \cite{Baader2005-ki}) or with concrete domains 
\cite{Baader-ijcai-2005}, but are, in our opinion, very natural and 
very likely to occur in ontologies. 
Moreover, we showed that the results in this 
paper can also be used for the extension ${\cal EL}^{++}$ introduced 
in \cite{Baader-ijcai-2005} (it seems that the results on ${\cal EL}^{++}$
can be extended to tackle also ABoxes). In the future we would like to also 
analyze generalizations of existential concept restrictions in ${\cal EL}$ 
to existential relation restrictions of the form
$\exists r.r1$ interpreted as 
$$ \{ x \mid \exists x_1, x_2: r(x, x_1, x_2) \wedge r_1(x_1, x_2) \},$$
implications of the form: 
$$ r_1(x, {\overline y}) \wedge r_2(x, {\overline y}) \rightarrow r_3(x, {\overline y})$$ 
and guarded role inclusions of the form:
$$ r(x_1, x_2) \wedge r_1(x, x_1, x_2) \rightarrow r_2(x, x_1, x_2).$$
We also showed that the results in \cite{Sofronie-ijcar-06} can be used to 
prove that the class of 
semilattices with monotone operations satisfying the types of axioms 
considered here allows ground (equational) 
interpolation. We used this for proving interpolation properties in 
extensions of $\mathcal{EL}^+$. We would like to further explore the area 
of applications of such results for efficient (modular) 
reasoning in combinations of ontologies based on 
extensions of $\mathcal{EL}$ and $\mathcal{EL}^+$.

\smallskip
\noindent {\bf Acknowledgments.} We thank St{\'e}phane Demri and 
Michael Zakharyaschev for asking the right questions 
and Carsten Ihlemann for his comments on a previous version of 
the paper.


\end{document}